\documentclass[12pt]{article}
\usepackage{amsmath,geometry,color}
\definecolor{grey}{rgb}{0.75,0.75,0.75}
\definecolor{orange}{rgb}{1.0,0.5,0.5}
\definecolor{brown}{rgb}{0.5,0.25,0.0}
\definecolor{pink}{rgb}{1.0,0.5,0.5}
\usepackage[english]{babel}
\usepackage{hyperref}
\usepackage{amsmath}
\usepackage{amsfonts}
\usepackage{amssymb}
\renewcommand{\theequation}{\thesection.\arabic{equation}}
\begin{document}
\title{
{\bf{} Constrained   BRST--BFV   Lagrangian Formulations   for
Higher Spin Fields in Minkowski Spaces}}
\author{\sc A.A.~Reshetnyak\thanks{reshet@ispms.tsc.ru}
\\
\small{\em Institute of
 Strength Physics and Materials Science of SB RAS,} \small{\em Tomsk 634055,
Russia}
  }
\date{}
\maketitle
\thispagestyle{empty}
\begin{abstract}

BRST--BFV method  to construct   constrained Lagrangian
formulations  for  (ir)reducible half-integer higher-spin Poincare
group representations  in Minkowski space is suggested. The
procedure is  derived by two ways: first, from  the unconstrained
BRST--BFV method for  mixed-symmetry higher-spin fermionic fields
subject to an arbitrary Young tableaux with $k$ rows (suggested in
Nucl. Phys. B 869   (2013) 523, arXiv:1211.1273[hep-th]) by
extracting  the second-class constraints  subsystem,
$\widehat{O}_\alpha=(\widehat{O}_a, \widehat{O}^+_a)$,  from a
total superalgebra of constraints, second,  in self-consistent way
by means of finding BRST-extended initial off-shell algebraic
constraints, $\widehat{O}_a$. In both cases, the latter
constraints supercommute on the constraint surface with
constrained BRST operator $Q_C$ and  spin operators $\sigma^i_C$.
The  closedness of the superalgebra   $\{Q_C, \widehat{O}_a,
\sigma^i_C\}$ guarantees   that the final gauge-invariant
Lagrangian formulation is compatible with the off-shell algebraic
constraints  $\widehat{O}_a$ imposed on the field  and gauge
parameter vectors   of the Hilbert space not depending from  the
ghosts and conversion auxiliary oscillators related to
$\widehat{O}_a$,  in comparison with the vectors  for
unconstrained BRST--BFV Lagrangian formulation. The suggested
constrained  BRST--BFV approach   is valid for both massive HS
fields and integer HS fields in the second-order formulation.  It
is shown that the respective constrained and unconstrained
Lagrangian formulations for (half)-integer HS fields with a given
spin are equivalent.    The constrained Lagrangians in
ghost-independent and component (for  initial spin-tensor field)
are obtained and shown to coincide with the Fang--Fronsdal
formulation for totally-symmetric HS field with respective
off-shell gamma-traceless constraints. The triplet and
unconstrained quartet Lagrangian formulations for the latter field
are  derived.  The constrained BRST--BFV methods without off-shell
constraints describe reducible half-integer HS Poincare group
representations with multiple spins as a  generalized triplet and
provide a starting point for constructing unconstrained Lagrangian
formulations by using the generalized quartet mechanism. A
gauge-invariant Lagrangian constrained description for a  massive
spin-tensor field of spin $n+1/2$  is obtained using a set  of
auxiliary Stueckelberg spin-tensors. A concept of BRST-invariant
second-class constraints for dynamical systems with mixed-class
constraints is suggested, leading to equivalent (w.r.t. the
BRST--BFV prescription) results of quantization both at the
operator level and in terms of the partition function.
\end{abstract}


\section{Introduction }\label{intr}

Higher-spin (HS) field theory, in its various reflections,  has
been under a long and intense study in order to re-analyse the
problems of a unified description for the variety of elementary
particles, thereby discovering approaches to a unification of the
known and new possible fundamental interactions, and remains part
of the LHC experimental program. HS field theory is in close
relation to superstring theory due to its tensionless limit
\cite{tensionlessl}, which uses a BRST operator to handle an
infinite set of HS fields with integer and half-integer
generalized spins in $d$-dimensional ($d\geq 4$) space-time and
incorporates HS field theory into superstring theory, providing
the consideration of HS theory as an instrument for investigating
the structure of superstring theory (for the current status of HS
field theory and recent advances in HS theory, see the reviews
\cite{reviews}, \cite{reviewsV}).  A powerful systematic tool used
to reconstruct Lagrangian formulations for HS fields described by
(spin-)tensor fields as elements of an irreducible representation
of the Poincare $ISO(1,d-1)$ or (anti)-de-Sitter ((A)dS) groups in
the respective $d$-dimensional Minkowski or (A)dS space-time by
means of second- or first-order non-Lagrangian equations for free
propagating HS integer and half-integer fields, respectively, is
based on a BRST construction, developed initially for HS fields
with lower spins \cite{firstBRST} in $\mathbb{R}^{1,d-1}$,
extended for higher integer spin fields in $\mathbb{R}^{1,d-1}$
\cite{Pashnev1}, \cite{Pashnev2}, \cite{Pashnev3} and
(A)dS${}_{d}$ \cite{Pashnev4}, \cite{BuchPashnev}, and for the
fields with higher half-integer spin\cite{totfermiMin},
\cite{totfermiAdS} (for review, see \cite{reviews3}). This
approach is based on the BRST--BFV method \cite{BFV}, \cite{bf},
originally developed to solve the problem of Hamiltonian
quantization for dynamical systems with first-class constraints in
Yang--Mills theories, suggested in \cite{FaddeevH}, and is usually
known as the BRST construction. The application of the BRST
construction to free HS field theory consists of four steps.
First, the conditions that determine representations with a given
spin are regarded as a topological (i.e., without Hamiltonian)
gauge system of first- and second-class constraints with a spin
operator in an auxiliary Fock space. Second, the subsystem of the
initial constraints, which contains only the second-class
constraints (for massless case), together with the spin operator,
is converted, while preserving the initial algebraic structure,
into a system of first-class constraints alone, in an enlarged
Fock space (for the conversion methods, see \cite{conversion},
\cite{conversion1}, \cite{conversion2}). Third, with respect to
the converted constraints, one constructs a BRST operator, which
is a more involved problem for HS fields in (A)dS spaces due to
quadratic constraint algebras \cite{Sevrin}, \cite{totBose},
\cite{ReshKuleshov}, \cite{BurdikResh}. Fourth, the Lagrangian and
the reducible gauge transformations for an HS field are
constructed in terms of the BRST operator in such a way that the
corresponding equations of motion reproduce the initial
constraints. We emphasize that the approach leads automatically to
a gauge-invariant Lagrangian description with all the necessary
auxiliary and Stuckelberg fields. Applying the BRST--BFV approach
to the HS field theory, one usually works within a metric-like
formulation due to Fronsdal's results in totally-symmetric HS
fields with integer \cite{Fronsdint} and half-integer spins
\cite{Fronsdalhalfint}, whereas in the frame-like formulation
\cite{frameVasiliev} the results for the Lagrangian dynamics of HS
fields \cite{masslessAdS} were obtained beyond this construction.

In constructing Lagrangian formulations for free and interacting
HS fields, one examines the cases of unconstrained or constrained
dynamics, which means, respectively, the absence or presence of
consistent usually off-shell holonomic (traceless, $\gamma
$-traceless, mixed-symmetry) constraints. As a rule, most of the
results in the metric- \cite{Metsaev}, \cite{AlkalaevGrig} and
frame-like formulations \cite{Skvortsov}, \cite{SkvortsovZinoviev}, \cite{Zinoviev}, \cite{BoulSkvZinint} were
obtained for off-shell constraints. There exists so-called Maxwell-like formulations for metric-like tensor fields (on flat and to some extent on (A)dS spaces)  developed originally with off-shell differential constraints on the gauge parameters and with their resolution  by means of a tower of reducible gauge transformations with higher derivatives  \cite{Maxwell-like}.  We recall that irreducible
Poincare or (A)dS group representations in constant curvature
space-times is described by mixed-symmetric (MS) HS fields with an
arbitrary Young tableaux of $k$ rows, $Y({s}_{1},...,{s}_{k})$
(symmetric basis), determined by more than one spin-like
parameters ${s}_{i}$ \cite{Labastida}, \cite{metsaevmixirrep},
and, equivalently, by mixed-antisymmetric (spin-~)tensor fields
with an arbitrary Young tableaux of $l$ columns,
$Y[\hat{s}_{1},...,\hat{s}_{l}]$ (antisymmetric basis), the
integers or half-integers $\hat{s}_{1}\geq \hat{s} _{2}\geq
...\geq \hat{s}_{l}$ having a spin-like interpretation
\cite{BurdikReshetnyak}. BRST--BFV Lagrangian formulations for
arbitrary free MS HS fields with integer and half-integer spins
were constructed in an unconstrained form (complete with all the
algebraic constraints that follow from the Lagrangian and the
tower of the respective gauge transformations after a partial
gauge fixing) in our papers \cite{BuchbinderRmix},
\cite{Reshetnyak2}. In turn, the notion \textquotedblleft
unconstrained\textquotedblright\ has numerous interpretations, and
was introduced originally as the  geometric formulations of the field equations and Lagrangians for totally-symmetric HS fields both in non-local and local (minimal) representations \cite{geomFrancia} within the metric-like formulation,  and in more symmetric form for Lagrangians
with totally-symmetric HS fields in \cite{quartmixbosemas} as a
\textquotedblleft quartet unconstrained
formulation\textquotedblright\ obtained from the triplet
formulation with off-shell algebraic constraints
\cite{FranciaSagnottitrip}. Another usage of this term is due to
Lagrangian formulations for MS HS fields in $\mathbb{R}^{1,d-1}$
\cite{Franciamix}. Constrained Lagrangians, together with minimal
BV actions for totally and mixed-symmetric HS fields with integer
spin, were studied within the so-called BRST--BV approach
\cite{Barnich}, \cite{BRST-BV2}, \cite{BRST-BV3}, which includes
the constrained BRST--BFV Lagrangian approach itself. At the same
time, no constrained BRST--BFV or BRST--BV constructions for
half-integer HS fields in constant curvature spaces have been
suggested so far, due to the yet inknown form of consistent (with
constrained first-order Lagrangians) holonomic constraints.
Moreover, explicit relations between the unconstrained and
constrained BRST--BFV approaches to Lagrangian formulations for
the same HS field with a given spin have not been established,
despite the fact that the respective Lagrangians are to describe
dynamics equivalent to the initial relations of an irreducible
representation for the HS field. The same problem (which is left
out of the paper's scope) arises as one examines unconstrained (so
far undeveloped) and constrained BRST--BV constructions for both
integer and half-integer HS fields.

The paper is devoted to solving the following problems:

\begin{enumerate}
\item Derivation of constrained BRST--BFV approaches to
constrained Lagrangian formulations for MS HS fields in
$\mathbb{R}^{1,d-1}$ with given integer and half-integer spins
from the respective unconstrained BRST--BFV approaches;

\item Derivation of constrained BRST--BFV approaches to
constrained Lagrangian formulations for integer and half-integer
MS HS fields in $\mathbb{R}^{1,d-1}$ in a self-consistent way;

\item Study of equivalence between unconstrained and constrained
BRST--BFV Lagrangian formulations for the same MS HS field;

\item Derivation of constrained gauge-invariant Lagrangian actions
for totally symmetric half-integer massless and massive HS field
in the metric formulation with a single spin-tensor, in the
triplet and unconstrained quartet forms;

\item Study of a BRST invariant extension of second-class
constraints for a general dynamical system with independent sets
of first- and second-class constraints and its application to the
quantization procedure within both the conventional path integral
approach and generalized canonical quantization.
\end{enumerate}

The organization of the paper is  as follows. In the
Section~\ref{BRST--BFVorig} we remind the crucial points of BFV
method application to the quantization of dynamical systems
subject to the first and second-class constraints  in operator and
functional integral  forms. In Section~\ref{superalg}, we briefly
review the ingredients of the unconstrained BRST--BFV method for
gauge-invariant Lagrangian formulations with free half-integer MS
HS fields in Minkowski space, which is the starting point to
obtain a constrained BRST--BFV operator, a spin operator and
off-shell algebraic constraints in
Subsection~\ref{constr-unconstr} of Section~\ref{constrBRST}.
 A self-consistent way to construct the basic elements of the
constrained BRST--BFV method is examined in
Subsection~\ref{constr-self}. Section~\ref {constr-Lagr} is
devoted to the construction of constrained Lagrangian formulations
for free massless and massive half-integer HS fields subject to
$Y(s_{1},...s_{k})$. The case of constrained Lagrangian
formulations for integer MS HS fields is examined in
Subsection~\ref{constr-Lagrint}. In Section~\ref{examples}, we
consider ghost-independent, component spin-tensor and triplet
forms of constrained Lagrangians for totally-symmetric fermionic
HS fields, in Subsection~\ref{totasymconbfv}, and quartet
unconstrained Lagrangians for the same fields, together with
massive fields, in Subsection~\ref{2totasymconbfv}.  In Conclusion,
we present a review of our basic results. Finally, in
Appendix~\ref{mixgen}, we suggest a new algorithm for quantizing
dynamical systems with mixed-class constraints.

We use the conventions of \cite{BuchbinderRmix}, \cite{Reshetnyak2}, which includes  the mostly minus signature for the metric tensor
$\eta_{\mu\nu} = diag (+, -,...,-)$, with Lorentz indices $\mu,
\nu = 0,1,...,d-1$, the relations $\{\gamma^{\mu},
\gamma^{\nu}\} = 2\eta^{\mu\nu}$ for the Dirac matrices
$\gamma^{\mu}$, the  notation $\varepsilon(A)$, $gh(A)$ for
the respective values of Grassmann parity and BFV ghost number of a
quantity $A$, and denote by $[A,\,B\}$ the supercommutator of
quantities $A, B$, which, in case they possess definite values of
Grassmann parity, is given by $[A\,,B\}$ = $AB -
(-1)^{\varepsilon(A)\varepsilon(B)}BA$.

\section{On  BRST--BFV method for dynamical systems subject to constraints}
\label{BRST--BFVorig}

Here, we briefly consider some specific points of the BRST--BFV
construction (following in part to \cite{BFV}, \cite{bf}, see as
well \cite{ff1}) as applied to the solution of the direct  problem
of generalized canonical quantization of the dynamical systems
subject to the first  and second-class  constraints in order to
calculate average expectation values  of the physical quantities
on appropriate Hilbert space in gauge-invariant way and to the
inverse problem of  reconstruction of the Lagrangian formulation
for initial non-Lagrangian equations on the (spin)- tensor  fields
when applying to the HS field theory, firstly on the free and then
on the interacting levels.

The constrained dynamical system   is described, according to Dirac proposal \cite{Dirac}, \cite{Dirac2}  by the  Hamiltonian, $H_0(\Gamma)$, and finite set of the first-class $T_A(\Gamma)=0$ and second-class constraints, $\Theta_\alpha(\Gamma)=0$, $A=(A_+,A_-)=1,...,M; M=(M_+,M_-)$, $\alpha=(\alpha_+,\alpha_-) =1,...,m;  m=(m_+,m_-)$ with Grassman gradings $\varepsilon(H_0;T_A;\Theta_\alpha)=(0; \varepsilon_A; \varepsilon_\alpha ) $ depending on the phase-space coordinates $\Gamma^p = (q^i, p_i)$, $i=1,...,n,$ $n=(n_+,n_-)$ for $(m+2M)\leq 2n$  in the $2n$-dimensional phase-space $(\mathcal{M}, \omega)$ with Grassmann-even  non-degenerate closed  2-form $\omega $, $d \omega =0$,   and  fundamental  Poisson superbrackets $\{\Gamma^p,\Gamma^q\} = \omega^{pq}$   at a fixed time instant $t$ for constant  $\omega^{pq} = -(-1)^{\varepsilon(\Gamma^p)\varepsilon(\Gamma^q)}\omega^{qp}$ and
 \begin{eqnarray}\label{firstclassconstr}
 && \{T_A,\, T_B\} = f^C_{AB}(\Gamma)T_C +  f^{\alpha\beta}_{AB}(\Gamma)\Theta_\alpha \Theta_\beta,  \quad  \{T_A, \Theta_\alpha\} =  f^C_{A\alpha}(\Gamma)T_C, \\
 \label{secclassconstr}
    && \{\Theta_\alpha,\, \Theta_\beta \} = \Delta_{\alpha\beta}(\Gamma)+f_{\alpha\beta}^\gamma(\Gamma)\Theta_\gamma, \ \ \mathrm{for} \  \ \mathrm{sdet}\|\Delta_{\alpha\beta}(\Gamma)\|\big|_{\Theta_\alpha = 0} \ne 0,  \\
   && \label{secclassconstr1}
    \ \Big(f^C_{AB}, f^{\alpha\beta}_{AB}\Big)=-(-1)^{\varepsilon_A\varepsilon_B}\Big(f^C_{BA}, f^{\alpha\beta}_{BA}\Big), \quad \Big(\Delta_{\alpha\beta}, f_{\alpha\beta}^\gamma\Big)=-(-1)^{\varepsilon_\alpha\varepsilon_\beta}\Big(\Delta_{\beta\alpha}, f_{\beta\alpha}^\gamma\Big),\\
    &&\label{hamconstr}  \{H_0,\,\Theta_\alpha\} = V^\beta_{\alpha}(\Gamma)\Theta_\beta +V^A_{\alpha}(\Gamma)T_A  ,  \quad \{H_0,\,T_A\} =V^B_A(\Gamma)T_B+V^{\alpha}_A(\Gamma)\Theta_\alpha,
  \end{eqnarray}
  with some functions  $f^C_{AB}$,  $ f^{\alpha\beta}_{AB}$,  $f^C_{A\alpha}$, $\Delta_{\beta\alpha}, f_{\beta\alpha}^\gamma$, $V^\beta_{\alpha}$, $V^A_{\alpha}$, $V^B_A$, $V^{\alpha}_A$ given on $\mathcal{M}$ and for $\varepsilon_A\equiv \varepsilon(T_A)$, $\varepsilon_\alpha \equiv \varepsilon(\Theta_\alpha)$.
The dynamics and gauge transformations   are determined by the equations,
\begin{equation}\label{Eqham}
  \partial_t \Gamma^p = \{\Gamma^p,\, H_0\}, \ T_A=0,\,  \Theta_\alpha = 0; \  \quad    \delta_\xi \Gamma^p = \{\Gamma^p,\,  T_A(\Gamma)\}\xi^A(t)
\end{equation}
with arbitrary  functions $\xi^A(t)$ (functionally independent for linear independent set of  $ T_A(\Gamma)$) meaning the necessity to introduce gauge conditions $\chi^B(\Gamma)=0$ such that:
\begin{equation}\label{propchi0}
\{\chi^A,\,\chi^B\}\vert_{T=\Theta=0}=0,\quad  \mathrm{sdet}\|\{T_A,\,\chi^B\}\|\vert_{T=\Theta=0} \ne 0
\end{equation}
   and appropriate Cauchy problem:   $\Gamma^p(t)_{\vert t=t_0} =  \Gamma^p_0$.

There exists representation for the system of mixed constraints $(T_A, \Theta_\alpha)$ in the form  $(T_{c|A}$, $\Phi_\alpha)(\Gamma_c) $ given  in the extended phase-space $\mathcal{M}_c$ with coordinates
$\Gamma^P_{c}= (\Gamma^p, \zeta^\alpha)$ for  $\varepsilon(\zeta^\alpha) = \varepsilon(\Theta_\alpha)$ being subject to  new fundamental Poisson superbrackets,
\begin{equation}\label{newoscill}
  \{\zeta^\alpha,\zeta^\beta\} = \widetilde{\omega}^{\alpha\beta}:  \ \{\zeta^\alpha,\Gamma^p\} = \{\widetilde{\omega}^{\alpha\beta},\Gamma^p\}=0; \  \widetilde{\omega}{}^{\alpha\beta} = -(-1)^{\varepsilon_\alpha \varepsilon_\beta}\widetilde{\omega}{}^{\beta\alpha},   \mathrm{sdet}\|\widetilde{\omega}{}^{\alpha\beta}\| \ne 0,
\end{equation}
such that the set of  $(T_{c|A}, \Phi_\alpha)(\Gamma_c) $  appears by the first-class constraints system, known as the \emph{converted}   constraints  in $\mathcal{M}_c$ with deformed Hamiltonian ${H}_{c|0}(\Gamma_c)$. The {converted} functions satisfy to the involution relations in  $\mathcal{M}_c$ for the system of the first-class constraints :
 \begin{eqnarray}\label{confirstclassconstr}
 &\hspace{-0.7em}& \hspace{-0.7em}\{T_{c|A},\, T_{c|B}\} = F^C_{AB}(\Gamma_c)T_{c|C} +  F^{\alpha}_{AB}(\Gamma_c)\Phi_\alpha ,  \quad \{T_{c|A}, \Phi_\alpha\} =  F^C_{A\alpha}(\Gamma_c)T_{c|C}, \\
 \label{consecclassconstr}
    &\hspace{-0.7em}&\hspace{-0.7em} \{\Phi_\alpha,\, \Phi_\beta \} = F_{\alpha\beta}^\gamma(\Gamma_c)\Phi_\gamma,   \\
    &\hspace{-0.7em}&\hspace{-0.7em}\label{conhamconstr}  \{{H}_{c|0},\,\Phi_\alpha\} = {V}^\beta_{c|\alpha}(\Gamma_c)\Phi_\beta +{V}^A_{c|\alpha}(\Gamma_c)T_{c|A}  ,  \quad \{{H}_{c|0},\,T_{c|A}\} ={V}^B_{c|A}(\Gamma_c)T_{c|B}+{V}^{\alpha}_{c|A}(\Gamma_c)\Phi_\alpha,
  \end{eqnarray}
 with, in general,   new structure functions, $F^C_{c|AB}, F^{\alpha}_{c|AB}$, $ F^C_{c|A\alpha}$, $F_{c|\alpha\beta}^\gamma$, ${V}^\beta_{c|\alpha}, {V}^A_{c|\alpha}$, ${V}^B_{c|A}, {V}^{\alpha}_{c|A}$  subject to  analogous symmetry properties as for unconverted ones  in (\ref{firstclassconstr})--(\ref{hamconstr}) and with boundary conditions:
 \begin{equation}\label{boundarycond}
   \big(T_{c|A}, \Phi_\alpha, {H}_{c|0}\big)(\Gamma_c)\vert_{\zeta = 0} = \big({T}_A, {\Theta}_\alpha, {H}_0\big)(\Gamma).
 \end{equation}
   The dynamics and gauge transformations for converted system   are determined by the equations,
\begin{equation}\label{conEqham}
  \partial_t \Gamma^P_{c} = \{\Gamma^P_{c},\, H_{c|0}\},  \quad    \delta_{\xi,\varpi} \Gamma^P_{c} = \{\Gamma^P_{c},\,  T_{c|A}(\Gamma_c)\}\xi^A(t)+\{\Gamma^P_{c},\,  \Phi_\alpha(\Gamma_c)\}\varpi^\alpha(t)
\end{equation}
with arbitrary functions $\xi^A(t), \varpi^\alpha(t)$, so that  after   introduction of the gauge conditions $\big({\chi}_c^B$, $\Xi^\beta \big)(\Gamma_c)=0$ such that: $\{C,\,D\}\vert_{{T_c}=\Phi=0}=0$, for any $C,\,D \in  ({\chi}_c^B,\,\Xi^\beta)$:
\begin{equation}\label{MFP}
\mathrm{sdet}\|\{T_{c|A},\,{\chi}_c^B\}\|\vert_{{T_c}=\Phi=0} \ne 0, \quad \mathrm{sdet}\|\{\Phi_\alpha,\,\Xi^\beta\}\|\vert_{T_c=\Phi=0} \ne 0,
\end{equation}
it leads  to solution of appropriate Cauchy problem:   $\Gamma^P_{c}(t_0) =  \Gamma^P_{c|0}$ with admissible $\Gamma^P_{c|0}$.
The crucial moment in the conversion procedure that the dynamics  of initial and converted dynamical systems are equivalent.

The quantization problem  for the dynamical system in terms of the functional integrals in both forms: initial  and converted  should  be equivalent when calculating vacuum average expectation values for the quantities  given on original phase-space  $\mathcal{M}$.

We recall that the total phase space $\mathcal{M}_{\mathrm{tot}}$,
($\mathcal{M} \subset \mathcal{M}_{\mathrm{tot}}$)  underlying the
BRST--BFV generalized canonical Hamiltonian quantization
\cite{BFV} is parameterized (for linearly independent  constraints
$T_A, \Theta_\alpha$) by the canonical
phase-space variables, $\Gamma^{\mathbf{P}}_{T}$, $\varepsilon(\Gamma^{\mathbf{P}}_{T})=\varepsilon_{\mathbf{P}}%
$,
\begin{eqnarray}
&& \Gamma^{\mathbf{P}}_{T} = \left(  \Gamma^{p}, \Gamma_{\mathrm{gh}
}\right)   ,\  \mathrm{with}  \    \Gamma_{\mathrm{gh}
} =   \left( C^{A},\,\overline{\mathcal{P}}_{A
};  \overline{C}{}_{A},\,\mathcal{P}^{A
}; \pi_{A},\,\lambda^{A} \right),\label{tpsv}\\
&& \label{tpsv1} \begin{array}{|c|cccccc|} \hline
                   &   C^{A} & \overline{\mathcal{P}}_{A
} &  \overline{C}_{A} & {\mathcal{P}}^{A
} & \pi_{A}  & \lambda^{A} \\
                    \hline
                   \varepsilon  &\varepsilon_A+1 & \varepsilon_A+1 & \varepsilon_A+1 & \varepsilon_A+1 & \varepsilon_A & \varepsilon_A \\
                   gh  & 1 & -1 & -1 & 1 & 0 & 0\\
                  \hline \end{array}
\end{eqnarray}
with canonical pairs of ghost, $C^{A},\, \overline{\mathcal{P}}_{A
} $, antighost, $\overline{C}_{A},\,\mathcal{P}^{A
}$ and   Lagrangian multipliers, $\pi_{A}, \lambda^{A}\ $  for  $\chi^A$ and $T_A$  respectively with non-vanishing fundamental Poisson superbrackets:
\begin{equation}\label{ghPbr}
  \big\{C^{A},\, \overline{\mathcal{P}}_{B
}\big\}=\delta^A_B, \ \ \big\{\overline{C}{}_{A},\, \mathcal{P}^{B
}\big\}=\delta_A^B, \ \ \big\{\pi_{A},\, \lambda^{B}\big\}=\delta^B_A.
\end{equation}
The generating functional of Green's functions for a dynamical system in
question has the form%
\begin{eqnarray}
&& Z_{\Psi}\left(  I\right)  =\int d \mu({\Gamma}_{T})\exp\left\{  \frac{i}{\hbar}\int
dt\left[  \frac{1}{2}\Gamma^{\mathbf{P}}_{T}(t){\omega}_{T|\mathbf{P}\mathbf{Q}}\dot{\Gamma}{}_{T}^{\mathbf{Q}}(t)-H_{\Psi
}(t)+I(t){\Gamma}_{T}(t)\right]  \right\}, \label{ZPI}\\
 && \  d \mu({\Gamma}_T)  = d {\Gamma}_T  \delta(\Theta) \mathrm{sdet}^{\frac{1}{2}}\| \{\Theta_\alpha, \Theta_\beta\}\|, \ d \mu({\Gamma}_T) = \prod_t d \mu({\Gamma}_T(t)),  \label{ZPI1}%
\end{eqnarray}
with functional measure $d \mu$  introduced according to \cite{Fradkinmes}, \cite{Senjonovic}  and determines the partition function $Z_{\Psi}=Z_{\Psi}\left(  0\right)  $ at
the vanishing external sources $I_{\mathbf{P}}(t)$ to ${\Gamma}_T^{\mathbf{P}}$. In (\ref{ZPI}),
integration over time is taken over the range $t_{\mathrm{in}}\leq t\leq
t_{\mathrm{out}}$; the functions of time ${\Gamma}_T^{\mathbf{P}}(t)\equiv {\Gamma}_{T}^{\mathbf{P}}$
for $t_{\mathrm{in}}\leq t\leq t_{\mathrm{out}}$\ are trajectories,
$\dot{\Gamma}{}_T^{P\mathbf{}}(t)\equiv d{\Gamma}{}_T^{\mathbf{P}}(t)/dt$; the quantities $\omega
_{T|\mathbf{P}\mathbf{Q}}=(-1)^{(\varepsilon_{\mathbf{P}}+1)(\varepsilon_{\mathbf{Q}}+1)}{\omega}_{T|\mathbf{Q}\mathbf{P}}$ compose an
even supermatrix inverse to that with the (constant) elements ${\omega}_T^{\mathbf{P}\mathbf{Q}} = \{ \Gamma_T^{\mathbf{P}}(t), \Gamma_T^{\mathbf{Q}}(t)\}_t$; the
unitarizing Hamiltonian $H_{\Psi}(t)=H_{\Psi}({\Gamma}_T(t))$ is determined by
three $t$-local functions: even-valued  $\mathcal{H}(t)$ with ${gh}(\mathcal{H})=0$,
odd-valued functions  $\Omega(t)$, with ${gh}(\Omega)=1$, and  $\Psi(t)$, with ${gh}(\Psi)=-1$, known as
the BRST--BFV operator and gauge-fixing Fermion, given by the equations in terms of Dirac superbracket \cite{Dirac}, \cite{Dirac2} constructed with help of the second-class constraints $\Theta_\alpha$: %
\begin{align}
&  H_{\Psi}(t)=\mathcal{H}(t)+ \left\{
\Omega(t), \Psi(t),\right\} _{\mathcal{D}{}t}%
 ,\ \mathrm{with}\ \ \left\{  A(t),B(t)\right\} _{\mathcal{D}{}t}=\left.  \left\{
A(\widehat{\Gamma}),B(\widehat{\Gamma})\right\}_{\mathcal{D}}  \right\vert _{\widehat{\Gamma}=\widehat{\Gamma}(t)}%
\,\   \forall  A,B\ ,\label{Hphi}\\
&  \left\{  \Omega,\Omega\right\}_{\mathcal{D}}  \simeq  0\ ,\ \ \left\{  \mathcal{H}%
,\Omega\right\}_{\mathcal{D}}  \simeq 0\ , \label{HOmega}\\
& \left\{
A({\Gamma}_T),B({\Gamma}_T)\right\}_{\mathcal{D}} = \left\{
A({\Gamma}_T),B({\Gamma}_T)\right\} - \left\{
A({\Gamma}_T),  \Theta_\alpha \right\} (\Delta^{-1})^{\alpha\beta}\left\{
  \Theta_\beta , \, B({\Gamma}_T) \right\}, \label{DiracBracket} %
\end{align}
with the boundary conditions%
\begin{equation}
\left.  \mathcal{H}\right\vert _{\Gamma_{\mathrm{gh}}=0}\simeq H_{0}\left(
\Gamma \right)  \ ,\ \ \ \left.  \frac{\delta\Omega}{\delta C^{A}%
}\right\vert _{\Gamma_{\mathrm{gh}}=0}\simeq T_A \left(
\Gamma\right), \ \ \left.  \frac{\delta\Omega}{\delta \mathcal{P}^{A}%
}\right\vert _{\Gamma_{\mathrm{gh}}=0}=  \pi_A \left(
\Gamma\right) ,  \    \frac{\delta\Omega}{\delta \overline{C}_{A}%
}= 0  . \label{bcond}%
\end{equation}
The sign  $" \simeq" $  means weak equality, modulo arbitrary linear combination of the second-class constraints $\Theta_\alpha$ \cite{Dirac}.
From equations (\ref{HOmega}) and the Jacobi identities for the Dirac
superbracket, it follows that%
\begin{equation}
\left\{  {H}_{\Psi},\Omega\right\}_{\mathcal{D}}  \simeq 0\ . \label{HunitOm}%
\end{equation}
The solutions for the generating equations (\ref{HOmega})  exist \cite{conversion1}, \cite{ff1},  \cite{battyutin_conv} in the form of series in powers of \emph{ minimal} ghost coordinates and momenta $C^A, \overline{\mathcal{P}}_A$ with use of $C \overline{\mathcal{P}}$-ordering  up to the second order in $\Gamma_{\mathrm{gh}}$:
\begin{eqnarray}\label{Hrep}
 \mathcal{H}  &\simeq & H_0 + (-1)^{\varepsilon_C} C^AV^C_{A}(\Gamma)\overline{\mathcal{P}}_C + O(C^2),   \\
   \Omega &\simeq&  C^A\Big(T_A+ \frac{1}{2}(-1)^{\varepsilon_C+\varepsilon_A} C^Bf^C_{BA}(\Gamma)\overline{\mathcal{P}}_C + O(C^2)\Big) + \pi_A \mathcal{P}^A \equiv \Omega_{\min}+ \pi_A \mathcal{P}^A ,\label{Orep}
\end{eqnarray}
which encode in $\mathcal{H}$ and
$\Omega_{\min}=\Omega_{\min}(\Gamma, \Gamma_{gh|m})$ the structure
functions with the terms  proportional to only $T_A$  from the
algebra of the constraints
(\ref{firstclassconstr})--(\ref{hamconstr}). The BRST--BFV
operator $\Omega_{\min}$ depends only on ghost coordinates and
momenta from minimal sector $\Gamma_{gh|m} = \big(C^A,
\overline{\mathcal{P}}_A\big)$ and satisfies to the equation
(\ref{HOmega}). In turn, the simple form for the quadratic in
powers of  ${\Gamma}_T^{\mathbf{P}}$ gauge fermion to be
sufficient for existence of  $Z_{\Psi}\left(  I\right)$ looks as
\begin{equation}\label{gaugefer}
  \Psi =  \overline{C}_A \chi^A(\Gamma) + \lambda^A \overline{\mathcal{P}}_A
\end{equation}
with the gauge conditions $\chi^A(\Gamma(t))=0$  satisfying to (\ref{propchi0}).

The integrand in (\ref{ZPI})  for $I=0$ is invariant with respect to the infinitesimal
BRST  transformations,  for  odd-valued  $\mu$, $\mu^2=0$ %
\begin{equation}
{\Gamma}_T^{\mathbf{P}}\rightarrow {\Gamma}_T^{\prime \mathbf{P}}={\Gamma}_T^{\mathbf{P}}\left(1+  \overleftarrow{s}\mu\right)  \ ,\ \ \ \mathrm{with}\ \ \ \ \overleftarrow{s}=\left\{
\bullet,\Omega \right\}_{\mathcal{D}}  \ , \label{Binf}%
\end{equation}
with nilpotent generator $\overleftarrow{s}$: $\overleftarrow{s}{}^2=0$ (by  virtue of Jacobi identity for Dirac superbracket and equation (\ref{HOmega})),   realized on the phase-space trajectories ${\Gamma}_T^{\mathbf{P}}(t)$ (to be solutions of  the first equations in (\ref{Eqham}), but with Hamiltonian $H_{\Psi}$, with respective Poisson bracket  and on the surface $\Theta_\alpha = 0$)  as%
\begin{equation}
{\Gamma}_T^{\mathbf{P}}(t) \rightarrow {\Gamma}_T^{\prime \mathbf{P}}(t)={\Gamma}_T^{\mathbf{P}}(t)\left(1+  \overleftarrow{s}\mu\right)  \ ,\ \ \ \mathrm{with}\ \ \ \ {\Gamma}_T^{ \mathbf{P}}(t) \overleftarrow{s}=\left\{{\Gamma}_T^{ \mathbf{P}}(t) , \Omega \right\}_{\mathcal{D}{}t}\ .\label{Binftr}%
\end{equation}
The invariance of  the integrand is due to additional (as compared for  the Poisson superbracket) property for  Dirac superbracket:  $\{F, \Theta_\alpha \}_{\mathcal{D}} \simeq 0$  for any  function $F$ given on ${M}_{\mathrm{tot}}$.

For the  converted dynamical system with   first-class constraints  $(T_{c|A}, \Phi_\alpha)(\Gamma_c) $ given  in  $\mathcal{M}_c$ the  total phase space $\mathcal{M}_{c|\mathrm{tot}}$, ($\mathcal{M}_c \subset {\mathcal{M}}_{\mathrm{tot}}$ and ${\mathcal{M}}_{\mathrm{tot}} \subset \mathcal{M}_{c|\mathrm{tot}}$)  is parameterized by the canonical  phase-space variables, $\Gamma^{\mathfrak{P}}_{c|T}$,
\begin{eqnarray}
&& \Gamma^{\mathfrak{P}}_{c|T} = \left(  \Gamma^{P}_{c}, \Gamma_{\mathrm{gh}}, \Gamma_{2|\mathrm{gh}
}\right)   ,\  \mathrm{with}  \    \Gamma_{2|\mathrm{gh}
} =   \left( C^{\alpha},\,\overline{\mathcal{P}}_{\alpha
};  \overline{C}{}_{\alpha},\,\mathcal{P}^{\alpha
}; \pi_{\alpha},\,\lambda^{\alpha} \right),\label{contpsv}\\
&& \label{tpsv2} \begin{array}{|c|cccccc|} \hline
                   &   C^{\alpha} & \overline{\mathcal{P}}_{\alpha
} &  \overline{C}_{\alpha} & {\mathcal{P}}^{\alpha
} & \pi_{\alpha}  & \lambda^{\alpha} \\
                    \hline
                   \varepsilon  &\varepsilon_\alpha+1 & \varepsilon_\alpha+1 & \varepsilon_\alpha+1 & \varepsilon_\alpha+1 & \varepsilon_\alpha & \varepsilon_\alpha\\
                   gh  & 1 & -1 & -1 & 1 & 0 & 0\\
                  \hline \end{array}
\end{eqnarray}
with account for the representation (\ref{tpsv}), (\ref{tpsv1}) and properties  (\ref{ghPbr})  for $ \Gamma_{\mathrm{gh}}$   which correspond  to only initial first-class constraints $T_A$ subsystem,   with canonical pairs of ghost, $C^{\alpha},\, \overline{\mathcal{P}}_{\alpha
} $, antighost, $\overline{C}_{\alpha},\,\mathcal{P}^{\alpha
}$ and   Lagrangian multipliers, $\pi_{\alpha}, \lambda^{\alpha} $  for  $\Xi^\alpha$ and $\Phi_\alpha$  respectively with the same  non-vanishing fundamental Poisson superbrackets as in (\ref{ghPbr}) determined now in  ${\mathcal{M}}_{c|\mathrm{tot}}$,  with respect to the coordinates $\Gamma^{\mathfrak{P}}_{c|T}$ so that $\{\Gamma_{\mathrm{gh}}, \Gamma_{2|\mathrm{gh}
}\}_{T}=0$.

The generating functional of Green's functions for the dynamical system with converted constraints has  the representation%
\begin{eqnarray}
&& \hspace{-1em} Z_{\Psi_c}\left(  I_{T}\right)  =\int d {\Gamma}_{c|T}\exp\left\{  \frac{i}{\hbar}\int
dt\left[  \frac{1}{2}\Gamma^{\mathfrak{P}}_{c|T}(t){\omega}_{cT|\mathfrak{P}\mathfrak{Q}}\dot{\Gamma}_{c|T}^{\mathfrak{Q}}(t)-
H_{{\Psi}_c
}(t)+I_{T}(t){\Gamma}_{c|T}(t)\right]  \right\}, \label{ZPIcov}%
\end{eqnarray}
and determines the partition function $Z_{\Psi_c}=Z_{\Psi_c}\left(  0\right)  $ at
the vanishing external sources ${I}_{T|\mathfrak{P}}(t)$ to $\Gamma_{c|T}^{\mathfrak{P}}$ with the same
properties as for   (\ref{ZPI}), whereas  the quantities $\omega_{cT|\mathfrak{P}\mathfrak{Q}}=(-1)^{(\varepsilon_{\mathfrak{P}}+1)(\varepsilon_{\mathfrak{Q}}+1)}{\omega}_{cT|\mathfrak{Q}\mathfrak{P}}$ compose an
even supermatrix inverse to that with  (constant) elements ${\omega}_{c|T}^{\mathfrak{P}\mathfrak{Q}} = \{ \Gamma_{c|T}^{\mathfrak{P}}(t), \Gamma_{c|T}^{\mathfrak{Q}}(t)\}_T$. Again  the
unitarizing Hamiltonian $H_{{\Psi}_c}(t)=H_{{\Psi}_c}(\Gamma_{c|T}(t))$ is determined by
three $t$-local functions:  $\mathcal{H}_c(t)$ with $(\varepsilon,{gh})(\mathcal{H}_c)=(0,0)$,
  $\Omega_c(t)$, with $(\varepsilon,{gh})(\Omega_c)=(1,1)$, and  ${\Psi}_c(t)$, with $(\varepsilon,{gh})(\Psi_c)=(1,-1)$,  defined  for  dynamical system with the  converted constraints $T_{c|A},  \Phi_\alpha$: %
\begin{align}
&  H_{\Psi_c}(t)=\mathcal{H}_c(t)+ \left\{
\Omega_c(t), \Psi_c(t),\right\} _{T|t}%
  ,\label{Hphic}\\
&  \left\{  \Omega_c,\Omega_c\right\}_{T} =  0\ ,\ \ \left\{  \mathcal{H}_c%
,\Omega_c\right\}_{T}  =  0\ , \label{HOmegac} %
\end{align}
subject to the boundary conditions%
\begin{eqnarray}
&& \left.  \mathcal{H}_c\right\vert _{\Gamma_{\mathrm{gh}}=\Gamma_{2|\mathrm{gh}}=0} =  {H}_{c|0}\left(
\Gamma_c \right)   ,\ \ \left. \left(\frac{\delta\Omega_c}{\delta C^{A}%
}, \frac{\delta\Omega_c}{\delta C^{\alpha}%
} \right)\right\vert _{\Gamma_{\mathrm{gh}}=\Gamma_{2|\mathrm{gh}}=0}= \left(T_{c|A} , \Phi_\alpha\right), \\
 \label{bcondc}\\
 && \left.  \left(\frac{\delta\Omega_c}{\delta \mathcal{P}^{A}
}, \frac{\delta\Omega_c}{\delta \mathcal{P}^{\alpha}
}\right)\right\vert _{\Gamma_{\mathrm{gh}}=\Gamma_{2|\mathrm{gh}}=0}=  \left(\pi_A  , \pi_\alpha\right), \  \    \frac{\delta\Omega_c}{\delta \overline{C}_{A}%
}=\frac{\delta\Omega_c}{\delta \overline{C}_{\alpha}%
} =0  . \nonumber%
\end{eqnarray}
From the generating equations (\ref{HOmegac})  it follows that total Hamiltonian commutes with BRST charge $\Omega_c$: $\{H_{\Psi_c}, \Omega_c\}_{T} =  0$.  The solutions for the  equations (\ref{HOmegac})  exist  in the form of series in powers of \emph{ minimal} ghost coordinates and momenta $C^A, \overline{\mathcal{P}}_A, C^\alpha, \overline{\mathcal{P}}_\alpha$ and up to the second order in $\Gamma_{\mathrm{gh}}$ looks as:
\begin{eqnarray}\label{Hrepconv}
 \mathcal{H}_c \hspace{-0.35em} &\hspace{-0.35em}=\hspace{-0.35em}& {H}_{c|0} +  \left(C^A, \,C^\alpha\right)  \left(\begin{array}{cc}
                                                                            {V}^C_{c|A} & {V}^C_{c|\alpha} \\
                                                                            {V}^\beta_{c|A} & {V}^\beta_{c|\alpha}
                                                                          \end{array}
 \right)\left(\begin{array}{c}
                                                (-1)^{\varepsilon_C} \overline{\mathcal{P}}_C           \\
                                                                            (-1)^{\varepsilon_\beta}\overline{\mathcal{P}}_\beta
                                                                          \end{array}
 \right)+ O(C^2),   \\
   \Omega_c\hspace{-0.35em} &\hspace{-0.35em}=\hspace{-0.35em} &  \hspace{-0.25em}\left(\hspace{-0.15em}C^A, C^\alpha \hspace{-0.15em}\right)\hspace{-0.25em}\Bigg[\hspace{-0.25em}\left(\hspace{-0.25em}\begin{array}{c}
                                                T_{c|A}          \\
                                                                           \Phi_\alpha
                                                                          \end{array}
 \hspace{-0.25em}\right)+ \frac{1}{2} \hspace{-0.15em}\left(\hspace{-0.25em}\begin{array}{cc}
                                                                           \big[C^B F{}^C_{BA}+C^\beta F^C_{\beta A}\big](-1)^{\varepsilon_A}\hspace{-0.25em} & \hspace{-0.25em} \big[C^B F^\gamma_{BA}+C^\beta F^\gamma_{\beta A}\big](-1)^{\varepsilon_A}   \\
                                                                             C^B F^C_{B\alpha}(-1)^{\varepsilon_\alpha} \hspace{-0.25em} & \hspace{-0.25em} C^\beta F^\gamma_{\beta \alpha}(-1)^{\varepsilon_\alpha}
                                                                          \end{array}\hspace{-0.45em}
 \right) \label{Orepconv} \\
 \phantom{\Omega_c }& \times&   \left(\begin{array}{c}
                                                (-1)^{\varepsilon_C} \overline{\mathcal{P}}_C           \\
                                                                            (-1)^{\varepsilon_\gamma}\overline{\mathcal{P}}_\gamma
                                                                          \end{array}
 \right) + O(C^2)\Bigg] + \pi_A \mathcal{P}^A+ \pi_\alpha \mathcal{P}^\alpha \ \equiv \ \Omega_{c|\min}+\pi_A \mathcal{P}^A+ \pi_\alpha \mathcal{P}^\alpha, \nonumber
\end{eqnarray}
which  encode in $\mathcal{H}_c$ and BRST--BFV operator
$\Omega_{c|\min}\equiv \Omega_{c|\min}(\Gamma_c, \Gamma_{gh|m})$
(depending  on the  ghost coordinates and moments in minimal
sector  $\Gamma_{gh|m}\equiv \left(C^A,
\overline{\mathcal{P}}_A\right.$;  $\left.C^\alpha,
\overline{\mathcal{P}}_\alpha\right)$)  the structure functions
from the   algebra of the converted constraints
(\ref{confirstclassconstr})--(\ref{conhamconstr}).

The  quadratic in powers of  ${\Gamma}^{\mathfrak{P}}_{c|T}$ gauge fermion $\Psi_c$ to be sufficient for existence of  $Z_{\Psi_c}\left(  I_{T}\right)$ may be chosen  as
\begin{equation}\label{gaugefercon}
  \Psi_c =  \left(\overline{C}_A,\, \overline{C}_\alpha\right) \left({\chi}^A_c(\Gamma_c),\, \Xi^\alpha(\Gamma_c)\right)^T
  + \left(\lambda^A,\, \lambda^\alpha\right)   \left(\overline{\mathcal{P}}_A,\,  \overline{\mathcal{P}}_\alpha\right)^T,
\end{equation}
where the upper sign "$T$" denotes the matrix  transposition and  the gauge conditions ${\chi}_c^A(\Gamma_c(t))=0$, $\Xi^\alpha(\Gamma_c(t))=0$  should satisfy to (\ref{MFP}).

The integrand in (\ref{ZPIcov})  for $I_{T}=0$ is invariant with respect to the infinitesimal
BRST  transformations,  %
\begin{equation}
{\Gamma}_{c|T}^{\mathfrak{P}}\rightarrow {\Gamma}_{c|T}^{\prime \mathfrak{P}}=\Gamma_{c|T}^{\mathfrak{P}}\left(1+  \overleftarrow{s}_c\mu\right)  \ ,\ \ \ \mathrm{with}\ \ \ \ \overleftarrow{s}_c=\left\{
\bullet,\Omega_c \right\}_{T}  \ , \label{Binfcon}%
\end{equation}
with nilpotent generator $\overleftarrow{s}_c$: $\overleftarrow{s}{}_c^2=0$,   realized on phase-space trajectories $\Gamma_{c|T}^{\mathfrak{P}}(t)$ (to be solutions of  the first equations in (\ref{Eqham}), but with Hamiltonian $H_{\Psi_c}$ and respective Poisson bracket)  analogously to the rule  (\ref{Binftr}) but with Poisson superbracket determined on $\mathcal{M}_{c|\mathrm{tot}}$ instead of Dirac one.

The equivalence of both  quantizations  for the initial  with $({T}_A, {\Theta}_\alpha , H_0)(\Gamma) $ in  $\mathcal{M}$  and converted   with $(T_{c|A}, \Phi_\alpha , {H}_{c|0})(\Gamma_c) $   in  $\mathcal{M}_{c}$  dynamical systems means  that the vacuum average expectation values for any  quantity $A(\Gamma)$  determined on the initial  phase-space  in  $\mathcal{M}$  coincide when calculating with respect to
path integrals $Z_{\Psi}$  (\ref{ZPI}) and $Z_{\Psi_c}$ (\ref{ZPIcov}):
\begin{eqnarray}
  \label{eqvacexp}
  \left\langle A\right\rangle _{\Psi }
  &=& \left\langle\left\langle A\right\rangle\right\rangle_{\Psi_c} \ \   \mathrm{ where}\\
  \label{expval1} \left\langle A\right\rangle _{\Psi }  &=& Z_{\Psi}^{-1}
  \int d \mu({\Gamma}_T) \ A(\Gamma) \exp\left\{  \frac{i}{\hbar}\int
dt\left[  \frac{1}{2}\Gamma_T^{\mathbf{P}}(t){\omega}_{T|\mathbf{P}\mathbf{Q}}\dot{\Gamma}_T^{\mathbf{Q}}(t)-H_{\Psi
}(t)\right]  \right\}, \\
  \label{expval12} \left\langle\left\langle A\right\rangle\right\rangle_{\Psi_c}  &=&    Z_{\Psi_c}^{-1}\int d {\Gamma}_{c|T}\ A(\Gamma) \exp\left\{  \frac{i}{\hbar}\int
dt\left[  \frac{1}{2}{\Gamma}_{c|T}^{\mathfrak{P}}(t){\omega}_{cT|\mathfrak{P}\mathfrak{Q}}\dot{\Gamma}_{c|T}^{\mathfrak{Q}}(t)-
H_{{\Psi}_c
}(t)\right]  \right\}.
\end{eqnarray}

On the operator level for the quantization problem let us consider, first,  a constrained dynamical system with only  second-class constraints $\Theta_\alpha$ satisfying to the relations (\ref{secclassconstr}) and (\ref{hamconstr}) for $T_A\equiv 0$.
In case of existing the splitting of $\Theta_\alpha$ (at least, locally) on two subsystems $\Theta_\alpha(\Gamma)$ $\rightarrow$  $\Theta^{\prime}_\alpha(\Gamma) = \Lambda^\beta_\alpha (\Gamma) \Theta_\beta(\Gamma)$= $\big(\theta_{\bar{\alpha}}, \theta_{\underline{\alpha}}\big)$ with non-degenerate (on the surface $\Theta_\alpha = 0 $) supermatrix of rotation of the constraints: $\mathrm{sdet}\|\Lambda^\beta_\alpha \|\ne 0 $,   for the index $\alpha$ division:  $\alpha = (\bar{\alpha}, \underline{\alpha})$ for $\bar{\alpha} = 1,...,\frac{1}{2}m$ and $\underline{\alpha}=\frac{1}{2}m+1,...,m$ such that  each subsystems $\theta_{\bar{\alpha}}$, $\theta_{\underline{\alpha}}$
  appear by the first-class constraints  ones:
  \begin{eqnarray}\label{split}
    &\hspace{-0.5em}& \hspace{-0.5em} \{\theta_{\bar{\alpha}},\, \theta_{\bar{\beta}}\} = \bar{f}{}^{\bar{\gamma}}_{{\bar{\alpha}}{\bar{\beta}}}(\Gamma) \theta_{\bar{\gamma}},\quad \{\theta_{\underline{\alpha}},\, \theta_{\underline{\beta}}\} = \underline{f}{}^{\underline{\gamma}}_{{\underline{\alpha}}{\underline{\beta}}}(\Gamma) \theta_{\underline{\gamma}}, \\
    &\hspace{-0.5em}& \hspace{-0.5em} \label{split1}\{\theta_{\bar{\alpha}},\, \theta_{\underline{\beta}}\} = \check{\Delta}_{\bar{\alpha}\underline{\beta}}(\Gamma)+\check{f}_{\bar{\alpha}\underline{\beta}}^{\bar{\gamma}}(\Gamma)\theta_{\bar{\gamma}} + \check{f}_{\bar{\alpha}\underline{\beta}}^{\underline{\gamma}}(\Gamma)\theta_{\underline{\gamma}}, \quad \{H_0, \, \theta_{\bar{\alpha}}\} = \bar{V}{}^{\bar{\beta}}_{\bar{\alpha}} \theta_{\bar{\beta}}, \quad \{H_0, \, \theta_{\underline{\alpha}}\} = \underline{V}^{\underline{\beta}}_{\underline{\alpha}} \theta_{\underline{\beta}}.
  \end{eqnarray}
In (\ref{split}), (\ref{split1})  the structure functions $\bar{f}{}^{\bar{\gamma}}_{{\bar{\alpha}}{\bar{\beta}}}$, $\underline{f}{}^{\underline{\gamma}}_{{\underline{\alpha}}{\underline{\beta}}}$, $\check{f}_{\bar{\alpha}\underline{\beta}}^{\bar{\gamma}}$, $ \check{f}_{\bar{\alpha}\underline{\beta}}^{\underline{\gamma}} $, $ \bar{V}{}^{\bar{\beta}}_{\bar{\alpha}} $, $\underline{V}^{\underline{\beta}}_{\underline{\alpha}}$ are related to ones  in (\ref{secclassconstr}), (\ref{hamconstr}) for only  subsystem of  $\Theta_\alpha$  with  representation for invertible $\|\check{\Delta}_{\bar{\alpha}\underline{\beta}}(\Gamma)\|_{\Theta = 0}$ with use of the Leibnitz rule for the Poisson brackets and  for $\varepsilon(\Lambda^\beta_{\underline{\beta}})=\varepsilon_\beta +\varepsilon_{\underline{\beta}}$:
\begin{equation}\label{split2}
 \check{\Delta}_{\bar{\alpha}\underline{\beta}}(\Gamma)  =  \Lambda^\alpha_{\bar{\alpha}} (\Gamma)  \Delta_{\alpha\beta}(\Gamma)\Lambda^\beta_{\underline{\beta}} (\Gamma)(-1)^{\varepsilon_\beta(\varepsilon_{\underline{\beta}}+1)}  \Rightarrow    \mathrm{sdet}\|\check{\Delta}_{\bar{\alpha}\underline{\beta}}(\Gamma)\|\big|_{\Theta^{\prime}_\alpha = 0} \ne 0.
\end{equation}
  In the corresponding Hilbert space $H_{\Gamma}$  [for the correspondence $\Gamma^p \to \hat{\Gamma}{}^p $: $[\hat{\Gamma}{}^p, \hat{\Gamma}{}^q \} = \imath\hbar \omega^{pq} $, $\omega^{pq}=\mathrm{const}$,    with representation respecting the division for $\Theta_\alpha$, but without
exception of the degrees of freedom  related to the second-class constraints  and with  choice of some $qp$-ordering for $\Theta_\alpha(\hat{\Gamma})$]  according to Dirac approach \cite{Dirac}, \cite{Dirac1}  should be realized by means of only the first-class operator constraints imposing to extract the physical states $|\psi\rangle \in  H^{phys}_{\Gamma}$ from Hilbert subspace  of physical vectors  $H^{phys}_{\Gamma}\subset H_{\Gamma}$:
 \begin{equation}\label{hphys}
   \theta_{\bar{\alpha}}(\hat{\Gamma}) |\psi\rangle  =0 , \  \forall |\psi \rangle  \in  H^{phys}_{\Gamma}, \ \mathrm{where} \   [\theta_{\bar{\alpha}},   \theta_{\bar{\beta}}\}\big|_{  \theta_{\bar{\alpha}}=0}=0.
 \end{equation}
The physical states should satisfy to the Schrodinger equation  with Hamiltonian not depending on  the rest constraints $\theta_{\underline{\alpha}}$, playing the role of the gauge conditions for $\theta_{\bar{\alpha}}$:
\begin{equation}\label{quantev}
  \left(\imath\hbar \partial_t -H_0(\hat{\Gamma})\vert_{\theta_{\underline{\alpha}}=0}\right) |\psi\rangle  =0.
\end{equation}

In turn, for the classically  equivalent dynamical system of converted operator first-class constraints  $( \Phi_\alpha)(\hat{\Gamma}_{c}) $   [for the correspondence  $\Gamma^P_{c} \to \hat{\Gamma}{}^P_{c}$= $(\hat{\Gamma}{}^p, \hat{\zeta}{}^\alpha)$:   with  choice of some  ordering  for the products in powers of  ${\zeta}{}^\alpha$ additional to  $qp$-ones for $\Phi_\alpha(\hat{\Gamma}_{c})$ and without $T_{c|A}$]  in $\mathcal{M}_c$ with operator of  Hamiltonian ${H}_{c|0}(\hat{\Gamma}_{c})$  (\ref{consecclassconstr}), (\ref{conhamconstr}) it is valid the

\noindent
 \textbf{Statement 1:} The  second-class constraints system $\Theta_\alpha(\hat{\Gamma})$ converted into first-class constraints one   $\Phi_\alpha(\hat{\Gamma}, \hat{\zeta})  $ with additional to $\hat{\Gamma}{}^p$ operators $\hat{\zeta}{}^{\alpha}$ (\ref{newoscill}),  whose number coincides with one of $\Theta_\alpha$,   satisfying  to the  superalgebra:    $[\hat{\zeta}{}^\alpha, \hat{\zeta}{}^\beta \} = \imath\hbar \widetilde{\omega}{}^{\alpha\beta}$ with constant   $\widetilde{\omega}^{{\alpha}\beta} $:
 \begin{eqnarray}\label{convfirclassconstr}
    \big[\Phi_\alpha(\hat{\Gamma},\hat{\zeta}),\, \Phi_\beta(\hat{\Gamma},\hat{\zeta}) \big\}  =  F_{\alpha\beta}^\gamma(\hat{\Gamma}, \hat{\zeta})\Phi_\gamma(\hat{\Gamma}, \hat{\zeta})\footnotemark
, \  \mathrm{with} \  \Phi_\alpha(\hat{\Gamma},0) = \Theta_\alpha(\hat{\Gamma}) ,
     \end{eqnarray}\footnotetext{Due to the  correspondence: $\lim_{\hbar \to 0}(\imath \hbar)^{-1}\big[\Phi_\alpha(\hat{\Gamma},\hat{\zeta}),\, \Phi_\beta(\hat{\Gamma},\hat{\zeta}) \big\}= \big\{\Phi_\alpha({\Gamma},{\zeta}),\, \Phi_\beta({\Gamma},{\zeta}) \big\}$, between the Poisson superbracket  and  the supercommutator, the operators  $F_{\alpha\beta}^\gamma(\hat{\Gamma}, \hat{\zeta})$ in (\ref{convfirclassconstr}) is proportional to $(\imath \hbar)$, that we always imply in the rest text when one passes  from the classical Hamiltonian description to the operator one.}selects the same set of the physical states in $H_{\Gamma}$  as  the converted constraints $\Phi_\alpha(\hat{\Gamma},\hat{\zeta}) $ select from  $H_{\Gamma}\otimes H_{\zeta}$:
    \begin{eqnarray}\label{equiv21}
      H_{\Gamma}^{phys}& =& \big\{|\psi\rangle| \  \theta_{\bar{\alpha}}(\hat{\Gamma}) |\psi\rangle =0 , \  |\psi\rangle\in H_{\Gamma} \big\} \\
       &=&  \big\{|\chi\rangle| \  \Phi_{{\alpha}}(\hat{\Gamma},\hat{\zeta}) |\chi\rangle =0 , \  |\chi\rangle\in H_{\Gamma}\otimes H_{\hat{\zeta}}\big\}. \nonumber
    \end{eqnarray}
       \vspace{1ex}

Note, first, the   operator functions $F_{\alpha\beta}^\gamma(\hat{\Gamma}, \hat{\zeta})$ in (\ref{convfirclassconstr})  obey to the  properties analogous to ones for $f_{\alpha\beta}^\gamma(\hat{\Gamma})$ (\ref{secclassconstr1}) and there are no anomalies in the right-hand side of (\ref{convfirclassconstr}) which  in opposite case should be proportional to $\hbar^2 D_{\alpha\beta}(\hat{\Gamma},\hat{\zeta})$. Second,  for  special (but interesting) cases the additional phase-space operators $\hat{\zeta}^{\alpha}$ may be chosen,
as respecting the division of the second-class constraints: $\Theta^{\prime}_\alpha = (\theta_{\bar{\alpha}},   \theta_{\underline{\alpha}})$ as follows $\hat{\zeta}^\alpha = (  q^{\underline{\alpha}},\,p^{\bar{\alpha}})$: $[q^{\underline{\alpha}}, p^{\bar{\alpha}}   \}=\imath\hbar\delta^{\bar{\alpha}\underline{\alpha}}$,  in particular, as for the  additional operators (case of additive conversion) $ \Phi_\alpha(\hat{\Gamma}_c)=  \Theta_\alpha(\hat{\Gamma})+ \vartheta_\alpha(\hat{\zeta}) $ for $[\Theta_\alpha,\, \vartheta_\beta \}=0$,   for Fock space $H_{\hat{\zeta}}$  with oscillators
\begin{equation}\label{auxosc}
\left( B^{a},   B^{a+}\right) \equiv  \frac{1}{\sqrt{2}}\left( p^{\bar{\alpha}}- \frac{\imath}{\hbar} q^{\underline{\alpha}},\,  p^{\bar{\alpha}}+ \frac{\imath}{\hbar} q^{\underline{\alpha}}\right) \ \mathrm{for }\  a = 1,..., \frac{1}{2}m,
\end{equation}
    satisfying to  $\big[B^{a},   B^{b+}\big\}=\delta^{ab}$.  Third, the presentation (\ref{equiv21}) permits  the boundary conditions for any $ |\chi\rangle \in H_{\Gamma}\otimes H_{\zeta}$:  $ |\chi\rangle\big|_{\zeta=0}= |\psi\rangle \in H_{\Gamma} $. Fourth, the quantum evolution of the converted constrained system
is described by the Schrodinger equation analogous to (\ref{quantev})
\begin{equation}\label{quantev1}
  \left(\imath\hbar \partial_t -H_{c|0}(\hat{\Gamma})\right) |\chi\rangle  =0.
\end{equation}
Fifth, the Statement 1 guarantees a preservation of explicit Poincare covariance for field-theoretic models,  like  QED, gravity,  models with  HS  fields when working with converted constraints.

For proper gauge dynamical system with some operator first-class constraints system $T_A(\hat{\Gamma})  $ only  (without reference to any  second-class constraints)  it is valid
the following
\vspace{1ex}

\noindent
 \textbf{Statement 2:} Nilpotent  Grassman-odd BRST--BFV operator $Q=\hat{C}{}^A T_A(\hat{\Gamma}) + "more"$, $Q\equiv \Omega_{\min}(\hat{\Gamma},  \hat{\Gamma}_{gh|m})$ from (\ref{Orep})  with   $gh(Q)=1$  constructed with respect to the system of $T_A(\hat{\Gamma})  $ in the Hilbert space $H_{Q}=H_{\Gamma}\otimes H_{gh|m}$  admitting  $Z$-grading: $H_{Q}= \sum_k H^k_{Q}$,  $gh(|\chi^k\rangle) =-k, $ for any $|\chi^k\rangle \in H^k_{Q})$   with Hilbert space $H_{gh|m}$ generated by the operators   $\hat{C}^A, \hat{\overline{P}}_A$  from the minimal sector subject to  (\ref{tpsv1})  and $[\hat{C}{}^A, \hat{\overline{P}}_{B}\}=\imath \hbar \delta^A_B$ permits to find
  the physical Hilbert subspace $\mathcal{H}_1^{phys}$    as follows:
    \begin{eqnarray}\label{equiv32}
      \mathcal{H}_1^{phys} &=& \big\{|\psi\rangle| \, T_A(\hat{\Gamma})  |\psi\rangle =0 , \  |\psi\rangle\in  H_{\Gamma} \big\} \\
       &=&  \big\{|\Psi\rangle| \,gh(|\Psi\rangle)=0,\, \   |\Psi\rangle\in   \ker {Q}\diagup {Im }\, Q \big\}. \nonumber
    \end{eqnarray}
    for the quotient of the subspace of $Q$-closed vectors ($\ker Q \in H_{Q}$) with respect to subspace of $Q$-exact ones (${Im }\, Q \in H_{Q}$). The evolution of the system is described by the Schrodinger equations (\ref{quantev}) with Hamiltonian $H_0(\hat{\Gamma})$ acting  in Hilbert space  $H_{\Gamma}$  and one with Hamiltonian $\mathcal{H}(\hat{\Gamma}, \hat{\Gamma}_{gh|m})$  (\ref{Hrep}) but in $H^0_{Q}$.
       \vspace{1ex}

When presenting operator $Q(\hat{\Gamma}, \hat{\Gamma}_{gh|m})$, $\mathcal{H}(\hat{\Gamma}, \hat{\Gamma}_{gh|m})$ one should use some ordering for ghost operators in (\ref{Hrep}), (\ref{Orep}), e.g. $\hat{C} \hat{\overline{\mathcal{P}}}$ ordering.  The  representation for $\mathcal{H}_1^{phys}$ in the second row, equivalently, in terms of BRST $Q$-complex (equivalent to chain of reducible gauge transformations for state vector: $|\chi^0\rangle$, and gauge parameters:  $|\chi^k\rangle$  $k\geq 1$) in   $H_{Q}$, may be equivalently rewritten as follows
    \begin{eqnarray}\label{equiv32f}
      \mathcal{H}_1^{phys} &=&   \big\{|\chi^0\rangle|\,   \delta|\chi^0\rangle=Q|\chi^1\rangle, ...,  \delta|\chi^{M-1}\rangle=Q|\chi^M\rangle, \delta|\chi^M\rangle=0,\, \   |\chi^k\rangle \in   H^k_{Q}\big\}
    \end{eqnarray}
 for $ k=0,...,M $ with $M$ being finite for finite number of the constraints $T_A(\hat{\Gamma})  $: $[\Phi_A] = M_++M_- $, and therefore finite degrees of $\hat{C}^A,  \hat{\overline{P}}_A$ in decomposition of arbitrary vector $|\Psi\rangle \in H(Q)$ in powers of ghosts operators. Thus, $H^{k+M}_{Q}\equiv 0$ for $k\in \mathbb{N}$, for  $M=M_++M_- $.
\vspace{1ex}

From the Statements 1, 2 it follows the (modulo description the evolution problem)

\noindent \textbf{Corollary 1:}  The physical  states   in
$H_{\Gamma}$ for the dynamical system with second-class
constraints, permitting the  division  on two sets with only
first-class constraints, $\Theta_\alpha(\hat{\Gamma}) \to
\Theta^{\prime}_\alpha(\hat{\Gamma})= \big(\theta_{\bar{\alpha}},
\theta_{\underline{\alpha}}\big)$, maybe equivalently presented
by nilpotent   BRST--BFV operator $Q_{c|2}=\hat{C}^\alpha
\Phi_\alpha(\hat{\Gamma}_c) + "more" $, $Q_c\equiv
\Omega_{c|\min}(\hat{\Gamma}_c,  \hat{\Gamma}_{gh|m})$
(\ref{Orepconv}) for $T_A\equiv 0$,  constructed with respect to
the system of converted  second-class constraints
$\Phi_\alpha(\hat{\Gamma}, \hat{\zeta})  $ in the Hilbert space
$H(Q_{c|2})=H_{\Gamma}\otimes H_{\zeta}\otimes H_{2gh|m}$ in the
form:
    \begin{eqnarray}\label{equiv31sec}
        \mathcal{H}_2^{phys}\hspace{-0,3em}& \hspace{-0,3em}=\hspace{-0,3em}& \hspace{-0,3em}\big\{|\psi\rangle| \, \theta_{\bar{\alpha}}(\hat{\Gamma}) |\psi\rangle =0 , \  |\psi\rangle\in H_{\Gamma} \big\} \\
      \hspace{-0,3em}& \hspace{-0,3em}=\hspace{-0,3em}& \hspace{-0,3em} \big\{|\chi^0\rangle|\,   \delta|\chi^0\rangle=Q_{c|2}|\chi^1\rangle, ...,  \delta|\chi^{M-1}\rangle=Q_{c|2}|\chi^M\rangle, \delta|\chi^M\rangle=0,\, \   |\chi^k\rangle \in   H^k(Q_{c|2}) \big\} ,\label{equiv31Brst}
    \end{eqnarray}
for $ k=0,...,M$, $(\varepsilon, gh)(|\chi^k\rangle)=(k\mod 2, -k)$.
\vspace{1ex}

Finally, for initial dynamical  system with mixed-class constraints $T_A, \Theta_\alpha$ and Hamiltonian $H_0$, satisfying to the operator analog of the superalgebra (\ref{firstclassconstr})--(\ref{hamconstr}), we get to

\noindent \textbf{Corollary 2:} The physical  states   in
$H_{\Gamma}$ for the dynamical system with first -
$T_A(\hat{\Gamma})$ and second-class  $\Theta_\alpha$ constraints,
permitting for the latter the  division  on two sets with only
first-class constraints, $\Theta_\alpha(\hat{\Gamma}) \to
\Theta^{\prime}_\alpha(\hat{\Gamma})= \big(\theta_{\bar{\alpha}},
\theta_{\underline{\alpha}}\big)$ subject to operator analog of
the relations (\ref{split}), maybe equivalently presented  by
nilpotent   BRST--BFV operator $Q_c=\hat{C}^\alpha
\Phi_\alpha(\hat{\Gamma}_c) + \hat{C}^A T_{c|A}(\hat{\Gamma}_c) +
"more" $, $Q_c\equiv \Omega_{c|\min}(\hat{\Gamma}_c,
\hat{\Gamma}_{gh|m})$ (\ref{Orepconv}),  constructed with respect
to the system of converted first $T_{c|A}(\hat{\Gamma}_c)$ and
second-class constraints $\Phi_\alpha(\hat{\Gamma}_c)  $ in the
Hilbert space $H(Q_c)=H_{\Gamma}\otimes H_{\zeta}\otimes
H_{gh|m}\otimes H_{2|gh|m}$ in the form:
    \begin{eqnarray}\label{equiv41sec}
        \mathcal{H}_{1,2}^{phys}& =& \big\{|\psi\rangle| \, \big(T_A(\hat{\Gamma}),\, \theta_{\bar{\alpha}}(\hat{\Gamma})\big) |\psi\rangle =(0 , 0),\  |\psi\rangle\in H_{\Gamma} \big\} \\
       &=&   \big\{|\chi^0\rangle|\,   \delta|\chi^0\rangle=Q_c|\chi^1\rangle, ...,  \delta|\chi^{N-1}\rangle=Q_c|\chi^N\rangle, \delta|\chi^N\rangle=0,\, \   |\chi^k\rangle \in   H^k(Q_c) \big\} \label{equiv41Brst}
    \end{eqnarray}
for $ k=0,...,N$.

The above Statements and Corollaries are the crucial results in
the application of the BRST--BFV method to a canonical
quantization of any dynamical system with finite degrees of
freedom subject to first- and second-class constraints. In fact,
Statement 2 was proved in the excellent textbook
\cite{Henneauxteitelboim} (see Chapter 14 and Theorem 14.7
therein). The correctness of Statement 1 is based, first of all,
on the previously established (see \cite{conversion1,conversion2})
classical equivalence of a dynamical system with second-class
constraints to the same dynamical  system with converted
first-class constraints. Secondly, the Statement is based on the
representation given by the first line of (\ref{equiv21}) for the
physical space $H_{\Gamma}^{phys}$ of a dynamical system with
second-class constraints known from Dirac's work \cite{Dirac1}.
Therefore, under the absence of anomalies in
(\ref{convfirclassconstr}) a classically equivalent dynamic al
system with converted first-class constraints leads to the same
physical space $ H_{\Gamma}^{phys}$ given by the second line of
(\ref{equiv21}), according to the same Dirac quantization concept
for a first-class constraints system, also presented in
\cite{Henneauxteitelboim} (see Chapter 13 and  Section 13.3
therein).

In case of the first-class constraints subsystem $\{T_A(\hat{\Gamma})\}$ to be closed with respect to Hermitian conjugation: $(T_A)^+ \in  \{T_A(\hat{\Gamma})\}$, the results of the Statement 2 and Corollary 2 can be refined. Namely, the property above means that the presentation
\begin{equation}\label{hermpres}
  T_A(\hat{\Gamma}) = \big(t_{\bar{a}},\,t_{\underline{a}};\, t_{e}  \big)\ \texttt{for} \ \big(t_{\bar{a}},\, t_{e}\big)^+ =  \big(t_{\underline{a}},\, t_{e}\big)
\end{equation}
holds with division of the index $A$: $A=(\bar{a}, \underline{a}, {e})$ for $\bar{a}=1,...,\frac{1}{2}(M-p)$, $\underline{a}=\frac{1}{2}(M-p)+1,...,M-p$ and $e=M-p+1,...,M$.
Therefore, the only zero-mode constraints $t_{e}$ and half from the  pairs $\big(t_{\bar{a}},\,t_{\underline{a}}\big)$ (e.g. $t_{\bar{a}}$) should be imposed to select  the physical state vectors in $H_{\Gamma}$:
  \begin{equation}\label{equiv32m}
 \mathcal{H}_1^{phys} = \big\{|\psi\rangle| \, \big(t_{\bar{a}}(\hat{\Gamma}),\, t_{e}(\hat{\Gamma})  \big)  |\psi\rangle =0 , \  |\psi\rangle\in  H_{\Gamma} \big\}
\end{equation}
  instead of the whole set of $T_A(\hat{\Gamma})$ in (\ref{equiv32}) and  then in (\ref{equiv41sec}).
The argumentation to prove the last condition is analogous to one considered for the case of conditions (from the Virasoro algebra) which have been used to select  physical states from the total Hilbert space for bosonic string or for the superstring models \cite{BrinkHenneaux}, \cite{GreenSchwarz}, but now for the finite set of constraints.

  We stress that when analyzing the presentation the
structure of the  physical states we did not considered the BRST
operator, $\Omega(\hat{\Gamma}_T)$, $\Omega_c(\hat{\Gamma}_{c|T})$
and unitarizing  Hamiltonian, $H_{\Psi}(\hat{\Gamma}_T)$,
$H_{\Psi_c}(\hat{\Gamma}_{c|T})$ for the dynamical system in
question, which act on more wider respective Hilbert spaces (than
$ H(Q),  H(Q_c)$) extended by additional operators
$\hat{\overline{C}}, \hat{\mathcal{P}} $, $\hat{\pi},
\hat{\lambda}$ from non-minimal sectors.  Therefore, it is not
necessary to operate with the gauge conditions
$\chi^A(\hat{\Gamma})$, $\Xi^\beta(\hat{\Gamma}_c)$ to select
physical states from $ H(Q)$,  $H(Q_c)$.      There exists the
question does it possible to select any physical vectors for the
dynamical system with mixed--class constraints by means of
intermediate procedure, with BRST--BFV operator $Q$ for only
first-class constraints $T_A(\hat{\Gamma})$ imposed  on  such
state, $Q|\psi\rangle=0$,   with  $|\psi\rangle \in  H^0(Q)$
together with special extension of the maximal subsystem of the
first-class constraints $\theta_{\bar{\alpha}}(\hat{\Gamma})$ from
$\Theta_{\alpha}(\hat{\Gamma})$:
$\theta_{\bar{\alpha}}(\hat{\Gamma})|\psi\rangle =0 $? We consider
the solution for this problem within application of BRST--BFV
approach for the construction of the unconstrained and constrained
Lagrangian formulations for HS fields given on
$\mathbb{R}^{1,d-1}$.

\section{BRST operator for HS symmetry superalgebra for\newline
mixed-symmetric fermionic fields}\label{superalg}
\setcounter{equation}{0}

In this section, we shortly  repeating a basic results of our  research \cite{Reshetnyak2} consider a  half-integer spin irreducible
representation of Poincare group in a  Minkowski space $\mathbb{R}^{1,d-1}$ within metric-like formalism
which is to be described by a spin-tensor field:
$\Psi_{(\mu^1)_{n_1},(\mu^2)_{n_2},...,(\mu^k)_{n_k}}
\hspace{-0.2em}$ $\equiv \hspace{-0.2em}
\Psi_{\mu^1_1\ldots\mu^1_{n_1},\mu^2_1\ldots\mu^2_{n_2},...,}$
${}_{ \mu^k_1\ldots \mu^k_{n_k}{}A}(x)$, with the Dirac index $A$
(later suppressed) of rank $\sum_{j= 1}^k n_j$
and the generalized spin
 $\mathbf{s} = (n_1 +1/2, n_2+1/2,
 ... , n_k+1/2)$ ($n_1 \geq n_2\geq ... \geq n_k>0, k \leq [(d-1)/2])$,
 which corresponds to a Young tableaux $Y(s_1,s_2,...,s_k)$ with $k$
rows of length $n_1, n_2,  ..., n_k$, respectively,
\begin{equation}\label{Young k}
Y(s_1,s_2,...,s_k) = \begin{array}{|c|c|c c c|c|c|c|c|c| c|}\hline
  \!\mu^1_1 \!&\! \mu^1_2\! & \cdot \ & \cdot \ & \cdot \ & \cdot\  & \cdot\  & \cdot\ &
  \cdot\    &\!\! \mu^1_{n_1}\!\! \\
   \hline
    \! \mu^2_1\! &\! \mu^2_2\! & \cdot\
   & \cdot\ & \cdot  & \cdot &  \cdot & \!\!\mu^2_{n_2}\!\!   \\
 \cline{1-8} \cdots & \cdot &\cdot &\cdot & \cdot & \cdot &\cdot   \\
   \cline{1-7}
     \! \mu^k_1\! &\! \mu^k_2\! & \cdot\
   & \cdot\ & \cdot  & \cdot &   \!\!\mu^k_{n_k}\!\!   \\
   \cline{1-7}
\end{array}\ ,
\end{equation}
The field is symmetric with respect to permutations of each type
of Lorentz indices
 $\mu^i$ and
obeys, respectively, to the  Dirac,
gamma-tracelessness  and mixed-symmetry equations
[for $i,j=1,...,k;\, l_i,m_i=1,...,n_i$],
\begin{eqnarray}
\label{Eq-0} &&
\imath\gamma^{\mu}\partial_{\mu}\Psi_{(\mu^1)_{n_1},(\mu^2)_{n_2},...,(\mu^k)_{n_k}}
=0\,,
\\
&& \gamma^{\mu^i_{l_i}}\Psi_{
(\mu^1)_{n_1},(\mu^2)_{n_2},...,(\mu^k)_{n_k}} =0\,,
 \label{Eq-1}\\
&& \Psi_{
(\mu^1)_{n_1},...,\{(\mu^i)_{n_i}\underbrace{,...,\mu^j_{1}...}\mu^j_{l_j}\}...\mu^j_{n_j},...(\mu^k)_{n_k}}=0,\quad
i<j,\ 1\leq l_j\leq n_j,
 \label{Eq-2}
\end{eqnarray}
and, in addition for any neighbour rows with equal length: $n_i=n_{i+1}=...=n_{i+m}, m=1,...,k-i$   the field should be identical with respect to permutation  of any two groups from the total group of the respective indices $(\mu^i)_{n_i}, (\mu^{i+1})_{n_{i+1}},...,(\mu^{i+m})_{n_{i+m}}$.
The underlined figure bracket in (\ref{Eq-2}) denotes that the indices included
in it do not take part in symmetrization, which thus concerns only
the indices $(\mu^i)_{n_i}, \mu^j_{l_j} $ in
$\{(\mu^i)_{n_i}\underbrace{,...,\mu^j_{1}...}\mu^j_{l_j}\}$.

Equivalently, in terms of the  general state (being by a Dirac-like spinor) of the Fock space, $\mathcal{H}$, generated by the $k$ bosonic pairs  of creation and annihilation operators $a^{\mu^i_{l_i}}_i, a^{+\mu^i_{l_i}}_i$, $i=1,...,k$:
\begin{eqnarray}
\label{PhysState} |\Psi\rangle &=&
\sum_{n_1=0}^{\infty}\sum_{n_2=0}^{n_1}\cdots\sum_{n_k=0}^{n_{k-1}}
\frac{\imath^{\sum_in_i}}{n_1!\times...\times n_k!}\Psi_{(\mu^1)_{n_1},(\mu^2)_{n_2},...,(\mu^k)_{n_k}}\,
\prod_{i=1}^k\prod_{l_i=1}^{n_i} a^{+\mu^i_{l_i}}_i|0\rangle, \\
& \mathrm{for} & \label{comrels}
[a^i_{\mu^i}, a_{\nu^j}^{j+}]=-\eta_{\mu^i\nu^j}\delta^{ij}\,,
\qquad \delta^{ij} = diag(1,1,\ldots 1)\,,
\end{eqnarray}
providing the symmetry property of
$\Psi_{(\mu^1)_{n_1},(\mu^2)_{n_2},...,(\mu^k)_{n_k}}$
the set of the relations (\ref{Eq-0})--(\ref{Eq-2}) are equivalent for any spin $\mathbf{s}$ to set of $(\frac{1}{2}k(k+1)+1)$ operator equations:
\begin{eqnarray}\label{t0t1t}
&&    \tilde{t}_0|\Psi\rangle =
\tilde{t}^i|\Psi\rangle = t^{i_1j_1}|\Psi\rangle =  0, \\
&& \mathrm{for} \ \Big(\tilde{t}_0\,, \tilde{t}^i \,,
   t^{i_1j_1} \Big)=\Big(-i\gamma^{\mu}\partial_\mu\,,  \gamma^{\mu}a^i_\mu\,,  a^{i_1+}_\mu
a^{j_1\mu}\Big), \quad {i_1 < j_1}.\label{totit12}
\end{eqnarray}
Adding to the relations (\ref{t0t1t}), the generalized spin  constraints imposed on $|\Psi\rangle$ in terms of number particle operator $g_0^i$:
\begin{equation}\label{gocond}
g_0^i|\Psi\rangle =\textstyle (n_i+\frac{d}{2}) |\Psi\rangle\,, \ \mathrm{with} \ g_0^i=- \frac{1}{2}\{a^{\mu^i}_i, a^{+}_{i{}\mu^i}\}
\end{equation}
the irreducible massless of spin $\mathbf{s}=\mathbf{n}+\frac{1}{2}$ Poincare group representation is equivalently to the Eqs. (\ref{Eq-0})--(\ref{Eq-2}) given by (\ref{t0t1t})--(\ref{gocond}).

The bosonic character of the primary constraint operators $\tilde{t}_0, \tilde{t}^i$,
$\varepsilon(\tilde{t}_0) = \varepsilon(\tilde{t}^i)= 0$ does not permit to obtain the second-order operator $l_0=\partial^\mu\partial_\mu$ in terms of a commutator constructed from $\tilde{t}_0$, $l_0\stackrel{?}{=} -\frac{1}{2} [\tilde{t}_0,\tilde{t}_0]$, due to $\varepsilon({\gamma}^\mu)=\varepsilon(a^\mu_i)=0$ (see footnote~5 in \cite{Reshetnyak2}). These operators are transformed into fermionic ones (originally proposed in \cite{Klishevich}, Eqs. (16)--(19) for $d=2N$,  and partially following Ref. \cite{totfermiAdS}) by means of certain $d+1$ Grassmann-odd gamma-matrix-like objects $\tilde{\gamma}^\mu$, $\tilde{\gamma}$, $\varepsilon(\tilde{\gamma}^\mu)=.\varepsilon(\tilde{\gamma})=1$, whose explicit realization differs in even, $d=2N$, and odd, $d=2N+1$, $N\in \mathbb{N}$ dimensions. In the former case, we have the following definition \cite{totfermiAdS}:
\begin{eqnarray}
&& \{\tilde{\gamma}^\mu,\tilde{\gamma}^\nu\} = 2\eta^{\mu\nu}, \qquad
\{\tilde{\gamma}^\mu,\tilde{\gamma}\}=0, \qquad
\tilde{\gamma}^2=-1, \label{tgammas} \ \ \mathrm{so}\ \mathrm{that} \ \gamma^{\mu} = \tilde{\gamma}^{\mu} \tilde{\gamma},
\end{eqnarray}
with a non-trivial realization for $\tilde{\gamma}$,
\begin{equation}\label{evengamm}
  \tilde{\gamma} = \kappa_{d}\Pi  {\gamma}_{d+1} \ \ \mathrm{for}\ \  {\gamma}_{d+1}=\frac{1}{d!}\Big(\prod_{i=1}^{d}{\gamma}_{\mu_i}\Big)\epsilon^{\mu_1\ldots \mu_d} \ \mathrm{and} \ \kappa_{d} = \left\{\begin{array}{cc}
                                                                                                                                          1 , & d=4M ,\\
                                                                                                                                          \imath , & d=4M+2,
                                                                                                                                        \end{array}
  \right.
\end{equation}
{where the completely antisymmetric Levi-Civita tensor $\epsilon^{\mu_1\ldots \mu_d}$ is normalized by $\epsilon^{01\ldots d-1}=1$, and the odd unit matrix $\Pi$: $\Pi^2=1$,\footnote{$\Pi$ is similar to the odd supermatrix $\omega = \|\omega^{\mathbf{AB}}\|$ =
$\|(\Gamma^{\mathbf{A}},\Gamma^{\mathbf{B}})\| $, $\epsilon(\omega)=1$, resulting from the odd Poisson bracket $
(\bullet,\bullet)$ calculated with respect to the field-antifield variables $\Gamma^\mathbf{A}$ of the
field-antifield formalism \cite{bv}, which was also used in \cite{reshetnyakN34} to construct an $N=3$-BRST invariant quantum action for Yang--Mills theories in the minimal configuration space.} changes the Grassmann parity of columns (or rows) alone, in such a way that $\varepsilon(\tilde{\gamma})$ = $\varepsilon({\gamma}_{d+1})+\varepsilon(\Pi)$ = $0+1=1$. Indeed,  $\tilde{\gamma}{}^2=\kappa_{d}^2\Pi^2 (-1)^{d^2/4}\prod_{i=1}^{d}{\gamma}_{\mu_i}^2$ = $\kappa_{d}^2(-1)^{d^2/4+d-1}=-1$.
In odd space-time dimensions the matrix ${\gamma}_{d+1}$ is trivial,  but the definition of $ \tilde{\gamma}$ in (\ref{evengamm}) remains valid under the following modification of the commutation relations (\ref{tgammas}):
\begin{eqnarray}
&& [{\gamma}^\mu,\tilde{\gamma}]=0, \ \tilde{\gamma}=  \Pi   \ \ \mathrm{and} \ \ \gamma^{\mu} = \tilde{\gamma}^{\mu} \tilde{\gamma} \ \ \mathrm{so}\ \mathrm{that} \ \tilde{\gamma}^2=1, \ [\tilde{\gamma}^\mu,\tilde{\gamma}]=0, \  \{\tilde{\gamma}^\mu,\tilde{\gamma}^\nu\} = 2\eta^{\mu\nu} \label{tgammasodd} .
\end{eqnarray}
In both cases, we have a realization of the Clifford algebra for Grassmann-odd
gamma-matrix-like objects $\tilde{\gamma}^\mu$ in $\mathbb{R}^{1,d-1}$. \footnote{The final Lagrangian formulation in terms of the ghost-independent or spin-tensor forms depends only on the standard Grassmann-even matrices ${\gamma}^\mu$, and does not depend on $\tilde{\gamma}^\mu, \tilde{\gamma}$, due to the presence of the latter only as even degrees inside the Lagrangian, and due to the homogeneity of the gauge transformations w.r.t. $\tilde{\gamma}$, as shown by the totally-symmetric half-integer case in Sections~\ref{totasymconbfv}, \ref{2totasymconbfv}. Therefore, $\tilde{\gamma}^\mu, \tilde{\gamma}$ may be viewed as intermediate  objects as compared to ${\gamma}^\mu$.} The primary Grassman-odd constraints result in
\begin{eqnarray}
\label{t0ti} {t}_0 = -\imath\tilde{\gamma}^\mu \partial_\mu\,,
\qquad {t}^i =  \tilde{\gamma}^\mu a^i_\mu : \quad
 \left(t_0, t^i\right) =  \tilde{\gamma}\left(\tilde{t}_0,\tilde{t}^i\right) .
\end{eqnarray}
 The set of primary constraints $\{{t}_0, {t}_i, {t}_{ij}, g_0^i\}$ is closed with respect to the $[\ ,\ ]$--multiplication if we add to them divergentless, $l_i$, traceless, $l_{ij}$, $i\leq j$   and D'alamber  operators, $l_0$:
\begin{equation} \label{lilijpr}
\Big(l_0\,,  l^i\,,  l^{ij}\Big)=\Big( \partial^\mu\partial_\mu\,,- ia^i_\mu
\partial^\mu \,, {\textstyle\frac{1}{2}}\,a^{i{}\mu}
a^j_{\mu}\Big),
\end{equation}
but for the reality of the Lagrangian  we  need, in addition,  a closedness with respect to the appropriate hermitian conjugation
 defined by means of odd scalar product in $\mathcal{H}$:
 \begin{eqnarray}
 \langle\tilde{\Phi}|\Psi\rangle & =  & \int
d^dx\sum_{n_1=0}^{\infty}\sum_{n_2=0}^{n_1}\cdots\sum_{n_k=0}^{n_{k-1}}
         \sum_{p_1=0}^{\infty}\sum_{p_2=0}^{p_1}\cdots\sum_{p_l=0}^{p_{l-1}}\frac{\imath^{\sum_in_i}(-\imath)^{\sum_jp_j}}{n_1!\times...\times n_k!p_1!\times...\times p_l!}
\nonumber\\
&& \times \langle 0|\prod_{j=1}^l\prod_{m_j=1}^{p_j}a^{\nu^j_{m_j}}_j\Phi^+_{(\nu^1)_{p_1},...,(\nu^l)_{p_l}}\tilde{\gamma}_0
\Psi_{(\mu^1)_{n_1},...,(\mu^k)_{n_k}}\,
\prod_{i=1}^k\prod_{l_i=1}^{n_i} a^{+\mu^i_{l_i}}_i|0\rangle\nonumber\\
 &=& \hspace{-0.5em}  \sum_{n_1=0}^{\infty}\sum_{n_2=0}^{n_1}\cdots\sum_{n_k=0}^{n_{k-1}}\prod_{i=1}^k \frac{(-1)^{n_i}}{{n_i!}}\int d^dx
 \Phi^+_{(\mu^1)_{n_1},...,(\mu^k)_{n_k}}\tilde{\gamma}_0
\Psi^{(\mu^1)_{n_1},...,(\mu^k)_{n_k}}, \label{sproduct}
\end{eqnarray}
that means the set:
\begin{eqnarray} \label{oA}o_A=\{t_0\,,l_0\,,  l_i\,,l^+_i\}  \ \mathrm{for} \ l^+_i=(l_i)^+=-i a^{+}_{i{}\mu}
\partial^\mu
\end{eqnarray}  composes the first-class constraints subsystem.
As a result, the superalgebra $\mathcal{A}^f(Y(k),
\mathbb{R}^{1,d-1})$, known as the \emph{half-integer HS symmetry
algebra in Minkowski space with a Young tableaux having $k$ rows}
\cite{Reshetnyak2} and containing the extra operators
\begin{eqnarray} \label{lilijt+} \hspace{-2ex} && \hspace{-2ex}
o^+_a \equiv \bigl( t^{i+},\,  l^{ij+},\, t^{i_1j_1+} \bigr) =
\bigl(\tilde{\gamma}^\mu a^{i+}_\mu,\,  \textstyle\frac{1}{2}a^{i+}_\mu a^{j\mu+},\,
a^{i_1}_\mu a^{j_1\mu+}\bigr) ,\ i\leq j;\ i_1 < j_1,
\end{eqnarray}
in addition to $o_a \equiv \bigl( t^{i},\,  l^{ij},\, t^{i_1j_1}
\bigr)$, is closed w.r.t. the Hermitian conjugation and the $[\ ,\
\}$-multiplication. The complete $[\ ,\ \}$-multiplication table
for $\mathcal{A}^f(Y(k), \mathbb{R}^{1,d-1})$ is given by
Multiplication Table~\ref{table in}, with commutators present only
in the upper subtable, and with the odd-odd part containing only
anticommutators in the subtable below.
\hspace{-1ex}{\begin{table}[t] {{\footnotesize
\begin{center}
\begin{tabular}{||c||c|c|c|c|c|c|c||c||}\hline\hline
$\hspace{-0.2em}[\; \downarrow, \rightarrow
]\hspace{-0.5em}$\hspace{-0.7em}&
 $t^{i_1j_1}$ & $t^+_{i_1j_1}$ &
$l_0$ & $l^i$ &$l^{i{}+}$ & $l^{i_1j_1}$ &$l^{i_1j_1{}+}$ &
$g^i_0$ \\ \hline\hline $t_0$
    & $0$ & $0$
   & $0$&\hspace{-0.3em}
    $0$\hspace{-0.5em} &
    \hspace{-0.3em}
    $0$\hspace{-0.3em}
    &\hspace{-0.7em} $0$ \hspace{-1.2em}& \hspace{-1.2em}$
    0$\hspace{-1.2em}& $0$ \\
\hline $t^{i_2}$
    & $-t^{j_1}\delta^{i_2i_1}$ & $-t_{i_1}\delta^{i_2}{}_{j_1}$
   & $0$&\hspace{-0.3em}
    $\hspace{-0.2em}0$\hspace{-0.5em} &
    \hspace{-0.3em}
    $-t_0\delta^{i_2i}$\hspace{-0.3em}
    &\hspace{-0.7em} $0$ \hspace{-1.2em}& \hspace{-1.2em}$
    -\frac{1}{2}t^{\{i_1+}\delta^{j_1\}i_2}\hspace{-0.9em}$\hspace{-1.2em}& $t^{i_2}\delta^{i_2i}$ \\
\hline$t^{i_2+}$
    & $t^{i_1+}\delta^{i_2j_1}$ & $t^+_{j_1}\delta_{i_1}{}^{i_2}$
   & $0$&\hspace{-0.3em}
    $\hspace{-0.2em}t_0\delta^{i_2i}$\hspace{-0.5em} &
    \hspace{-0.3em}
    $0$\hspace{-0.3em}
    &\hspace{-0.7em} $\hspace{-0.7em}\frac{1}{2}t^{\{i_1}\delta^{j_1\}i_2}
    \hspace{-0.9em}$ \hspace{-1.2em}& \hspace{-1.2em}$0\hspace{-0.9em}$\hspace{-1.2em}& $-t^{i_2+}\delta^{i_2i}$ \\
\hline\hline $t^{i_2j_2}$
    & $A^{i_2j_2, i_1j_1}$ & $B^{i_2j_2}{}_{i_1j_1}$
   & $0$&\hspace{-0.3em}
    $\hspace{-0.2em}l^{j_2}\delta^{i_2i}$\hspace{-0.5em} &
    \hspace{-0.3em}
    $-l^{i_2+}\delta^{j_2 i}$\hspace{-0.3em}
    &\hspace{-0.7em} $\hspace{-0.7em}l^{\{j_1j_2}\delta^{i_1\}i_2}
    \hspace{-0.9em}$ \hspace{-1.2em}& \hspace{-1.2em}$
    -l^{i_2\{i_1+}\delta^{j_1\}j_2}\hspace{-0.9em}$\hspace{-1.2em}& $F^{i_2j_2,i}$ \\
\hline $t^+_{i_2j_2}$
    & $-B^{i_1j_1}{}_{i_2j_2}$ & $A^+_{i_1j_1, i_2j_2}$
&$0$   & \hspace{-0.3em}
    $\hspace{-0.2em} l_{i_2}\delta^{i}_{j_2}$\hspace{-0.5em} &
    \hspace{-0.3em}
    $-l^+_{j_2}\delta^{i}_{i_2}$\hspace{-0.3em}
    & $l_{i_2}{}^{\{j_1}\delta^{i_1\}}_{j_2}$ & $-l_{j_2}{}^{\{j_1+}
    \delta^{i_1\}}_{i_2}$ & $-F_{i_2j_2}{}^{i+}$\\
\hline $l_0$
    & $0$ & $0$
& $0$   &
    $0$\hspace{-0.5em} & \hspace{-0.3em}
    $0$\hspace{-0.3em}
    & $0$ & $0$ & $0$ \\
\hline $l^j$
   & \hspace{-0.5em}$- l^{j_1}\delta^{i_1j}$ \hspace{-0.5em} &
   \hspace{-0.5em}$
   -l_{i_1}\delta_{j_1}^{j}$ \hspace{-0.9em}  & \hspace{-0.3em}$0$ \hspace{-0.3em} & $0$&
   \hspace{-0.3em}
   $l_0\delta^{ji}$\hspace{-0.3em}
    & $0$ & \hspace{-0.5em}$- \textstyle\frac{1}{2}l^{\{i_1+}\delta^{j_1\}j}$
    \hspace{-0.9em}&$l^j\delta^{ij}$  \\
\hline $l^{j+}$ & \hspace{-0.5em}$l^{i_1+}
   \delta^{j_1j}$\hspace{-0.7em} & \hspace{-0.7em}
   $l_{j_1}^+\delta_{i_1}^{j}$ \hspace{-1.0em} &
   $0$&\hspace{-0.3em}
      \hspace{-0.3em}
   $-l_0\delta^{ji}$\hspace{-0.3em}
    \hspace{-0.3em}
   &\hspace{-0.5em} $0$\hspace{-0.5em}
    &\hspace{-0.7em} $ \textstyle\frac{1}{2}l^{\{i_1}\delta^{j_1\}j}
    $\hspace{-0.7em} & $0$ &$-l^{j+}\delta^{ij}$  \\
\hline $l^{i_2j_2}$
    & \hspace{-0.3em}$\hspace{-0.4em}-l^{j_1\{j_2}\delta^{i_2\}i_1}\hspace{-0.5em}$
    \hspace{-0.5em} &\hspace{-0.5em} $\hspace{-0.4em}
    -l_{i_1}{}^{\{i_2+}\delta^{j_2\}}_{j_1}\hspace{-0.3em}$\hspace{-0.3em}
   & $0$&\hspace{-0.3em}
    $0$\hspace{-0.5em} & \hspace{-0.3em}
    $ \hspace{-0.7em}-\textstyle\frac{1}{2}l^{\{i_2}\delta^{j_2\}i}
    \hspace{-0.5em}$\hspace{-0.3em}
    & $0$ & \hspace{-0.7em}$\hspace{-0.3em}
L^{i_2j_2,i_1j_1}\hspace{-0.3em}$\hspace{-0.7em}& $\hspace{-0.7em}  l^{i\{i_2}\delta^{j_2\}i}\hspace{-0.7em}$\hspace{-0.7em} \\
\hline $l^{i_2j_2+}$
    & $ l^{i_1 \{i_2+}\delta^{j_2\}j_1}$ & $ l_{j_1}{}^{\{j_2+}
    \delta^{i_2\}}_{i_1}$
   & $0$&\hspace{-0.3em}
    $\hspace{-0.2em} \textstyle\frac{1}{2}l^{\{i_2+}\delta^{ij_2\}}$\hspace{-0.5em} & \hspace{-0.3em}
    $0$\hspace{-0.3em}
    & $-L^{i_1j_1,i_2j_2}$ & $0$ &$\hspace{-0.5em}  -l^{i\{i_2+}\delta^{j_2\}i}\hspace{-0.3em}$\hspace{-0.2em} \\
\hline\hline $g^j_0$
    & $-F^{i_1j_1,j}$ & $F_{i_1j_1}{}^{j+}$
   &$0$& \hspace{-0.3em}
    $\hspace{-0.2em}-l^i\delta^{ij}$\hspace{-0.5em} & \hspace{-0.3em}
    $l^{i+}\delta^{ij}$\hspace{-0.3em}
    & \hspace{-0.7em}$\hspace{-0.7em}  -l^{j\{i_1}\delta^{j_1\}j}\hspace{-0.7em}$\hspace{-0.7em} & $ l^{j\{i_1+}\delta^{j_1\}j}$&$0$ \\
   \hline\hline
\end{tabular}\end{center}}}
{\footnotesize
\hspace{1.6em}\begin{flushleft}\quad
\begin{tabular}{l}${}B^{i_2j_2}{}_{i_1j_1}  \  = \
  (g_0^{i_2}-g_0^{j_2})\delta^{i_2}_{i_1}\delta^{j_2}_{j_1} +
  (t_{j_1}{}^{j_2}\theta^{j_2j_1} + t^{j_2}{}^+_{j_1}\theta^{j_1j_2})\delta^{i_2}_{i_1}
  -(t^+_{i_1}{}^{i_2}\theta^{i_2i_1} + t^{i_2}{}_{i_1}\theta^{i_1i_2})
  \delta^{j_2}_{j_1}
\,, $  \\
 $ A^{i_2j_2, i_1j_1} \ = \  t^{i_1j_2}\delta^{i_2j_1}-
  t^{i_2j_1}\delta^{i_1j_2}  ,   \ \ \ \  F^{i_2j_2,i} \ =\
   t^{i_2j_2}(\delta^{j_2i}-\delta^{i_2i})\,, $\\
$ L^{i_2j_2,i_1j_1} \ = \   \textstyle\frac{1}{4}\Bigl\{\delta^{i_2i_1}
\delta^{j_2j_1}\Bigl[2g_0^{i_2}\delta^{i_2j_2} + g_0^{i_2} +
g_0^{j_2}\Bigr]  - \delta^{j_2\{i_1}\Bigl[t^{j_1\}i_2}\theta^{i_2j_1\}} +t^{i_2j_1\}+}\theta^{j_1\}i_2}\Bigr]  $
   \\
    $\phantom{L^{i_2j_2,i_1j_1} \ = \ } - \delta^{i_2\{i_1}\Bigl[t^{j_1\}j_2}\theta^{j_2j_1\}}
+t^{j_2j_1\}+}\theta^{j_1\}j_2}\Bigr] \Bigr\}$;
   \\
\end{tabular}
\end{flushleft}}
{\footnotesize
\begin{flushleft}\qquad
\begin{tabular}{||c||c|c|c||}\hline\hline
$\hspace{-0.2em}[\; \downarrow, \rightarrow
\}\hspace{-0.5em}$\hspace{-0.7em}&
 $t_0$ & $t^{i}$ &
$t^{i{}+}$  \\
\hline\hline $t_0$
    & $-2l_0$ & $2l^{i}$
   & $2l^{i{}+}$\\
   \hline $t^{j}$
    & $2l^{j}$ & $4l^{ji}$
   & $2\bigl(-g_0^{i}\delta^{ij} +
   t^{ij}\theta^{ji} +
   t^{ji+}\theta^{ij} \bigr)$
   \\
   \hline
   $t^{j+}$
    & $2l^{j{}+}$ & $2\bigl(-g_0^{i}\delta^{ij} +
   t^{ji}\theta^{ij} +
   t^{ij+}\theta^{ji} \bigr)$
   & $4l^{ji+}$
   \\ \hline\hline
\end{tabular}
\end{flushleft}}
 \vspace{-2ex}\caption{Even-even, odd-even and odd-odd parts
 of HS symmetry  superalgebra  $\mathcal{A}^f(Y(k),
\mathbb{R}^{1,d-1})$.\label{table in} }\end{table} The figure
brackets for the indices $i_1$, $i_2$ of Table~\ref{table in} in
the quantity $C^{\{i_1}D^{i_2\}i_3}\theta^{i_3i_2\}}$ imply
symmetrization, $C^{\{i_1}D^{i_2\}i_3}\theta^{i_3i_2\}}$ =
$C^{i_1}D^{i_2i_3}\theta^{i_3i_2}+
D^{i_2}C^{i_1i_3}\theta^{i_3i_1}$; these indices are raised and
lowered using the Euclidian metric tensors $\delta^{ij}$,
$\delta_{ij}$, $\delta^{i}_{j}$. The quantity $\theta^{ji} =
1(0)$ for $j>i(j\leq i)$ is the Heaviside $\theta$-symbol
$\theta^{ji}$. The products $B^{i_2j_2}_{i_1j_1}, A^{i_2j_2,
i_1j_1}, F^{i_1j_1,i}, L^{i_2j_2,i_1j_1}$ are given explicitly by
the relations of Table~\ref{table in} (for details, see
\cite{BuchbinderRmix,Reshetnyak2}).
Representing the basis elements of $\mathcal{A}^f(Y(k),
\mathbb{R}^{1,d-1})$, with allowance for the definitions
(\ref{gocond}), (\ref{t0ti}), (\ref{lilijpr}), (\ref{oA}) and in
agreement with (\ref{lilijt+}), as follows:
\begin{eqnarray}\label{repsconst}
\hspace{-0.8em} &\hspace{-0.8em}&\hspace{-0.7em}  o_I=\big\{t_0,\,l_0,\,{l}^{i},\,{l}^{i+}; \,o^+_a, \,  o_a; g_0^i\big\} \equiv \big\{o_A; \,o^+_a, \,  o_a; g_0^i\big\}, \ \ \mathrm{for } \ \ g_0^i =  - \textstyle\frac{1}{2}\{a^{i{}\mu^i}, a^{i+}_{\mu^i}\} \ \ \mathrm{ and} \ \ \\
\hspace{-1.0em}&\hspace{-1.0em}&\hspace{-1.0em} \begin{array}{|c|c|c|c|c|c|c|c|c|c|}\hline
     t_0 & t^i & t^{i+} & l_0 & l^i & l^{i+} & l^{ij} & l^{ij+} & t^{rs} & t^{rs+}  \\ \hline
     -\imath\tilde{\gamma}^\mu \partial_\mu &  \tilde{\gamma}^\mu a^i_\mu  & \tilde{\gamma}^\mu a^{i+}_\mu & \partial^\mu\partial_\mu & - ia^i_\mu
\partial^\mu & - ia^{i+}_\mu
\partial^\mu & {\textstyle\frac{1}{2}}\,a^{i{}\mu}
a^j_{\mu} & {\textstyle\frac{1}{2}}\,a^{i{}\mu+}
a^{j+}_{\mu} & a^{r+}_\mu
a^{s\mu} & a^{r{}\mu}
a^{s+}_{\mu} \\
   \hline \end{array} \label{explsalg}
\end{eqnarray}
the table of basis $[\ ,\ \}$-products can be presented as a Lie superalgebra:
\begin{eqnarray}
&& \label{geninalg}
    [o_I,\ o_J\}= f^K_{IJ}o_K, \  f^K_{IJ}= - (-1)^{\varepsilon(o_I)\varepsilon(o_J)}f^K_{JI},\\
&& [o_a,\; o_b^+\} = \underline{f}^c_{ab} o_c+\bar{f}^c_{ab} o^+_c +\Delta_{ab}(g_0^i),\ \  [o_A,\;o_B\}
= f^C_{AB}o_C, \ \  [o_a,\; o_B\} = f^C_{aB}o_C
\label{inconstraintsd},\\
&&\label{inconstraintsd1} [o_a,\; o_b\} = f^c_{ab} o_c, \ \ \ [o^+_a,\; o^+_b\} = f^{+c}_{ab} o^+_c, \ \mathrm{with} \ (f^c_{ab})^+= f^{+c}_{ab},
\end{eqnarray}
for $\Delta_{ab}(g_0^i) = \sum_i f^i_{ab}g_0^i$. From the Hamiltonian analysis   of
dynamical systems it follows the operators $o_a,\; o_a^+ $ are respective operator-valued $2k^2$ bosonic and
$2k$ fermionic second-class, as well as $o_A$  are  $(2k+1)$ bosonic and $1$
fermionic first-class constraint subsystems among $\{o_I\}$ for a
topological gauge system (one with vanishing  Hamiltonian $H_0 \equiv 0$) .
The symmetry properties of
 the real constants $f^c_{ab}, f^C_{AB}, f^C_{aB}$ (being another than general ones in (\ref{firstclassconstr}) , (\ref{secclassconstr}), (\ref{secclassconstr1})  of  the Section~\ref{BRST--BFVorig}) are described in \cite{Reshetnyak2}. We only remind  the quantities $\Delta_{ab}(g_0^i)$ form a
non-degenerate $(k\times k; k^2\times k^2)$ supermatrix,
$\|\Delta_{ab}\|$, in the Fock space $\mathcal{H}$ on the surface
$\Sigma \subset \mathcal{H}$: $\|\Delta_{ab}\|_{|\Sigma} \ne 0 $,
which is determined by the equations $(o_a, t_0,\ l_0,\
l^i)|\Psi\rangle = 0$.  The only operators
$g_0^i$  are not the constraints in $\mathcal{H}$, due to Eqs.
(\ref{gocond}),  but they select a vector from $|\Psi\rangle$ with fixed value of spin.
The second-class constraints $o_a,\; o_a^+ $ have already splitted on 2 first-class constraints subsystems due to (\ref{inconstraintsd1}).

In \cite{Reshetnyak2} it is shown the subsuperalgebra of $\{o_a,\,o_a^+,\, g_0^i\}$ is isomorphic to orthosymplectic superalgebra $osp(1|2k)$, thus realizing the generalization of  Howe duality \cite{Howe1} among whole the set of unitary half-integer  HS representations of  Lorentz subalgebra $so(1,d-1)$ and  $osp(1|2k)$.
The rest elements $\{o_A\}$ from
$\mathcal{A}^f(Y(k), \mathbb{R}^{1,d-1})$  forms the
subsuperalgebra  of Minkowski space $\mathbb{R}^{1,d-1}$ isometries, which has  the form of direct sum:
\begin{equation}\label{TkTk}
    \{o_A\} = (T^k \oplus T^{k*}\oplus [T^k,
    T^{k*}]),   \quad [T^k,
    T^{k*}] \sim l_0 = - t_0^2,
\end{equation}
of $k$-dimensional commutative algebra $T^k = \{l_i\}$ and its dual
$T^{k*}=\{l^{i+}\}$.
Finally,  HS symmetry  superalgebra $\mathcal{A}^f(Y(k),
\mathbb{R}^{1,d-1})$ represents the semidirect sum of  $osp(1|2k)$ [as an algebra of
internal derivations of $(T^k \oplus T^{k*})]$ with $(T^k \oplus
T^{k*}\oplus [T^k,
    T^{k*}])$,
\begin{equation}\label{identalg}
    \mathcal{A}^f(Y(k), \mathbb{R}^{1,d-1}) = \left(T^k \oplus T^{k*}\oplus [T^k,
    T^{k*}]\right) + \hspace{-1em} \supset  osp(1|2k).
\end{equation}
The nilpotent BRST--BFV  operator [having a matrix-like
$2^{\left[\frac{d}{2}\right]}\times 2^{\left[\frac{d}{2}\right]}$
structure] $Q'$: $(Q')^2=0$,  for Lie superalgebra
$\mathcal{A}^f(Y(k), \mathbb{R}^{1,d-1})$ was constructed  in
\cite{Reshetnyak2} with account of (\ref{geninalg})  according to
general formula \cite{BFV} (with the
$\hat{C}\hat{\overline{\mathcal{P}}}$-ordering\footnote{Further
on, we omit the symbol ``$\hat{\phantom{C} }$'' over an operator
for the Hamiltonian ghost coordinates and momenta
$\hat{C}{}^I,\hat{\overline{\mathcal{P}}}_I$, according to our
previous notation \cite{Reshetnyak2}.}):
\begin{equation}\label{generalQ'}
    Q'  = {C}^I{o}_I + \frac{1}{2}
    {C}^I {C}^Jf^K_{JI}\overline{\mathcal{P}}_K (-1)^{\varepsilon({o}_K) + \varepsilon({o}_I)} , \quad \big[\varepsilon, gh\big]{C}^I = \big[\varepsilon, -gh\big]\overline{\mathcal{P}}_I=\big[\varepsilon(o_I)+1, 1\big]
\end{equation}
 for ghost coordinates, ${C}^I$,  and momenta, $\overline{\mathcal{P}}_I$ from the minimal sector, subject to   $[{C}^I, \overline{\mathcal{P}}_J\}=\delta^I_J$ as for $Q=\Omega_{c|\min}(\hat{\Gamma}, \hat{\Gamma}_{gh|m})$ in (\ref{Orepconv}), but for $\zeta^\alpha\equiv 0$ and with $g_0^i$  considered as the constraints.  Explicitly, it has the form,
\begin{eqnarray}
\label{Q'k} {Q}' \hspace{-0.4em} &=&\hspace{-0.4em}
\frac{1}{2}q_0t_0+q_i^+t^i + \frac{1}{2}\eta_0l_0+\eta_i^+l^i
+\sum\limits_{l\leq m}\eta_{lm}^+l^{lm} + \sum\limits_{l<
m}\vartheta^+_{lm}t^{lm} + \frac{1}{2}\eta^i_{{g}}{g}_i
\\
&&  + \Bigl[\frac{1}{2}\sum_{l,m}(1+\delta_{lm})\eta^{lm}q_l^+
-\sum_{l<m}
q_l\vartheta^{lm}-\sum_{m<l}q_l\vartheta^{ml+}\Bigr]p_m^+ +
\frac{1}{2}
\sum_m\eta^m_{g}(q_mp_m^++q^+_mp_m)\nonumber\\
&& +
\imath\sum_l\Bigr[\frac{1}{2}\eta_l^+\eta^l{\cal{}P}_0
+\eta_l^+q^lp_0-  q^lq_l^+ {\cal{}P}^l_{g}\Bigr]- \frac{\imath}{2}q_0^2{\cal{}P}_0 \nonumber
\\\hspace{-0.4em}&&
{}\hspace{-0.4em} -\sum\limits_{i<l<j}
\vartheta^+_{lj}\vartheta^+_{i}{}^l\lambda^{ij} +
\frac{\imath}{2}
\sum\limits_{l<m}\vartheta_{lm}^+\vartheta^{lm}({\cal P}_g^m-{\cal
P}_g^l)-
\sum\limits_{l<n<m}\vartheta_{lm}^+\vartheta^{l}{}_n\lambda^{nm}
\nonumber\\&& +
\sum\limits_{n<l<m}\vartheta_{lm}^+\vartheta_{n}{}^m\lambda^{+nl}
- \sum_{n,l<m}(1+\delta_{ln})\vartheta_{lm}^+\eta^{l+}{}_{n}
\mathcal{P}^{mn}+
\sum_{n,l<m}(1+\delta_{mn})\vartheta_{lm}^+\eta^{m}{}_{n}
\mathcal{P}^{+ln}\nonumber\\
\phantom{Q'}&&  + \frac{\imath}{8}\sum_{l\leq m}(1+\delta_{lm})
\eta_{lm}^+\eta^{lm}({\cal{}P}^l_{{g}}+{\cal{}P}^m_{{g}})+\frac{1}{2}\sum_{l\leq
m}(1+\delta_{lm})\eta^l_{{g}}\bigl(
\eta_{lm}^+{\cal{}P}^{lm}-\eta_{lm}{\cal{}P}^{lm+}\bigr)
\nonumber\\
\hspace{-0.4em} && \hspace{-0.4em} +
\frac{1}{2}\sum\limits_{l<m,n\le
m}\eta^+_{nm}\eta^{n}{}_l\lambda^{lm}
 -2 \sum_{l<pm}q_lq_m^+\lambda^{lm}  +
\frac{1}{2}\sum\limits_{l<m}(\eta^m_{{g}}-
\eta^l_{{g}})\bigl(\vartheta^+_{lm}\lambda^{lm}- \vartheta_{lm}\lambda^{lm+}\bigr) \nonumber\\
&& - \Bigl[\frac{1}{2}\sum\limits_{l,
m}(1+\delta_{lm})\eta^m\eta_{lm}^+ +
\sum\limits_{l<m}\vartheta_{lm} \eta^{+m}
+\sum\limits_{m<l}\vartheta^+_{ml} \eta^{+m} +2\sum_lq_0q_l^+
\Bigr]\mathcal{P}^l \nonumber\\
&& -2\sum_{l,m}
q^+_lq^+_m\mathcal{P}^{lm} +
\frac{1}{2}\sum_l\eta^l_{{g}}\bigl(\eta_l^+\mathcal{P}^l-\eta_l\mathcal{P}^{l+}\bigr)+h.c. \nonumber
\end{eqnarray}
Here, the set of ghost operators $C^I
= (q_0, q_i, q_i^+ ; \eta_0, \eta^i, \eta^+_i, \eta^{ij},
\eta^+_{ij}$, $\vartheta_{rs}$, $\vartheta^+_{rs}$, $\eta^i_{g})$
 with the Grassmann parity and ghost number given according to (\ref{generalQ'}) respectively for the elements $o_I =
(t_0, t^+_i, t_i; l_0, l^+_i$, $l_i$, $l_{ij}$, $l^+_{ij}$,
$t_{ij}$, $t^+_{ij}$, $g_0^i)$, subject to the properties
\begin{equation}\label{propgho}
    (\eta^{ij}, \eta^+_{ij})\ =\ (\eta^{ji} , \eta^+_{ji}),\qquad (\vartheta_{rs}, \vartheta^+_{rs})\ = \ (\vartheta_{rs},
 \vartheta^+_{rs})\theta^{sr},
\end{equation}
and their conjugated ghost
momenta $\overline{\mathcal{P}}_I$ (composing   Wick pairs of ghost operators for the  $\{o_I\}\setminus \{t_0, l_0, g_0^i\}$ constraints) with the same properties as those for $C^I$
in (\ref{propgho}) with the only nonvanishing commutation
relations for bosonic \begin{align}\label{fghosts}
 & [q_i, p^{+}_j] = [p_i, q^{+}_j] =\delta_{ij}\,, &&  [q_0, p_0]=\imath ;
\end{align}
and the anticommutation ones for fermionic ghosts
\begin{align}\label{bghosts}
& \{\vartheta_{rs},\lambda^+_{tu}\}= \{{\lambda_{tu}},
\vartheta_{rs}^+\}= \delta_{rt}\delta_{su}, &&
\{\eta_i,{\cal{}P}_j^+\}= \{{\cal{}P}_j, \eta_i^+\}=\delta_{ij}\,, \nonumber \\
& \{\eta_{lm},{\cal{}P}_{ij}^+\}= \{{\cal{}P}_{ij}, \eta_{lm}^+\}
=\delta_{li}\delta_{jm}\,,  &&  \{\eta_0,{\cal{}P}_0\}= \imath,\
\{\eta^i_{{g}}, {\cal{}P}^j_{{g}}\}
 = \imath\delta^{ij}.
\end{align}
By construction the property $gh({Q}')$ = $1$ holds, whereas the
Hermitian conjugation  of zero-mode pairs:
\begin{eqnarray}\label{Hermnull}
 \left( q_0, \eta_0, \eta^i_{{g}}, p_0,  {\cal{}P}_0,
{\cal{}P}^i_{g} \right)^+ & = & \left(q_0, \eta_0, \eta^i_{g},
p_0,  - {\cal{}P}_0, -{\cal{}P}^i_{g}\right)
\end{eqnarray}
provides for $Q'$ to be hermitian $(Q')^+=Q'$.

Decomposing $Q'$ in powers of zero-mode ghosts $\eta^i_{{g}}, {\cal{}P}^i_{g} $  corresponding to the operators $g_0^i$ not entered into the  first-class  $\{o_A\}$ and second-class  $\{o_a, o^+_a \}$ subsystems of constraints
\begin{eqnarray} \label{Q'decomp}
{Q}'  &=& Q +
\eta^i_{g}\sigma^i(g) +\mathcal{B}^i \mathcal{P}^i_{g}\,, \\
\label{sigmai}
  \sigma^i(g) &=& g_0^i   - \eta_i \mathcal{P}^+_i +
   \eta_i^+ \mathcal{P}_i + \sum_{
m}(1+\delta_{im})(
\eta_{im}^+{\cal{}P}^{im}-\eta_{im}{\cal{}P}^+_{im})\\
   &&  + \sum_{l<i}[\vartheta^+_{li}
\lambda^{li} - \vartheta^{li}\lambda^+_{li}]-
\sum_{i<l}[\vartheta^+_{il} \lambda^{il} -
\vartheta^{il}\lambda^+_{il}] +   q_ip_i^+ + q_i^+p_i\,, \nonumber\\
\label{Q} {Q} \hspace{-0.4em} &=&\hspace{-0.4em}
 \frac{1}{2}q_0t_0+q_i^+t^i+\frac{1}{2}\eta_0l_0+\eta_i^+l^i
+\sum\limits_{l\leq m}\eta_{lm}^+l^{lm} + \sum\limits_{l<
m}\vartheta^+_{lm}t^{lm}
 + \frac{\imath}{2}\bigl(\sum_l\eta_l^+\eta^l-q_0^2\bigr){\cal{}P}_0
\\
\hspace{-0.4em}
&&\hspace{-0.4em}  + \Bigl[\frac{1}{2}\sum_{l,m}(1+\delta_{lm})\eta^{lm}q_l^+
-\sum_{l<m}
q_l\vartheta^{lm}-\sum_{m<l}q_l\vartheta^{ml+}\Bigr]p_m^+  +
\imath\sum_l
\eta_l^+q^lp_0  -2 \sum_{l<m}q_lq_m^+\lambda^{lm} \nonumber
\\
\hspace{-0.4em}&& {}\hspace{-0.4em}  -2\sum_{l,m}
q^+_lq^+_m\mathcal{P}^{lm} -\sum\limits_{i<l<j}
\vartheta^+_{lj}\vartheta^+_{i}{}^l \lambda^{ij}-
\sum\limits_{l<n<m}\vartheta_{lm}^+\vartheta^{l}{}_n\lambda^{nm} +
\sum\limits_{n<l<m}\vartheta_{lm}^+\vartheta_{n}{}^m\lambda^{+nl}
\nonumber\\&& -
\sum_{n,l<m}(1+\delta_{ln})\vartheta_{lm}^+\eta^{l+}{}_{n}
\mathcal{P}^{mn}+
\sum_{n,l<m}(1+\delta_{mn})\vartheta_{lm}^+\eta^{m}{}_{n}
\mathcal{P}^{+ln}+ \textstyle\frac{1}{2}\sum\limits_{l<m,n\leq
m}\eta^+_{nm}\eta^{n}{}_l\lambda^{lm}
\nonumber\\
\hspace{-0.4em} && \hspace{-0.4em}  
 - \Bigl[\frac{1}{2}\sum\limits_{l,
m}(1+\delta_{lm})\eta^m\eta_{lm}^+ +
\sum\limits_{l<m}\vartheta_{lm} \eta^{+m}
+\sum\limits_{m<l}\vartheta^+_{ml} \eta^{+m} +2\sum_lq_0q_l^+ \Bigr]\mathcal{P}^l
+h.c. ,\nonumber
\end{eqnarray}
\begin{eqnarray}
\label{Bi}
\mathcal{B}^i & =& -2{\imath} q^iq_i^+
   - {\imath}
\sum\limits_{l<m}\vartheta_{lm}^+\vartheta^{lm}(\delta^{mi}-\delta^{li})
+ \frac{\imath}{4}\sum_{l\leq m}(1+\delta_{lm})
\eta_{lm}^+\eta^{lm}(\delta^{il}+\delta^{mi}).
\end{eqnarray}
where the  operator $\vec{\sigma}(g) =
(\sigma^1,\sigma^2,..., \sigma^k)(g)$ playing the role of "ancient" of generalized spin operator $\vec{\sigma}$ \cite{Reshetnyak2}  is Hermitian, $({\sigma}^{i})^+ = {\sigma}^i$, as well as $Q$: $Q^+=Q$.

From the nilpotency condition for $Q'$ it follows the operator equations in powers of $\eta^i_{{g}}, {\cal{}P}^i_{g} $:
\begin{align}\label{eqQ1}
  &(\eta^i_{{g}})^0 ({\cal{}P}^j_{g})^0: \  Q^2 + \imath \sum_j\mathcal{B}^j {\sigma}^j (g) = 0, & (\eta^i_{{g}})^1 ({\cal{}P}^j_{g})^0:\  \eta^i_{{g}} [Q,\, {\sigma}^i(g)\}= 0, \\
  & (\eta^i_{{g}})^0 ({\cal{}P}^i_{g})^1:\  [Q,\, \mathcal{B}^i\}{\cal{}P}^i_{g}= 0, & (\eta^i_{{g}})^1 ({\cal{}P}^j_{g})^1:\   \sum_{i,j}\eta^i_{{g}}[\mathcal{B}^j,\,{\sigma}^i(g)\}{\cal{}P}^j_{g} = 0,\label{eqQ2}
\end{align}
where the both the operator  nilpotency  of the operator $Q$ and its weak nilpotency on some subspaces $\tilde{\mathcal{H}}^{{\sigma}(g)}_{tot|N}$ of a total Hilbert space ${\mathcal{H}}_{tot}=\mathcal{H}\otimes \mathcal{H}_{gh|m}$ can not be realized, except for the case:
${\sigma}^i (g){\mathcal{H}}^{{\sigma}(g)|0}_{tot|N}=0$. The latter variant does  work for only HS fields with lowest exceptional generalized spin related to the space-time dimension. The second equation  in (\ref{eqQ1})  and equations (\ref{eqQ2}) vanish identically, due to operator identities: $ [Q,\, {\sigma}^i(g)\}= [\mathcal{B}^j,\,{\sigma}^i(g)\}= 0$ and $[Q,\, \mathcal{B}^i\}=0$.
The non-trivial solution  of the first equations in (\ref{eqQ1}) , (\ref{eqQ2}) requires choosing the representation in ${\mathcal{H}}_{tot}$ and conversion  the second-class constraints subsystem $\{o_a,\,o_a^+\}$ into first-class one.

\subsection{Unconstrained Lagrangian formulation }\label{unconstLF}

To construct  Lagrangian formulation it is impossible to use BRST operator $Q'$  for the  initial HS superalgebra $\mathcal{A}^f(Y(k), \mathbb{R}^{1,d-1})$ of the elements $\{o_I\}$. Instead,  the additively converted first-class constraints, $O_I$: $O_I=o_I+o'_I$ and $[o_I,\,o'_J\}=0$ with unchanged first-class constraints, $O_A=o_A$, i.e.   $l'_0=t'_0=l'_i=(l'_i)^+ = 0$, are  found in \cite{Reshetnyak2}  for additional parts of the second-class constraints $\{o'_a,\, o_a^{\prime +}\}$ and $g_0^{\prime i}$  as the polynomials  in auxiliary creation and annihilation operators $B^{a+}, B^a $  in new Fock space $\mathcal{H}'  \equiv \mathcal{H}_{B} $: $\mathcal{H} \cap\mathcal{H}'  = \varnothing$. The number  of new oscillators (with $gh(B^{a+})=gh(B^a)=0 $) coincides respectively with one for  the second-class constraints ($\varepsilon(B_a)=\varepsilon(o_a)$).  The supercommutativity  requirement, $[o_I,\,o'_J\}=0$, permits to preserve both  for converted  $O_I$ and for additional $o'_I$  operators
the same superalgebra relations as for $\mathcal{A}^f(Y(k), \mathbb{R}^{1,d-1})$ given therefore by the Eqs. (\ref{geninalg})--(\ref{inconstraintsd1})  and Multiplication Table~\ref{table in}, with obvious respective changes: $o_I\to O_I$ and $o_I\to o'_I$ with unchanged $o_A$.
The explicit form of $o'_I(B^{a+}, B^a)$ has sufficiently non-trivial form and found from the procedure of generalized Verma module construction for subsuperalgebra of  $\{o'_a, o^{\prime +}_a, g_0^i\}$ isomorphic to $osp(1|2k)$ presented in the Appendix~A \cite{Reshetnyak2} as well as, so that after oscillator realization  over  the Heisenberg-Weyl superalgebra, parameterized by  $\{B^{a+}; B^a\}=\{f_i, b_{ij}, d_{rs}\theta^{sr}; f^+_i, b^+_{ij}, d^+_{rs}\theta^{sr}\}$ with non-vanishing anticommutators, $\{f_i,f_j^+\}=\delta_{ij}$, for odd oscillators and commutators, $[b_{ij},b^+_{lm}]= \frac{1}{2}\delta_{i\{l}\delta_{j\}m} $, $[d_{qt},d^+_{rs}]=\delta_{qr}\delta_{ts}$ for even.
Important for further construction additional number particle operators, $g_0^{\prime i}$,
\begin{eqnarray}
   g_0^{\prime i}& = & h^i+ f_i^+f_i + \sum_{l\leq m}
 b_{lm}^+b_{lm}(\delta^{il}+\delta^{im}) + \sum_{r< s}d^+_{rs}d_{rs}(\delta^{is}-
 \delta^{ir})
 \,,\label{g'0iFf}
  \end{eqnarray}
contain real-valued quantities $h^i$, $i=1,\ldots,k$ being by
the arbitrary dimensionless constants,  introduced in the process of generalized Verma module construction \cite{Reshetnyak2}, which permits to make  the generalized spin equations for proper
 eigen-vectors  in $\mathcal{H}_{c|tot}$: $\mathcal{H}_{c|tot} =\mathcal{H}\otimes \mathcal{H}'\otimes\mathcal{H}_{gh|m}$ :for the spin operator, ${\sigma}^i (G) \equiv {\sigma}^i+h^i  $, instead of ${\sigma}^i(g)$:
 \begin{equation}\label{genspineq}
   {\sigma}^i (G)|\chi\rangle = ({\sigma}^i +h^i) |\chi\rangle = 0
 \end{equation}
 with spectra of proper eigen-values for $h^i$. On the Hilbert subspace ${\mathcal{H}}^{{\sigma}(G)}_{c|tot|N}$  from $\mathcal{H}_{c|tot}$  consisting  from the  proper
 eigen-vectors satisfying to the equations (\ref{genspineq}) the BRST--BFV operator $Q$ (\ref{Q}),  but for converted first-class constraints, $O_I$,  without converted number particle operators $G_0^i=g_0^i+g^{\prime i}_0$ is nilpotent. Namely, the latter operator generates the right BRST complex, whose cohomology in Hilbert subspace ${\mathcal{H}}^{{\sigma}(G) k}_{c|tot|N}$ with respective  ghost number values spectrum, starting from, $k=0$  generates correct Lagrangian dynamic for initial spin-tensor with appropriate set of auxiliary spin-tensor fields.

The unconstrained gauge-invariant Lagrangian  formulation for HS field $\Psi_{(\mu^1)_{n_1},(\mu^2)_{n_2},...,(\mu^k)_{n_k}}
$ is derived from  the sequence of the equations, following from BRST-like equations for $Q'(O)|\chi^0\rangle =0 $, $\delta|\chi^0\rangle = Q'(O)|\chi^1\rangle $, ..., $\delta|\chi^{k-1}\rangle = Q'(O)|\chi^k\rangle$, with  $|\chi^k\rangle \in \mathcal{H}^k_{tot}$, due to existence of $\mathbb{Z}$-grading in  $\mathcal{H}_{c|tot}$: $\mathcal{H}_{c|tot} = \bigoplus_{k}\mathcal{H}^k_{c|tot}$   for $gh(|\chi^k\rangle)=-k$, $k=0,1,2,...$.   Indeed, decomposing  in ghosts  $\eta^i_g$ operator $Q'(O)$ (\ref{Q'decomp}) and vector  $|\chi\rangle$ with choice  the standard  representation in $\mathcal{H}_{c|tot}$:
\begin{eqnarray}
(q_i, \eta_i, \eta_{ij},  \vartheta_{rs}, p_0, p_i, \mathcal{P}_0, \mathcal{P}_i,
\mathcal{P}_{ij}, \lambda_{rs},
\mathcal{P}^{i}_g)|0\rangle=0,\qquad |0\rangle\in
\mathcal{H}_{c|tot},
\end{eqnarray}
not depending on
$\eta^{i}_g$,
\begin{eqnarray}
|\chi \rangle &=& \sum_n \prod_{c}^k ( f_c^+ )^{n^0_{c}}\prod_{i\le
j, r<s}^k( b_{ij}^+ )^{n_{ij}}( d_{rs}^+ )^{p_{rs}}q_0^{n_{b{}0}}\eta_0
^{n_{f 0}}\nonumber
\\
&&{}\times  \prod_{e,g, i, j, l\le m, n\le o}(q_e^+)^{n_{a{}e}}(p_g^+)^{n_{b{}g}}( \eta_i^+ )^{n_{f i}} (
\mathcal{P}_j^+ )^{n_{p j}} ( \eta_{lm}^+ )^{n_{f lm}} (
\mathcal{P}_{no}^+ )^{n_{pno}} \prod_{r<s, t<u}(
\vartheta_{rs}^+)^{n_{f rs}} ( \lambda_{tu}^+ )^{n_{\lambda tu}}
\nonumber
\\
&&{}\times |\Psi(a^+_i)^{n_{b{}0} n_{f 0};  (n)_{a{}e} (n)_{b{}g} (n)_{f i}(n)_{p j}(n)_{f lm}
(n)_{pno}(n)_{f rs}(n)_{\lambda
tu}}_{(n^0)_{c};(n)_{ij}(p)_{rs}}\rangle \,, \label{chif}
\end{eqnarray}
the following  spectral
problem  is derived %
\begin{align}
\label{Qchi} & Q(O)|\chi\rangle=0, && (\sigma^i+h^i)|\chi\rangle=0,
&& \left(\varepsilon, {gh}_H\right)(|\chi\rangle)=(1,0),
\\
& \delta|\chi\rangle=Q(O)|\chi^1\rangle, &&
(\sigma^i+h^i)|\chi^1\rangle=0, && \left(\varepsilon,
{gh}_H\right)(|\chi^1\rangle)=(0,-1), \label{Qchi1}
\\
& \delta|\chi^1\rangle=Q(O)|\chi^2\rangle, &&
(\sigma^i+h^i)|\chi^2\rangle=0, && \left(\varepsilon,
{gh}_H\right)(|\chi^2\rangle)=(1,-2),\\
& \delta|\chi^s\rangle=Q(O)|\chi^{s+1}\rangle, &&
(\sigma^i+h^i)|\chi^{s+1}\rangle=0, && \left(\varepsilon,
{gh}_H\right)(|\chi^{s+1}\rangle)= (s\mod{}2,-s-1). \label{Qchii}
\end{align}
Note , the brackets $(n^0)_{c}, (n)_{f i},(n)_{p j}, (n)_{ij}$ in  (\ref{chif}) imply, for instance, for $(n^0)_{c}$
and $(n)_{ij}$, the sets of indices $(n^0_{1},...,n^0_{k})$ and
$(n_{11},...,n_{1k},..., n_{k1},..., n_{kk})$. The above sum is
taken over $n_{b{}0}$, $n_{a{}e}$, $ n_{b{}g}$,  $h_{l}$,
$n_{ij}$, $p_{rs}$, running from $0$ to infinity, and over the
remaining $n$'s from $0$ to $1$.  Thus, the physical state  $|\chi^0\rangle$  for the vanishing of all auxiliary
operators $B^{a+}$ and ghost variables $q_0, q_i^+, \eta_0,
\eta^+_i, p_i^+, \mathcal{P}^+_i,...$,
 contains only the physical string-like vector $|\Psi\rangle
= |\Psi(a^+_i)^{0_{b{}0} 0_{f 0};  (0)_{a{}e} (0)_{b{}g} (0)_{ f
i}(0)_{p j}(0)_{f lm} (0)_{pno}(0)_{f rs}(0)_{\lambda
tu}}_{(0^0)_{c}; (0)_{ij}(0)_{rs}}\rangle$, so that
\begin{eqnarray}\label{decomptot}
|\chi^0\rangle&=&|\Psi\rangle+  |\Psi_A\rangle ,\qquad |\Psi_A\rangle\Big|_{\textstyle (B^{a+}, q_0, q_i^+, \eta_0, \eta^+_i, p_i^+, \mathcal{P}^+_i,...)=0}=0.
\end{eqnarray}
The middle set of
equations (\ref{Qchi})--(\ref{Qchii}), has the general solution for the set of proper
eigenvectors $|\chi^0\rangle_{(n)_k}$, $|\chi^1\rangle_{(n)_k}$,
 $\ldots$, $|\chi^{s}\rangle_{(n)_k}$, with $(n)_k=(s)_k-\frac{1}{2}$ , and the set of the corresponding eigenvalues
for  values of the parameters $h^i$,
\begin{eqnarray}
\label{hi} -h^i &=& n^i+\frac{d-4i}{2} \;, \quad
i=1,..,k,
\end{eqnarray} so that the vector $|\chi^0\rangle_{(n)_k}$ contains the physical field
(\ref{PhysState}) and all of its auxiliary fields.
Explicitly, the spin and ghost number $gh$ distributions are given by the respective relations being valid for a general case of HS fields subject to $Y(s_1,..., s_k)$ and for the subset of ``ghost'' numbers
 in
(\ref{chif}) and (\ref{PhysState}) for fixed values of $n_i$ for $|\chi^l\rangle_{(n)_k}$, $l
= 0,\ldots , \sum_{o=1}^k n_o + k(k-1)/2$:
\begin{eqnarray}\label{nidecomposf}
n_i &= & p_i+ n_{a{}i}+ n^0_{i}+
\sum_{j=1}(1+\delta_{ij})(n_{ij}+n_{f{}ij}+n_{p{}ij} )
+n_{f{}i}+n_{p{}i} \nonumber \\
&{}& +\sum_{r<i} (p_{ri} + n_{f{}ri}+ n_{\lambda{}ri}) -\sum_{r>i}
(p_{ir}+ n_{f{}ir}+ n_{\lambda{}ir} )\,,\  i=1,\ldots,k,\\
\label{ghnumf}
  \hspace{-0.9em} |\chi^l\rangle_{(n)_k}  \hspace{-0.9em}&:&\hspace{-0.9em} n_{b{}0}+ n_{f{}0}+\hspace{-0.2em}
     \sum_{i}\hspace{-0.2em}\bigl(n_{f{}i}- n_{p{}i} + n_{a{}i}- n_{b{}i}\bigr)+ \hspace{-0.2em}
\sum_{i\leq j}\hspace{-0.1em}\bigl(n_{f{}ij}-n_{p{}ij} \bigr) + \hspace{-0.2em}\sum_{r<s}\hspace{-0.1em} \bigl(
n_{f{}rs}- n_{\lambda{}rs}\bigr)\hspace{-0.1em} =\hspace{-0.1em} -l .
\end{eqnarray}
For fixed spin $(s)_k=(n)_k+(\frac{1}{2},...,\frac{1}{2})$  the value of $h^i$ is fixed by (\ref{hi}) as $h^i(n)$ and  $Q(O)|_{[h^i=h^i(n)]} = Q_{(n)_k}$ The second-order equations of motion (\ref{Qchi}) and sequence of the reducible gauge transformations (\ref{Qchi1})--(\ref{Qchii}) on the solutions for middle set of the equations therein has the form:
\begin{eqnarray}
\hspace{-0.7em}Q_{(n)_k}|\chi^0\rangle_{(n)_k}=0, \ \delta|\chi^0 \rangle_{(n)_k}
=Q_{(n)_k}|\chi^1\rangle_{(n)_k} \,, ...\,,
\delta|\chi^{s-1} \rangle_{(n)_k}
=Q_{(n)_k}|\chi^{s}\rangle_{(n)_k},\, \delta|\chi^{s}\rangle_{(n)_k} =0\label{Q12}
\end{eqnarray}
for $s=\sum_{o=1}^k n_o +
k(k-1)/2$ with nilpotent  operator  $Q_{(n)_k}$ on any $|\chi^m\rangle_{(n)_k}$, $m=0,...,s$: $Q^2_{(n)_k}|\chi^m\rangle_{(n)_k}=0$.
The corresponding formal BRST-like (as for integer HS fields \cite{BuchbinderRmix}) gauge-invariant action
\begin{equation}\label{secorder}
  {\cal{}S}^{(2)}_{(n)_k}= \int d \eta_0 {}_{(n)_k}\langle\tilde{\chi}^0|K_{(n)_k}Q_{(n)_k} |\chi^0\rangle_{(n)_k}, \ \ K=1\otimes K'\otimes 1_{gh}:\ K_{(n)_k}Q_{(n)_k}=Q^+_{(n)_k}K_{(n)_k}
\end{equation}
    first, contains second order derivative $l_0$, second
    the operator $K'$ realizing hermitian conjugation
    for additional parts to the constraints
    $o'_I$.\footnote{$K'$, $K'=K^{\prime +}$,
    is determined explicitly by Eq. (3.16)
    of \cite{Reshetnyak2} so that the set
    of additional parts
    $o'_a(B,B^+), o^{\prime +}_a(B,B^+)$, $g_0^{\prime i}$
    should be closed with respect to the new  Hermitian
    conjugation:
    $\langle\tilde{\Psi}_1|K'o^{\prime}_a|\Psi_2\rangle =
\langle\tilde{\Psi}_2|K'o^{\prime}_a|\Psi_1\rangle^* $,
$\langle\tilde{\Psi}_1|K'g_0^{\prime i}|\Psi_2\rangle =
\langle\tilde{\Psi}_2|K'g_0^{\prime i}|\Psi_1\rangle^*. $} The
dependence on $L_0, \eta_0$ from the BRST operator $Q_{(n)_k}$
(\ref{Q}) and from the whole set of the vectors
$|\chi^l\rangle_{(n)_k}$ maybe removed  by means of partial
gauge-fixing procedure, based on the
 extraction  of the zero-mode ghosts $q_0, \eta_0$ from
$Q$ :
\begin{eqnarray}
\label{strQ} Q &=& q_0 \overline{t}_0+\eta_0{l}_0 +
\imath(\eta_i^+q_i-\eta_iq_i^+)p_0 -
\imath(q_0^2-\eta_i^+\eta_i){\cal{}P}_0 +\Delta{}Q,
\end{eqnarray}
where
\begin{eqnarray}
\label{tildeT0} \overline{t}_0 &=& t_0 -2q_i^+{\cal{}P}_i
-2q_i{\cal{}P}_i^+\,,
\\
\label{deltaQ} \Delta{}Q & = & q_i^+T^i+\eta_i^+l^i
+\sum\limits_{l\leq m}\eta_{lm}^+L^{lm} + \sum\limits_{l<
m}\vartheta^+_{lm}T^{lm}+ \Bigl[\frac{1}{2}\sum_{l, m}(1+\delta_{lm})\eta^{lm}q_l^+
 \nonumber
\\
\hspace{-0.4em}
&&\hspace{-0.4em}  -\sum_{l<m}
q_l\vartheta^{lm}-\sum_{m<l}q_l\vartheta^{ml+}\Bigr]p_m^+  -2 \sum_{l<m}q_lq_m^+\lambda^{lm}-2\sum_{l,m}
q^+_lq^+_m\mathcal{P}^{lm}-\sum\limits_{i<l<j}
\vartheta^+_{lj}\vartheta^+_{i}{}^l \lambda^{ij} \nonumber
\\
\hspace{-0.4em}&& {}\hspace{-0.4em}   -
\sum\limits_{l<n<m}\vartheta_{lm}^+\vartheta^{l}{}_n\lambda^{nm} +
\sum\limits_{n<l<m}\vartheta_{lm}^+\vartheta_{n}{}^m\lambda^{+nl}-
\sum_{n,l<m}(1+\delta_{ln})\vartheta_{lm}^+\eta^{l+}{}_{n}
\mathcal{P}^{mn}з
\nonumber\\&& +
\sum_{n,l<m}(1+\delta_{mn})\vartheta_{lm}^+\eta^{m}{}_{n}
\mathcal{P}^{+ln}+ \frac{1}{2}\sum\limits_{l<m,n\leq
m}\eta^+_{nm}\eta^{n}{}_l\lambda^{lm}
\nonumber\\
\hspace{-0.4em} && \hspace{-0.4em}  
 - \Bigl[\frac{1}{2}\sum\limits_{l,
m}(1+\delta_{lm})\eta^m\eta_{lm}^+ +
\sum\limits_{l<m}\vartheta_{lm} \eta^{+m}
+\sum\limits_{m<l}\vartheta^+_{ml} \eta^{+m}  \Bigr]\mathcal{P}^l
+h.c.
\,,
\end{eqnarray}
%
%
%
%
and for  the state vector and gauge parameters, for $s=0, \ldots, \sum_{o=1}^k n_o + k(k-1)/2$:
\begin{align}
\label{0chi} |\chi^s\rangle &=\sum_{l=0}^{\sum_{o=1}^k n_o + k(k-1)/2-1}q_0^l(
|\chi_0^{s(l)}\rangle +\eta_0|\chi_1^{s(l)}\rangle), &
&gh(|\chi^{s(l)}_{m}\rangle)=-(s+l+m+l), \ m=0,1.
\end{align}
As the results of gauge-fixing procedure with help of  the equations of motion  and the set of gauge transformations
(\ref{Q12})  all the components in powers of $q_0$, $\eta_0$ vector  are removed except for two fields for each level,
$|\chi^{(s(0)}_0\rangle_{(n)_k}$, $|\chi^{(s(1)}_0\rangle_{(n)_k}$, for
 $s=0, \ldots, \sum_{o=1}^k n_o + k(k-1)/2$.
Therefore, the representation holds:
\begin{equation}\label{gfvectors}
  |\chi^{s} \rangle_{(n)_k} =  |\chi^{s (0)}_{0} \rangle_{(n)_k} + q_0|\chi^{s (1)}_{0} \rangle_{(n)_k} - i\eta_0 \overline{t}_0 |\chi^{s (1)}_{0} \rangle_{(n)_k},
\end{equation}
 which leads to the first-order independent equations of motion for the rest  vectors
\begin{eqnarray}
&& \left(\begin{array}{lr}
\overline{t}_0 & \Delta{}Q \\
\Delta{}Q  &  \frac{1}{2}\bigl\{\overline{t}_0,\eta_i^+\eta_i\bigr\}\end{array}
\right) \left(\begin{array}{c}   | \chi^{0}_{0}\rangle_{(n)_k} \\
|\chi^{1}_{0}\rangle_{(n)_k}  \end{array}
\right) =\left(\begin{array}{c}  0\\
0 \end{array}
\right) , \ \mathrm{for} \  \bigl\{A,B\bigr\} =  AB+BA, \label{EofM1}
\end{eqnarray}
which  follows from the   Lagrangian action with help of  supermatrix multiplication
\begin{eqnarray}
{\cal{}S}_{(n)_k} &=& \left({}_{(n)_k}\langle\tilde{\chi}^{0}_{0}|\  {}_{(n)_k}\langle\tilde{\chi}^{1}_{0}|\right) K_{(n)_k} \left(\begin{array}{lr} \overline{t}_0 & \Delta{}Q \\
\Delta{}Q  &  \frac{1}{2}\bigl\{\overline{t}_0,\eta_i^+\eta_i\bigr\} \end{array}
\right) \left(\begin{array}{c}   | \chi^{0}_{0}\rangle_{(n)_k} \\
|\chi^{1}_{0}\rangle_{(n)_k}  \end{array}
\right),\label{Lun}
\end{eqnarray}
where the standard odd scalar product for the creation and
annihilation operators is assumed, with the measure $d^dx$ over
the Minkowski space.     The vectors (Dirac spinors)
$|\chi^{s(l)}_{0}\rangle_{(n)_k}$   (\ref{0chi}) as the solution
of the spin distribution relations (\ref{nidecomposf})
are the respective vectors $|\chi^{l}_{0}\rangle$ in (\ref{chif})
for massless ($m=0$)  HS fermionic field
$\Psi_{(\mu^1)_{n_1},...,(\mu^k)_{n_k}}(x)$ with
the ghost number $gh(|\chi^{l}_{0}\rangle_{(n)_k})=-l$.

The action (\ref{Lun}) and the equations of motion (\ref{EofM1})
are invariant with respect to reducible gauge transformations:
\begin{eqnarray}
\hspace{-1em}\delta\left(\begin{array}{c}   | \chi^{s(0)}_{0}\rangle_{(n)_k} \\
|\chi^{s(1)}_{0}\rangle_{(n)_k}  \end{array}
\right) \hspace{-0.5em}&=&\hspace{-0.5em} \left(\begin{array}{lr}
\Delta{}Q  &  \frac{1}{2}\bigl\{\overline{t}_0,\eta_i^+\eta_i\bigr\} \\
\overline{t}_0 & \Delta{}Q \end{array}
\right) \left(\begin{array}{c}   | \chi^{s+1(0)}_{0}\rangle_{(n)_k} \\
|\chi^{s+1(1)}_{0}\rangle_{(n)_k}  \end{array}
\right) ,  \  \delta\left(\begin{array}{c}   | \chi^{s_{\mathrm{max}}(0)}_{0}\rangle_{(n)_k} \\
|\chi^{s_{\mathrm{max}}(1)}_{0}\rangle_{(n)_k}  \end{array}
\right) =0
\label{GT1}
\end{eqnarray}
for $s=0,1,...,s_{\mathrm{max}}= \sum_{o=1}^k n_o + k(k-1)/2$  with a finite number of reducibility stage
to be equal to $(s_{\mathrm{max}}-1)$.

In \cite{Reshetnyak2} it was shown (see Appendix~C for massive
case)  that  equivalence of the unconstrained BRST--BFV Lagrangian
formulation given by (\ref{Lun}), (\ref{GT1}) and by (\ref{EofM1})
to the set of the equations (\ref{Eq-0})--(\ref{Eq-2}) for
spin-tensor field, $\Psi_{(\mu^1)_{n_1},...,(\mu^k)_{n_k}}(x)$,
realizing initial irreducible Poincare group representation of
spin $\mathbf{s}=(n_1+\frac{1}{2},...,n_k+\frac{1}{2})$. Note,
that in fact,  for the massless case the above equivalence follows
from the Corollary 2 and comments with (\ref{hermpres}), (\ref{equiv32m}) applied for topological dynamical system,
where the selection of the solutions $\mathcal{H}_{1,2}^{phys}$
for the  equations (\ref{Eq-0})--(\ref{Eq-2})  in $\mathcal{H}$
are written in (\ref{equiv41sec}), whereas the second-order gauge
Lagrangian dynamics are described by the second row
(\ref{equiv41Brst}) with $Q_{(n)_k}$ acting on the Fock space
subspace, $\mathcal{H}{}^{{\sigma}^i (G)}_{c|tot}$   $\subset
\mathcal{H}_{c|tot}$,   playing the role of $Q_c$ in  $ H(Q_c)$.

Let us consider the HS symmetry  superalgebra  $\mathcal{A}^f(Y(k),
\mathbb{R}^{1,d-1})$ realization with use of the initial set of gamma-matrices $\gamma^\mu$ instead of Grassmann-odd ones $\tilde{\gamma}^\mu$ for massless case following to
\cite{totfermiMin}. Doing so, it is natural to assume, that, the Grassmann parity, $\varepsilon$,  of $\gamma^\mu$ is not vanishing but composed from two summands, $\varepsilon_{ISO}$,  $\varepsilon_{BFV}$  induced by the Poincare group $ISO(1,d-1)$ amd from the BRST-BFV method, respectively. It is explicitly  given by the rule
\begin{equation}\label{grassmapargam}
 \varepsilon(\gamma^\mu) = \varepsilon_{ISO}(\gamma^\mu)+\varepsilon_{BFV}(\gamma^\mu) = 0+1=1,
\end{equation}
so that $\gamma^\mu$ have the same properties as for $\tilde{\gamma}^\mu$ of anticommuting with Grassmann-odd ghost oscillators: $\{\gamma^\mu, A\} = 0$ for
\begin{equation}\label{grassmapargam1}
 \varepsilon(A) = \varepsilon_{ISO}(A)+\varepsilon_{BFV}(A) = 0+1=1, \ A\in \{ \eta_i^{(+)}, \eta_{ij}^{(+)},  \vartheta^{(+)}_{rs}, \eta_0, \mathcal{P}_0, \mathcal{P}^{(+)}_i,
\mathcal{P}^{(+)}_{ij}, \lambda^{(+)}_{rs},
\mathcal{P}^{i}_g, \eta^{i}_g\}
\end{equation}
(for $\big(\varepsilon_{ISO},\varepsilon_{BFV}\big) \Psi_{(\mu^1)_{n_1},...,(\mu^k)_{n_k}} = (1,0)$) and commuting with rest Grassmann-even ones. Thus, instead of $t_0, t_i, t_i^+$ the  superalgebra  $\mathcal{A}^f(Y(k),
\mathbb{R}^{1,d-1})$ will contain the  constraints $\big(\tilde{t}_0, \tilde{t}_i, \tilde{t}_i^+\big)$ = $\big(-i\gamma^{\mu}\partial_\mu\,,  \gamma^{\mu}a^i_\mu\,,\gamma^{\mu}a^{i+}_\mu \big)$ (\ref{totit12}) with unchanged multiplication table (\ref{table in}). Note, first, the Grassmann parities both of BRST-like second order  (un)constrained actions (\ref{secorder}), (\ref{2ordercon}) below and for its analogs constructed with use of $\gamma^{\mu}$ are equal to $1$ due to odd-scalar product nature (\ref{sproduct}), where $\varepsilon(\tilde{\gamma}^0)=\varepsilon({\gamma}^0)=1$. Second, when the whole ghost oscillators pairing have calculated, it is natural to factorize $\varepsilon$  with respect to $\varepsilon_{BFV}$-parity for $\gamma$-matrices passing to factor Grassmann parity: $ \varepsilon /  \varepsilon_{BFV}$. Third, the first order (un)constrained actions (\ref{Lun}), (\ref{Lcon}) appear by bosonic functionals with respect to $ \varepsilon$-parity (\ref{grassmapargam}). The (un)constrained Lagrangian formulations for massless half-integer HS fields in Minkowski space-time with use of $\gamma^\mu$ matrices are equivalent to ones constructed using $\tilde{\gamma}^\mu$ matrices for any dimension $d$.

Now, we have all the tools to pass to construction of the
constrained BRST--BFV approach for the same initial fermionic HS
field.
\section{Derivation of the constrained BRST operator,
spin operator and BRST-extended algebraic constraints}\label{constrBRST}
\setcounter{equation}{0}

Here, we consider the same main objects as for unconstrained BRST approach  but with extracted set of second-class constraints $o_a, o^+_a$  together with  $g_0^i$ (related to $\Delta_{ab}(g_0)$ in (\ref{inconstraintsd}))
from the HS symmetry superalgebra $\mathcal{A}^f(Y(k), \mathbb{R}^{1,d-1})$, which should be  imposed  on $\mathcal{H}_{tot}$ in compatible way  with the constrained  BRST and spin operator  for the superalgebra of the rest first-class constraints $o_A$ with space-time derivatives only.

\subsection{Reduction  from unconstrained BRST operator }\label{constr-unconstr}

Starting from the representation (\ref{Q'decomp})--(\ref{Q}) for
BRST--BFV operator $Q'$ (\ref{Q'k})  written for the  converted
constraints  $O_I$,  let us extract from $Q'$ (hence from $Q$,
$\sigma^i(G)$) the  terms corresponding to the  Minkowski space
$\mathbb{R}^{1,d-1}$ isometries subsuperalgebra (being the ideal
in $\mathcal{A}^f(Y(k), \mathbb{R}^{1,d-1})$) of  the  first-class
constraints subsystem $\{o_A\}$ and the summands  corresponding to
superalgebra isomorphic to $osp(1|2k)$ of the second--class
constraints $\{O_a,O_a^+\}$ only:\footnote{In this Section and
later on, we understand the subscript ``$C$'' in $\sigma^i_C(G)$,
$Q_C(o_A)$, $|\chi^l_c\rangle$ in the sense that the corresponding
objects belong to the constrained BRST--BFV approach unless stated
otherwise.}
\begin{eqnarray}
\label{sigmaidec}
  \sigma^i(G) &=& \sigma^i_C(G) + \sigma^i(O_a,O_a^+) , \quad
 \sigma^i_C(G) \ = \  G_0^i   - \eta_i \mathcal{P}^+_i +
   \eta_i^+ \mathcal{P}_i,\\
    \sigma^i(O_a,O_a^+) &=&   \sum_{
m}(1+\delta_{im})(
\eta_{im}^+{\cal{}P}^{im}-\eta_{im}{\cal{}P}^+_{im})+   q_ip_i^+ + q_i^+p_i + \sum_{l<i}[\vartheta^+_{li}
\lambda^{li} \label{sigmaisec} \\
   &&   - \vartheta^{li}\lambda^+_{li}]-
\sum_{i<l}[\vartheta^+_{il} \lambda^{il} -
\vartheta^{il}\lambda^+_{il}] ,\nonumber
 \\
\label{Qdec} {Q} \hspace{-0.4em} &=&\hspace{-0.4em} Q_C(o_A) +Q(\mathcal{O}_a,\mathcal{O}_a^+) ,\\
 \label{Qc}
 Q_C(o_A)\hspace{-0.4em} &=&\hspace{-0.4em}  q_0t_0+\eta_0l_0+\eta_i^+l^i+l^{i+}\eta_i
 + {\imath}\bigl(\sum_l\eta_l^+\eta^l-q_0^2\bigr){\cal{}P}_0,
\\
 \label{Qsec}
Q(\mathcal{O}_a,\mathcal{O}_a^+)
\hspace{-0.4em} &=&\hspace{-0.4em}  q_i^+\mathcal{T}^i
+\sum\limits_{l\leq m}\eta_{lm}^+\mathcal{L}^{lm} +
\imath\sum_l
\eta_l^+q^lp_0  +  \sum\limits_{l<
m}\vartheta^+_{lm}\mathcal{T}^{lm}+ \Bigl[\frac{1}{2}\sum_{l,m}(1+\delta_{lm})\eta^{lm}q_l^+
\\
\hspace{-0.4em}
&&\hspace{-0.4em} -\sum_{l<m}
q_l\vartheta^{lm}
 -\sum_{m<l}q_l\vartheta^{ml+}\Bigr]p_m^+   -2 \sum_{l<m}q_lq_m^+\lambda^{lm} -2\sum_{l,m}
q^+_lq^+_m\mathcal{P}^{lm} -\sum\limits_{i<l<j}
\vartheta^+_{lj}\vartheta^+_{i}{}^l \lambda^{ij} \nonumber
\\
\hspace{-0.4em}&& {}\hspace{-0.4em}  -
\sum\limits_{l<n<m}\vartheta_{lm}^+\vartheta^{l}{}_n\lambda^{nm} +
\sum\limits_{n<l<m}\vartheta_{lm}^+\vartheta_{n}{}^m\lambda^{+nl} -
\sum_{n,l<m}(1+\delta_{ln})\vartheta_{lm}^+\eta^{l+}{}_{n}
\mathcal{P}^{mn}
\nonumber\\&&+
\sum_{n,l<m}(1+\delta_{mn})\vartheta_{lm}^+\eta^{m}{}_{n}
\mathcal{P}^{+ln}+ \textstyle\frac{1}{2}\sum\limits_{l<m,n\leq
m}\eta^+_{nm}\eta^{n}{}_l\lambda^{lm}
+h.c. \nonumber
\end{eqnarray}
with unchanged operator $\mathcal{B}^i$ (\ref{Bi}).
The operator $Q(\mathcal{O}_a,\mathcal{O}_a^+)$ depends on the same, but  extended  by means of the ghost coordinates and momenta corresponding to the
first-class constraints  $\{o_A\}$ set of the second-class constraints:
\begin{align}\label{brstextended0}
 &\mathcal{T}_i = T_i(B, B^+)-\imath \eta_ip_0 -2q_0 \mathcal{P}_i,  & \mathcal{L}_{lm}= L_{lm}(B, B^+)+\frac{1}{2}\eta_{\{m} \mathcal{P}_{l\}},  \ l\leq m\\
  &\label{brstextended1} \mathcal{T}_{rs}= T_{rs}(B, B^+)-\eta^+_r\mathcal{P}_s -\mathcal{P}^+_r  \eta_s , \ r< s &
\end{align}
and respective Hermitian conjugated constraints $\mathcal{O}^+_a$.  The set of the constraints $\{ \mathcal{T}_i, \mathcal{L}_{lm}$, $\mathcal{T}_{rs};\mathcal{T}^+_i$, $\mathcal{L}^+_{lm}, \mathcal{T}^+_{rs}\} $ augmented by the constrained spin operator $\sigma^i_c(G)$
satisfies to the same algebraic relations as the superalgebra of $\{o_a,\,o_a^+; g_0^i\}$ given by Table~\ref{table in} under replacement $(o_a,\,o_a^+; g_0^i)$ $\to$ $\big(\mathcal{O}_a,\,\mathcal{O}_a^+; \sigma^i_C(G)\big)$.  For vanishing of the  additional parts $o'_a, o^{\prime +}_a$ (equivalently, quotient of  $\mathcal{H}_{c|tot}$ by Fock subspace generated by  $B^{a+}$)  the same superalgebra  for $\big([\mathcal{O}_a,\,\mathcal{O}_a^+]\big|_{(B=B^+=0)}; \sigma^i_C(g)\big)$ acting on $\mathcal{H}\otimes \mathcal{H}_{gh|m} $ holds.  We will call the set of the constraints $\{\mathcal{O}_a,\,\mathcal{O}_a^+\}\big|_{(B=B^+=0)}$ as the \emph{BRST extended constraints}.

From the supercommutator's relations it follows that algebraically independent set   for a half of the  BRST extended constraints, $\mathcal{O}_a$ is composed from
$k $ elements:
\begin{equation}\label{Obara}
   \{\mathcal{O}_{\check{a}}\}  = \{\mathcal{T}_1,\, \mathcal{T}_{12},\, \mathcal{T}_{23},\, \mathcal{T}_{34},\,\ldots , \mathcal{T}_{(k-1) k}\}.
\end{equation}
Indeed, the rest (dependent) elements $\{\mathcal{O}_{\mathbf{a}}\}$  from $\{\mathcal{O}_a\}= \{\mathcal{O}_{\check{a}},\,\mathcal{O}_{\mathbf{a}}\}$ are generated as follows:
\begin{align}\label{depel1}
&\mathcal{T}_i = \mathrm{ad}_{\mathcal{T}_{(i-1)i}}\ldots \mathrm{ad}_{\mathcal{T}_{12}}\mathcal{T}_{1}   &\mathcal{L}_{lm} = \frac{1}{4}   \mathrm{ad}_{\mathcal{T}_l} \mathrm{ad}_{\mathcal{T}_{(m-1)m}}\ldots \mathrm{ad}_{\mathcal{T}_{12}}\mathcal{T}_{1} \\
\label{depel2}  &\mathcal{T}_{rs} = (-1)^{s-1-r} \mathrm{ad}_{\mathcal{T}_{r (r+1)}}\ldots \mathrm{ad}_{\mathcal{T}_{(s-2)(s-1)}}\mathcal{T}_{(s-1)s}   &
\end{align}
for $\mathrm{ad}_BA \equiv [B,A\}$.

The nilpotency of unconstrained BRST--BFV operator $Q'$ for
$\mathcal{A}^f(Y(k), \mathbb{R}^{1,d-1})$ with following from it
equations  on unconstrained operators $Q$, $\sigma^i(g)$
(\ref{eqQ1}), but in $\mathcal{H}_{tot}$, leads to the nontrivial
equations for extracted operators
(\ref{sigmaidec})--(\ref{Qsec}):
\begin{align}\label{eqQc1}
  & Q_C^2 = 0, &    [Q_C,\, {\sigma}^i_C(g)\}= 0,   \\
   \label{eqQc2} &   [Q_C,\, Q(\mathcal{O}_a,\mathcal{O}_a^+)\}= 0, &  [Q(\mathcal{O}_a,\mathcal{O}_a^+),\, {\sigma}^i_C(G)\}=    [{\sigma}^i(O_a,O_a^+),\,Q(\mathcal{O}_a,\mathcal{O}_a^+)\} , \\
    & Q^2(\mathcal{O}_a,\mathcal{O}_a^+) + \imath \sum_j\mathcal{B}^j {\sigma}^j (G) = 0 .& \label{eqQc3}
\end{align}
Thus, we see from (\ref{eqQc1})  that  BRST--BFV operator $Q_C$ is
nilpotent and supercommute with constrained spin operator
${\sigma}^i_C(g)$ everywhere in $\mathcal{H}_{tot}$,  whereas the
BRST--BFV operator  $Q(\mathcal{O}_a,\mathcal{O}_a^+)$ according
to (\ref{eqQc3}) is nilpotent only on the proper eigen states
$|\chi\rangle$  for generalized spin operator  ${\sigma}^i
(G)|\chi\rangle=0$.  The set of $\mathcal{O}_a,\mathcal{O}_a^+$
when acting on such $|\chi\rangle$ is classified as the
first-class constraint system converted from the second-class
constraints $o_a,o_a^+$ acting on the Hilbert subspace
$\mathcal{H}\otimes H^{o_A}_{gh}$. From the supercommutativity of
both BRST--BFV operators $Q_C,\, Q(\mathcal{O}_a,\mathcal{O}_a^+)$
in  the left equation  in (\ref{eqQc2})  it follows at  the first
degree in  ghost operator $C^\alpha =(C^{a+}, C^a)$  the
supercommutativity of $Q_C$ and $\mathcal{O}_a,\mathcal{O}_a^+$:
\begin{equation}\label{supercommQcO}
  [Q_C,\, \mathcal{O}_a\big|_{(B=B^+=0)} \}= 0  \Rightarrow  \left( [Q_C,\, \mathcal{O}_a\big|_{(B=B^+=0)} \}\right)^+ =  [ \mathcal{O}^+_a\big|_{(B=B^+=0)},\,Q_C \}= 0.
\end{equation}
Finally, from the left-hand side of the  last equations in (\ref{eqQc2}) at  the first degree in  ghost operator $C^\alpha$ we have
 \begin{eqnarray}\label{commOsig}
  && [\mathcal{O}_a\big|_{(B=B^+=0)},  {\sigma}^i_C(g)\}= K_a^i \mathcal{O}_a\big|_{(B=B^+=0)}, \ \  K_a^i \in \mathbb{Z}: \\
  && \mathrm{for}    \  K_a^i\equiv \big( K_j^i,\,  K_{lm}^i,\, K_{rs}^i\big)=\big(\delta_j^i,\, \delta_{lm}(\delta_{l}^i+\delta_m^i),\,  \delta_{rs}(\delta_{s}^i-\delta^i_{r})\big) \nonumber
 \end{eqnarray}
respectively for $\mathcal{O}_a\big|_{(B=B^+=0)}=(\mathcal{T}_j, \mathcal{L}_{lm}, \mathcal{T}_{rs})$.
Therefore, the operators $Q_C, \mathcal{O}_a\big|_{(B=B^+=0)}, {\sigma}^i_C(g)$ compose the closed $[\ ,\ \}$-superalgebra acting in  $\mathcal{H}\otimes H^{o_A}_{gh}$.

For further research  we develop the results   known  from the
general theory of constrained system  exposed in the
Section~\ref{BRST--BFVorig}.

Doing so, the application of the Corollary 1  to  the BRST--BFV
operator, $ Q(\mathcal{O}_a,\mathcal{O}_a^+) \equiv
Q_{\mathcal{O}}$ (\ref{Qsec})  for the set of converted
second-class constraints ($Q_{c|2}$, $\Phi_\alpha(\hat{\Gamma}_c)$
in (\ref{equiv31sec}), (\ref{equiv31Brst})), $\mathcal{O}_\alpha =
\big(\mathcal{O}_a, \mathcal{O}_a^+\big)$  from the superalgebra
$\mathcal{A}^f(Y(k),\mathbb{R}^{1,d-1})$ means due to the validity
of the nilpotency for $ Q_{\mathcal{O}}$ only  on the Fock
subspace ${\mathcal{H}}^{{\sigma}(G) }_{tot}$ of
$\mathcal{H}_{c|tot}$ which contains only the  proper eigen
vectors for the spin operator ${\sigma}^i (G)$, that it is true
the

\vspace{1ex}

\noindent
 \textbf{Statement 3:} The Fock subspace,  $H^{\sigma_C^i}_{\mathcal{O}_a}$: $H^{\sigma_C^i}_{\mathcal{O}_a} \subset\mathcal{ H}\otimes H^{o_A}_{gh}$, being proper for the constrained spin operator $\sigma_C^i(g)= \sigma^i(G)\big|_{\mathcal{H}\otimes H^{o_A}_{gh}}$   and for the half of the BRST extended second-class constraints not depending  on the auxiliary (converted) oscillators: $\mathcal{O}_a\big|_{B=B^+=0}$ (\ref{brstextended0}), (\ref{brstextended1}) contains the same (equivalent) set of the states  as the Fock subspace in the quotient $\ker Q_{\mathcal{O}} \diagup {Im }\, Q_{\mathcal{O}}  $ with zero ghost number and being proper for generalized spin operator $\sigma^i(G)$ (\ref{sigmaidec}):
    \begin{eqnarray}\label{equivsec1}
        H^{\sigma_C^i}_{\mathcal{O}_a}& =& \big\{|\chi_c\rangle| \  \big(\mathcal{O}_a\big|_{B=B^+=0}, \,\sigma_C^i(G),\, gh\big) |\chi_c\rangle =(0,0 ,0),  \  \  |\chi_c\rangle\in \mathcal{H}\otimes H^{o_A}_{gh} \big\} \\
       &=&   \big\{|\chi^0\rangle|\    \delta|\chi^0\rangle=Q_{\mathcal{O}}|\chi^1\rangle, ...,  \delta|\chi^{s-1}\rangle=Q_{\mathcal{O}}|\chi^s\rangle, \delta|\chi^s\rangle=0,\, \   |\chi^k\rangle \in  \mathcal{ H}^{\sigma^i(G)|k}_{tot} \big\}, \label{equivsec2}
    \end{eqnarray}
 with $gh(|\chi^k\rangle)=-k$ for  $k=0,...,s$, $s= \sum_{o=1}^k n_o + k(k-1)/2$.
 \vspace{1ex}

 From the  Statement 3 it follows the important in practice

 \noindent
 \textbf{Corollary 3}: A set of the states  $H^{\sigma_C^i}_{\mathcal{O}_a, o_A}$: $H^{\sigma_C^i}_{\mathcal{O}_a,o_A} \subset \mathcal{H}\otimes H^{o_A}_{gh}$  with vanishing ghost number from the Fock subspace: $\ker Q_C \diagup {Im }\, Q_C$,  with nilpotent BRST--BFV operator $Q_C$ (\ref{Qc}) for the subsuperalgebra of constraints $o_A$ acting in   $\mathcal{H}\otimes H^{o_A}_{gh}$  being proper eigen-states both for constrained spin operator $\sigma^i_C(g)$  and  annihilated by  the half of the BRST extended second-class constraints $\mathcal{O}_a\big|_{B=B^+=0}$   is equivalent to the set of the states from the Fock subspace: $\ker Q \diagup {Im }\, Q$, with  BRST--BFV operator $Q$ (\ref{Qdec}) for only  system of constraints $\{O_I\setminus G_0^i\}$ in the  HS symmetry superalgebra $\mathcal{A}^f(Y(k),\mathbb{R}^{1,d-1})$ to be  nilpotent on the proper eigen-states  for generalized spin operator $\sigma^i(G)$
(\ref{sigmaidec}) acting in $\mathcal{H}_{tot} = \mathcal{H}\otimes H' \otimes H_{gh}$:
\begin{eqnarray}\label{equivsec11}
       H^{\sigma_C^i}_{\mathcal{O}_a, o_A}& =& \big\{|\psi_c\rangle| \  \big(\mathcal{O}_a\big|_{B=B^+=0}, \,\sigma_C^i(g),\, gh\big) |\psi_c\rangle =(0, 0 ,0),  \  \  |\psi_c\rangle \in   \ker {Q}_C\diagup {Im }\, Q_C \big\} \\
       &=&   \big\{|\psi\rangle|\ \big(\sigma^i(G),\, gh\big) |\psi\rangle =(0, 0 ),  \  \      |\psi\rangle \in \ker Q \diagup {Im }\, Q \big\}. \label{equivsec21}
    \end{eqnarray}

Equivalently, in terms of the respective $Q$- and    $Q_C$-complexes  the equivalence above means that  the found set of states,   $H^{\sigma_C^i}_{\mathcal{O}_a,o_A}$ may be presented according to (\ref{equivsec1}), (\ref{equivsec2}) as:
\begin{eqnarray}\label{equivsec1f}
       \hspace{-0.7em} H^{\sigma_C^i}_{\mathcal{O}_a,o_A}& \hspace{-0.7em}=&\hspace{-0.7em} \big\{|\chi^0_c\rangle| \  \big(Q_C,\mathcal{O}_a\big|_{B=B^+=0}, \sigma_C^i(g), gh\big) |\chi^{l}_c\rangle =(\delta|\chi^{l-1}_c\rangle ,0 ,0,-l) ,  |\chi^l_c\rangle\in \mathcal{H}\otimes H^{o_A|l}_{gh}   \big\} \\
       \hspace{-0.7em}&\hspace{-0.7em}=& \hspace{-0.7em}  \big\{|\chi^0\rangle|\ \big(Q,\,\sigma^i(G),\, gh\big),  |\chi^m\rangle = (\delta|\chi^{m-1}\rangle  ,0,-m) \, \   |\chi^m\rangle \in  \mathcal{ H}^m_{tot}\, m=0,...,s \big\}, \label{equivsec2f}
    \end{eqnarray}
 for $|\chi^{-1}_c\rangle =|\chi^{-1}\rangle=0$, $l=0,1,..., k$  and $s= \sum_{o=1}^k n_o + k(k-1)/2$.
 \vspace{1ex}

 Note, that due to algebraic independence of the set of $k$ constraints $\{\mathcal{O}_{\check{a}}\}  = \{\mathcal{T}_1, \mathcal{T}_{(m-1)m}\}, m=2,...,k$ (\ref{Obara}) among $\mathcal{O}_a$ it is sufficient to impose $\{\mathcal{O}_{\check{a}}\}$ instead of $\mathcal{O}_a$ in the (\ref{equivsec1}), (\ref{equivsec11})  and (\ref{equivsec1f}).

 Corollary 3 represents the basis for the equivalent description of the Lagrangian dynamics for free half-integer HS field on $\mathbb{R}^{1,d-1}$ subject to $Y(s_1,...,s_k)$  both on a base of unconstrained BRST method  and in terms  of  constrained gauge-invariant  Lagrangian formulation with use of constrained BRST approach. Note, earlier the Lagrangian dynamics for free half-integer HS field was developed only within unconstrained BRST method. Because of, HS symmetry algebra $\mathcal{A}(Y(k),\mathbb{R}^{1,d-1})$ \cite{BuchbinderRmix} for free integer HS field on $\mathbb{R}^{1,d-1}$ subject to $Y(s_1,...,s_k)$, in fact,  contains in the $\mathcal{A}^f(Y(k),\mathbb{R}^{1,d-1})$ the  Corollary 3 solves the same problem for free integer HS field.

Hence, in order to prove that the set of solutions for the irreducibility representations conditions (\ref{Eq-0})--(\ref{Eq-2}) [or in the form (\ref{t0t1t}), (\ref{gocond})] for half-integer HS  fields of fixed spin is equivalent to the Lagrangian equations of motion for more wider set of HS fields to be subject to the reducible gauge transformations, it is enough to  consider the constrained gauge-invariant  Lagrangian formulation.

\subsection{Self-consistent constrained BRST, spin operators,
off-shell constraints from HS symmetry superalgebra}\label{constr-self}

The constrained operator quantities $Q_C, \sigma^i_C(g) $ and algebraically independent BRST extended constraints $\{{O}_{\bar{a}}\}$ (\ref{Obara}) may be derived without appealing to the unconstrained BRST formulation starting explicitly from the equations  (\ref{Eq-0})--(\ref{Eq-2}), thus revelling the relation with so called tensionless limit of open superstring theory \cite{tensionlessl}, which contains many higher-spin excitations in its spectrum.

Starting from the Poincare group irreducibility conditions (\ref{Eq-0})--(\ref{Eq-2}) extracting the  massless HS field of spin  $\mathbf{s} = (n_1 +1/2, n_2+1/2,
 ... , n_k+1/2)$, realized  equivalently for Fock space vector (\ref{PhysState}) by the equations (\ref{t0t1t}), (\ref{gocond}), we consider only the
subsuperalgebra  of Minkowski space $\mathbb{R}^{1,d-1}$ isometries (\ref{TkTk}) $\{o_A\}=\{t_0, l_0, l_i, l_i^+\}$ as the dynamical constraints being used to construct Lagrangian formulation,  whereas the rest (algebraic) primary constraints $t_i$, $t_{rs}$ (\ref{t0ti}), (\ref{totit12}) should be imposed as off-shell (i.e. holonomic) constraints on the respective solutions of the Lagrangian equations   of motion, but in a \emph{consistent}  way. From the algebraic viewpoint, the set of operators   $t_i$, $t_{rs}$ appears by the derivations of the subalgebra  $\{o_A\}$ : $[A,  o_A\} \in  \{o_A\}$ for any $A \in \{t_i, t_{rs}\}$. The constraints $t_i$, $t_{rs}$ itself generates the superalgebra of total set of constraints $\{o_a\} = \{t_i, t_{rs}, l_{lm}\}$ being by the restriction of $\mathcal{O}_a$ (\ref{depel1}) on  the initial Fock space $\mathcal{H}$. The algebraically independent set of $\{o_a\}$ according to (\ref{Obara}), (\ref{depel1}) is given by:
\begin{equation}\label{obara}
   \{o_{\bar{a}}\}  = \{t_1,\, t_{12},\, t_{23},\, t_{34},\,\ldots , t_{(k-1) k}\}.
\end{equation}
The constrained BRST operator $Q_C$  for the system of the first-class constraints $\{o_A\}$ in the Hilbert space $\mathcal{H}_C$,  $\mathcal{H}_C =\mathcal{H}\otimes H^{o_A}_{gh}$ has the form (\ref{Qc}). The \emph{consistency} condition requires that the set of off-shell constraints $t_i$, $t_{rs}$ as well as the number particle operator being additively extended  in powers of the ghosts variables ${C}^A, \overline{\mathcal{P}}_A$ in  $\mathcal{H}_C$ to $\widehat{T}_i$, $\widehat{T}_{rs}$ and $\widehat{\sigma}{}^i_C(g)$:
\begin{equation}\label{extconstr}
  \Big(\widehat{T}_i,\,\widehat{T}_{rs},\, \widehat{\sigma}{}^i_C(g)\Big) =  \Big(t_i +  {{C}}^A t_{iA}^B \overline{\mathcal{P}}_B ,\,t_{rs}+{{C}}^A t_{rs A}^B \overline{\mathcal{P}}_B,\, g_0^i+{{C}}^A g_{A}^{iB} \overline{\mathcal{P}}_B\Big)+ o({C\overline{\mathcal{P}}})
\end{equation}
with preservation of the vanishing ghost number:  $gh\Big(\widehat{T}_i,\,\widehat{T}_{lm},\, \widehat{\sigma}{}^i_C(g)\Big)=(0,0,0)$ should satisfy to the  equations
\begin{align}\label{eqQctot}
  & [Q_C,\, \widehat{T}_i\}    = 0, &   [Q_C,\, \widehat{T}_{rs} \}= 0, &&   [Q_C,\, \widehat{\sigma}{}^i_C(g)\}= 0,
\end{align}
which we call by the \emph{generating equations} for superalgebra of the constrained BRST, $Q_C$ spin operators $\widehat{\sigma}{}^i_C(g)$ and extended in $\mathcal{H}_C$  off-shell constraints $\widehat{T}_i,\,\widehat{T}_{rs}$.

Explicit calculations show, that the solutions for unknown  $\widehat{\sigma}{}^i_C(g)$, $\widehat{T}_i,\,\widehat{T}_{rs}$  exists for quadratic  in powers of ghosts
${C}^A, \overline{\mathcal{P}}_A$ approximation in the form:
\begin{equation}\label{solextconstr}
 {\mathcal{C}}^A \Big(t_{iA}^B,\, t_{rsA}^B,\, g_{A}^{iB}\Big) \mathcal{P}_B =  \Big(-\imath \eta_ip_0 -2q_0 \mathcal{P}_i ,\, -\eta^+_r\mathcal{P}_s -\mathcal{P}^+_r  \eta_s,\,  \eta_i^+ \mathcal{P}_i- \eta_i \mathcal{P}^+_i
  \Big).
\end{equation}
Comparison with the structure of the BRST  extended constraints ${\mathcal{T}}_i,\,{\mathcal{T}}_{rs}$ (\ref{brstextended0})   and ${\sigma}^i_C(G)$ (\ref{sigmaidec})
restricted to $\mathcal{H}_C$, permits to state on their coincidence (for $h^i=0$ in case  ${\sigma}^i_C(g)={\sigma}^i_C(G)\big|_{B=0}$):
\begin{equation}\label{coinbrstcon}
  \Big(\widehat{T}_i,\,\widehat{T}_{rs},\,\widehat{\sigma}^i_C(g)\Big)\ = \ \Big({\mathcal{T}}_i,\,{\mathcal{T}}_{rs},\,{\sigma}^i_C(g)\big|_{h^i=0}\Big).
\end{equation}
Therefore, the superalgebra of the constrained  operators $\{Q_C, \widehat{T}_i,\,\widehat{T}_{rs},\,\widehat{\sigma}^i_C(g)\}$ derived in self-consistent way coincides with
the superalgebra $\{Q_C, {\mathcal{T}}_i,\,{\mathcal{T}}_{rs},\, {\sigma}^i_C(g)\big|_{h^i=0}\}$ of the operators  derived from unconstrained BRST formulation.

As to the difference of the  constrained spin operators: \emph{self-consistent}, $\widehat{\sigma}^i_C(g)$, and ${\sigma}^i_C(g)$
 then because of the coincidence the set of proper eigen-vectors, $|\chi^l_c\rangle_{(n)_k} \in \mathcal{H}_C$, ($gh(|\chi^l_c\rangle)=-l$)   (e.g.  due to its supercommutativity: $[\widehat{\sigma}^i_C(g),\,{\sigma}^i_C(g)\}=0$)
 the corresponding equations:
 \begin{equation}\label{spindiff}
  \widehat{\sigma}^i_C(g) |\chi^l_c\rangle_{(n)_k} = \left( n^i+\frac{d-2}{2}\right)|\chi^l_c\rangle_{(n)_k},\qquad {\sigma}^i_C(g)|\chi^l_c\rangle_{(n)_k}=0\ \ \mathrm{for} \ h^i_{C}= - n^i-\frac{d-2}{2}
 \end{equation}
with stabilized  $h^i_{C}\equiv  h_C$, $\forall i=1,...,k$,   in fact, determine identical spin value distribution:
\begin{eqnarray}\label{nidecomposfconst}
n_i &= & p_i+
n_{f{}i}+n_{p{}i} \,,\  i=1,\ldots,k,
\end{eqnarray}
for $|\chi_c\rangle$ having the representation
\begin{eqnarray}
|\chi_c \rangle &=& \sum_n q_0^{n_{b{}0}}\eta_0
^{n_{f 0}} \prod_{i, j}( \eta_i^+ )^{n_{f i}} (
\mathcal{P}_j^+ )^{n_{p j}} |\Psi(a^+_i)^{n_{b{}0} n_{f 0};   (n)_{f i}(n)_{p j}}\rangle \,,\label{chifconst}
\end{eqnarray}
following from (\ref{chif}) for $B^+=0$, $({C}^a, \overline{\mathcal{P}}_a) =0 $ and for
\begin{equation*}
|\Psi(a^+_i)^{n_{b{}0} n_{f 0};   (n)_{f i}(n)_{p j}}\rangle = |\Psi(a^+_i)^{n_{b{}0} n_{f 0};  (0)_{a{}e} (0)_{b{}g} (n)_{f i}(n)_{p j}(0)_{f lm}
(0)_{pno}(0)_{f rs}(0)_{\lambda
tu}}_{(0^0)_{c};(0)_{ij}(0)_{rs}}\rangle.
\end{equation*}
 From (\ref{chifconst}) [equivalently from (\ref{ghnumf})]
it follows the ghost number distribution for $n'$s:
\begin{eqnarray}\label{ghnumconnstr}
  |\chi^l_c\rangle_{(n)_k}  &:& n_{b{}0}+ n_{f{}0}+
     \sum_{i}\bigl(n_{f{}i}- n_{p{}i}\bigr)= -l .
\end{eqnarray}
Thus, we establish the fact, that  the same set of states  $H^{\sigma_c^i}_{(\mathcal{O}_a,o_A)}$ (\ref{equivsec1f}) from Corollary 3  is equivalently reproduced both by the superalgebra of constrained operators: $\{Q_C, {\mathcal{T}}_i,\,{\mathcal{T}}_{rs},\, {\sigma}^i_C(g)\}$ derived from unconstrained  superalgebra $\{Q, \, {\sigma}^i(G)\}$ and one of $\{Q_C, \widehat{T}_i,\,\widehat{T}_{rs},\, \widehat{\sigma}^i_C(g)\}$. Therefore, the two ways of derivation of the constrained BRST Lagrangian formulation  (self-consistently and from unconstrained Lagrangian formulation) are equivalent.

\section{Constrained gauge-invariant Lagrangian formulations}\label{constr-Lagr}
\setcounter{equation}{0}

In this section. we consider the  construction  of constrained gauge-invariant Lagrangian formulations for mixed-symmetric HS fields in case of massless field with fixed generalized half-integer spin, then in case of one with  fixed generalized integer spin, and in case of massive fields.

 \subsection{Constrained  Lagrangian formulation for half-integer HS fields}\label{constr-Lagrhalf-int}

From the nilpotent BRST operator $Q_C$, spin operator $\sigma^i_C$ ($\sigma^i_C \equiv \widehat{\sigma}^i_C(g)$), algebraically independent  BRST extended constraints  $\widehat{T}_i,\,\widehat{T}_{rs}$, $i, r, s=1,...,k$, $r<s$ we have the   spectral
problem, analogous to one for unconstrained case (\ref{Qchi})--(\ref{Qchii}), but  for $|\chi^l_c\rangle \in  \mathcal{H}^l_C$: %
\begin{align}
\label{Qchic} & Q_C|\chi_c\rangle=0, && \sigma^i_C|\chi_c\rangle=\left( n^i+\frac{d-2}{2}\right)|\chi_c\rangle,
&& \left(\varepsilon, {gh}_H\right)(|\chi_c\rangle)=(1,0),
\\
& \delta|\chi_c\rangle=Q_C|\chi^1_c\rangle, &&
 \sigma^i_C|\chi^1_c\rangle=\left( n^i+\frac{d-2}{2}\right)|\chi^1_c\rangle, && \left(\varepsilon,
{gh}_H\right)(|\chi^1_c\rangle)=(0,-1), \label{Qchi1c}
\\
& \ldots\ldots   &&
\ldots\ldots\ldots\ldots   && \ldots\ldots  \nonumber\\
& \delta|\chi^{s_c-1}_c\rangle=Q_C|\chi^{s_c}_c\rangle, &&
 \sigma^i_C|\chi^{s_c}_c\rangle=\left( n^i+\frac{d-2}{2}\right)|\chi^{s_c}_c\rangle, && \left(\varepsilon,
{gh}_H\right)(|\chi^{s_c}_c\rangle)= (s_c+1,-s_c), \label{Qchiic}\\
& \Big(\widehat{T}_i,\,\widehat{T}_{rs}\Big)|\chi^{l}_c\rangle =0,  &&  l=0,1,...,s_c.   &&\label{constBRST}
\end{align}
 Because of the representations (\ref{chifconst})  and (\ref{chif}) the physical state $|\Psi\rangle$ (\ref{PhysState})  contains in $|\chi^0_c\rangle=|\chi_c\rangle$ as it was
 for  $|\chi^0\rangle$ in (\ref{decomptot}) , but with $({C}^A, \overline{\mathcal{P}}_A)$-dependent vector $|\Psi_{Ac}\rangle$, $|\Psi_{Ac}\rangle$ =$ |\Psi_A\rangle$ when ${(B^+_a, \eta_{ij}^+, \mathcal{P}_{ij}^+, \vartheta_{rs}^+,\lambda^+_{rs})=0}$.

 The system (\ref{Qchic})--(\ref{constBRST}) is compatible, due to closedness of the superalgebra $\{Q_C,\, \sigma^i_C,\, \widehat{T}_i,\,\widehat{T}_{rs}\}$. Therefore, its resolution   for the joint set of proper eigen-vectors we start from the   middle set which determined by (\ref{spindiff}) with distributions (\ref{nidecomposfconst}), (\ref{ghnumconnstr}) for the proper eigen-vectors $|\chi^{l}_c\rangle_{(n)_k}$. For fixed spin $(s)_k=(n)_k+(\frac{1}{2},...,\frac{1}{2})$ the solution of the rest equations is written as the  second-order equations of motion  and sequence of the reducible gauge transformations (\ref{Qchic})--(\ref{Qchiic}) with off-shell constraints:
\begin{eqnarray}
&\hspace{-0.9em}& \hspace{-0.9em} Q_C|\chi^0_c\rangle_{(n)_k}=0, \ \delta|\chi^0_c \rangle_{(n)_k}
=Q_{C}|\chi^1_c\rangle_{(n)_k} \,, ...\,,
\delta|\chi^{s_c-1}_c \rangle_{(n)_k}
=Q_C|\chi^{s_c}_c\rangle_{(n)_k},\, \delta|\chi^{s_c}_c\rangle_{(n)_k} =0,\label{Q12c}\\
&& \Big(\widehat{T}_i,\,\widehat{T}_{rs}\Big)|\chi^{l}_c\rangle_{(n)_k} =0, \ \  l=0,1,...,s_c, \ \mathrm{for} \ s_c=
k. \label{Q12cc}
\end{eqnarray}
The corresponding BRST-like constrained  gauge-invariant action
\begin{equation}\label{2ordercon}
  {\cal{}S}^{(2)}_{C|(n)_k}= \int d \eta_0 {}_{(n)_k}\langle\tilde{\chi}{}^0_c|Q_C |\chi^0_c\rangle_{(n)_k}, \
\end{equation}
from which  follows the equations of motion $Q_C|\chi^0_c\rangle_{(n)_k}=0$,
  contains second order operator $l_0$, but less terms in comparison with its unconstrained analog ${\cal{}S}^{(2)}_{(n)_k}$ (\ref{secorder}).

 Again, repeating the procedure of the removing the dependence on $l_0, \eta_0, q_0$ from the BRST operator $Q_{C}$ (\ref{Qc})
and from the whole set of the vectors $|\chi^l_c\rangle_{(n)_k}$ as it was done in the Section~\ref{unconstLF}   by means of partial gauge-fixing
we come to the:
\vspace{1ex}

\noindent
 \textbf{Statement 4:} The first-order constrained gauge-invariant Lagrangian formulation for half-integer HS field, $\Psi_{(\mu^1)_{n_1},...,(\mu^k)_{n_k}}(x)$ with generalized spin $(s)_k=(n)_k+(\frac{1}{2},...,\frac{1}{2})$, is determined by the action,
  \begin{eqnarray}
{\cal{}S}_{C|(n)_k}\hspace{-0.3em} &=&\hspace{-0.3em} \left({}_{(n)_k}\langle\tilde{\chi}^{0}_{0|c}|\  {}_{(n)_k}\langle\tilde{\chi}^{1}_{0|c}|\right)  \hspace{-0.3em}\left(\begin{array}{lr} {t}_0 & \Delta{}Q_C \\
\Delta{}Q_C  &  {t}_0\eta_i^+\eta_i \end{array}
\right)\hspace{-0.3em} \left(\begin{array}{c}   | \chi^{0}_{0|c}\rangle_{(n)_k} \\
|\chi^{1}_{0|c}\rangle_{(n)_k}  \end{array}
\hspace{-0.3em}\right) \ \mathrm{for} \  \Delta{}Q_C= \eta_i^+l_i+ l_i^+\eta_i,\label{Lcon}
\end{eqnarray}
  invariant with respect to the sequence of the reducible gauge transformations (for $s_{c}-1=(k-1)$-being by the  stage of reducibility):
  \begin{eqnarray}
\delta\left(\begin{array}{c}   | \chi^{l(0)}_{0|c}\rangle_{(n)_k} \\
|\chi^{l(1)}_{0|c}\rangle_{(n)_k}  \end{array}
\right) &=& \left(\begin{array}{lr}
\Delta{}Q_C  &  {t}_0\eta_i^+\eta_i \\
{t}_0 & \Delta{}Q_C
\end{array}
\right) \left(\begin{array}{c}   | \chi^{l+1(0)}_{0|c}\rangle_{(n)_k} \\
|\chi^{l+1(1)}_{0|c}\rangle_{(n)_k}  \end{array}
\right) ,  \  \delta\left(\begin{array}{c}   | \chi^{k(0)}_{0|c}\rangle_{(n)_k} \\
|\chi^{k(1)}_{0|c}\rangle_{(n)_k}  \end{array}
\right) =0
\label{GTconst1}
\end{eqnarray}
(for $l =-1,0,...,k-1$ and $| \chi^{-1(m)}_{0|c}\rangle=0$, $m=0,1$)   with off-shell algebraically independent BRST-extended constraints   imposed on the whole set of field and gauge parameters:
\begin{equation}
\widehat{T}_i\Big(|\chi^{l (0)}_c\rangle_{(n)_k}+q_0|\chi^{l (1)}_c\rangle_{(n)_k}\Big)  =0, \  \widehat{T}_{rs}|\chi^{l (m)}_c\rangle_{(n)_k} =0, \  l=0,1,...,k; \ m=0,1 . \label{constralg}
\end{equation}
\vspace{1ex}

The proof is based on the
 extraction  of the zero-mode ghosts $q_0, \eta_0$ from
$Q_C$ and $|\chi^{l}_c\rangle_{(n)_k}$ :
\begin{eqnarray}
\label{strQc} Q_C &=& q_0{t}_0+\eta_0{l}_0 +
\imath(\eta_i^+q_i-\eta_iq_i^+)p_0 -
\imath(q_0^2-\eta_i^+\eta_i){\cal{}P}_0 +\Delta{}Q_C,\\
\label{0chif} |\chi^l_c\rangle &=& \sum_{m=0}^{ k}q_0^m(
|\chi_{0|c}^{l(m)}\rangle +\eta_0|\chi_{1|c}^{l(m)}\rangle), \quad gh(|\chi^{l(m)}_{p|c}\rangle)=-(p+l+m+l), \ p=0,1
\end{eqnarray}
(for $|\chi_{1|c}^{l(k)}\rangle \equiv 0$) which lead
after  gauge-fixing procedure with help of  the equations of motion  and the set of gauge transformations
(\ref{Q12cc})  to removing of all the components in powers of $q_0$, $\eta_0$ vector   except for two fields for each level, $l=0,...,k$.
As the result, the representation
\begin{equation}\label{gfvectorsc}
  |\chi^{l} \rangle_{(n)_k} =  |\chi^{l (0)}_{0} \rangle_{(n)_k} + q_0|\chi^{l (1)}_{0} \rangle_{(n)_k} - i\eta_0 {t}_0 |\chi^{l (1)}_{0} \rangle_{(n)_k}
\end{equation}
is true. Inserting (\ref{gfvectorsc}) in (\ref{constBRST})  the off-shell BRST-extended constraints will be presented by the system of the equations (with omitting spin index $(n)_k$):
\begin{eqnarray}
&\hspace{-1.3em}&\hspace{-1.3em} \begin{array}{lcl}\begin{array}{l}
   \widehat{T}_i|\chi^{l}_c \rangle \ =\              \\
    \widehat{T}_{rs}|\chi^{l}_c \rangle \, =\,
              \end{array} & \begin{array}{l}
\widehat{T}_i\Big(|\chi^{l (0)}_{0|c} \rangle + q_0|\chi^{l (1)}_{0|c} \rangle - i\eta_0 {t}_0 |\chi^{l (1)}_{0|c} \rangle\Big)     =     0   \\
    \widehat{T}_{rs}\Big(|\chi^{l (0)}_{0|c} \rangle + q_0|\chi^{l (1)}_{0|c} \rangle - i\eta_0 {t}_0 |\chi^{l (1)}_{0|c} \rangle \Big)= 0
              \end{array}& \Longleftrightarrow \begin{array}{l}
\widehat{T}_i\Big(|\chi^{l (0)}_{0|c} \rangle + q_0|\chi^{l (1)}_{0|c} \rangle \Big)=0             \\
    \widehat{T}_{rs}\Big(|\chi^{l (0)}_{0|c} \rangle + q_0|\chi^{l (1)}_{0|c} \rangle  \Big)=0
              \end{array}
              \end{array}
                 \label{offconsttic}
\\ \label{offconstti2c}
&\hspace{-1.3em}&\hspace{-1.3em} \quad \mathrm{and} \ t_it_0 |\chi^{l (1)}_{0|c}  \rangle + 2 q_0t_0\mathcal{P}_i |\chi^{l (1)}_{0|c}  \rangle  = 0,
\end{eqnarray}
which in terms of $q_0$-independent vectors is written as:
\begin{eqnarray}
&\hspace{-1em}&\hspace{-1em}  \begin{array}{lll}
t_i|\chi^{l (0)}_{0|c} \rangle - \eta_i|\chi^{l (1)}_{0|c} \rangle  = 0 ,    &         t_i|\chi^{l (1)}_{0|c}\rangle - 2\mathcal{P}_i|\chi^{l (0)}_{0|c} \rangle  = 0, & \mathcal{P}_i|\chi^{l (1)}_{0|c} \rangle =0,   \\
    \widehat{T}_{rs}|\chi^{l (0)}_{0|c} \rangle  =0, &   \widehat{T}_{rs}|\chi^{l (1)}_{0|c} \rangle = 0. &
              \end{array} .\label{compoffshelc}\\
&\hspace{-1em}&\hspace{-1em} \quad \mathrm{and} \      \ \          l_i |\chi^{l (1)}_{0|c}\rangle= - \mathcal{P}_it_0 |\chi^{l (0)}_{0|c}\rangle. \label{compoffshel11c}
\end{eqnarray}
To get (\ref{compoffshel11c})  from (\ref{offconstti2c}) we have used the last  equations in the first row of  (\ref{compoffshelc}) and operator identity, $t_it_0  = l_i-t_0t_i $.

 The solution for the gamma-traceless equations (\ref{compoffshelc}) leads to  $\eta_i^+$-independence of $|\chi^{l (1)}_{0|c} \rangle_{(n)_k} $ and, at most, linear in $\eta_i^+$-dependence of $|\chi^{l (1)}_{0|c} \rangle_{(n)_k}$, due to algebraic consequences from the second and third equations in (\ref{compoffshelc}): $\mathcal{P}_i\mathcal{P}_j|\chi^{l (0)}_{0|c} \rangle =0$ and $(\mathcal{P}_i)^2=0$.

 Consider, now the  half of the gauge transformations and equations of motion (for $l=-1$):
  \begin{eqnarray}
\delta  | \chi^{l(1)}_{0|c}\rangle_{(n)_k}  &=&  {t}_0| \chi^{l+1(0)}_{0|c}\rangle_{(n)_k} +  \Delta{}Q_C
|\chi^{l+1(1)}_{0|c}\rangle_{(n)_k}   ,  \ l=-1,0,...,k-1,
\label{GTconsthal}
\end{eqnarray}
and apply to them from the left the operator $\mathcal{P}_i$. Because of, $\eta_i^+$-independence of  $| \chi^{l(1)}_{0|c}\rangle_{(n)_k}$,  the constraints  (\ref{compoffshel11c}) coincide for each $l=-1,0,...,k-1$ with respective gauge transformations and equations of motion (for $l=-1$). Therefore, (\ref{compoffshel11c}) do not contain independent  [from the constraints (\ref{compoffshelc}) and sequence of the gauge transformations (\ref{GTconst1})] restrictions  on the vectors $|\chi^{l(m)}_{0|c}\rangle_{(n)_k}$. Because of the off-shell constraints (\ref{compoffshelc}) appear by unfolded decomposition in ghosts  of the BRST-extended   constraints (\ref{constralg})  the Statement~4 is completely proved.
\vspace{1ex}

From the Statement 4 and Corollary 3 it follows that the unconstrained BRST Lagrangian formulation for the fermionic field, $\Psi_{(\mu^1)_{n_1},...,(\mu^k)_{n_k}}(x)$, of spin $(s)_k=(n)_k+(\frac{1}{2},...,\frac{1}{2})$  given by the  action (\ref{Lun}) and sequence of the gauge transformations (\ref{GT1})  and constrained BRST Lagrangian formulation given by (\ref{Lcon})--(\ref{constralg}) for the same spin-tensor field determine the same dynamic, i.e.   equivalent. The respective values of the stage of reducibility are as follows, $s_{\mathrm{max}}-1=\sum_{o=1}^k n_o + k(k-1)/2-1$ and $s_{c}-1= k-1$.

 Hence, in order to prove that the set of solutions for the irreducibility representations conditions (\ref{Eq-0})--(\ref{Eq-2}) [or in the form (\ref{t0t1t}), (\ref{gocond}) ] for half-integer HS  fields of fixed spin is equivalent to the Lagrangian equations of motion for more wider set of HS fields to be subject to the reducible gauge transformations, it is enough to  consider the constrained gauge-invariant  Lagrangian formulation.
Thus, we come to the basic
\vspace{1ex}

\noindent
\textbf{Theorem}: The set of solutions, $H_{(m,(n)_k)}$,    for the equations (\ref{Eq-0})--(\ref{Eq-2}) [or in the form (\ref{t0t1t}), (\ref{gocond})] extracting the Poincare group massless ($m=0$) irreducibile  representation  of  spin $(n)_k+(\frac{1}{2},...,\frac{1}{2})$ in terms of spin-tensor  field, $\Psi_{(\mu^1)_{n_1},...,(\mu^k)_{n_k}}$  is equivalent to the solutions of the Lagrangian equations of motion, for $l=-1$  in (\ref{GTconst1})   subject to the reducible gauge transformations (\ref{GTconst1}) for $l=0,...,k-1$ and off-shell BRST-extended constraints (\ref{constralg}):
\begin{eqnarray}\label{equivconstinit}
        H_{(0,(n)_k)} &\hspace{-0.5em} =& \big\{ |\Psi\rangle |  \, \Big(t_0,\, t_i,\, t_{rs},\, g_0^i - [n_i+d/2]\Big) |\Psi\rangle =0   \big\} \\
       &\hspace{-0.5em}=&  \hspace{-0.5em} \left\{|\chi^{e}_{0|c}\rangle \Big|    \hspace{-0.2em}\left[\hspace{-0.2em}\left(\hspace{-0.5em}\begin{array}{lr} {t}_0 & \Delta{}Q_C \\
\Delta{}Q_C  &  {t}_0\eta_i^+\eta_i \end{array}
\hspace{-0.3em}\right),\  \Big\{\sigma^i_C - n^i-\frac{d-2}{2}\Big\}\mathbf{1}_{2}  \right] \hspace{-0.3em}\left(\hspace{-0.2em}\begin{array}{c}   | \chi^{0}_{0|c}\rangle \\
|\chi^{1}_{0|c}\rangle  \end{array}\hspace{-0.5em}\right) = 0, \right. \nonumber\\
&&  \left.  \delta\left(\begin{array}{c}   | \chi^{l(0)}_{0|c}\rangle \\
|\chi^{l(1)}_{0|c}\rangle \end{array}
\right) = \left(\begin{array}{lr}
\Delta{}Q_C  &  {t}_0\eta_i^+\eta_i \\
{t}_0 & \Delta{}Q_C
\end{array}
\right) \left(\begin{array}{c}   | \chi^{l+1(0)}_{0|c}\rangle \\
|\chi^{l+1(1)}_{0|c}\rangle  \end{array}
\right) ,  \  \delta\left(\begin{array}{c}   | \chi^{k(0)}_{0|c}\rangle \\
|\chi^{k(1)}_{0|c}\rangle  \end{array}
\right) =0
\right. \nonumber\\
&&  \left.\Big(\widehat{T}_i , \widehat{T}_{rs},  \Big\{\sigma^i_C - n^i-\frac{d-2}{2}\Big\} \Big) \Big(|\chi^{(l)0}_{0|c}\rangle+q_0|\chi^{(l)1}_{0|c}\rangle\Big) = 0
\right\},\label{equivconstinit2}
    \end{eqnarray}
where, $e=0,1$; $l=0,...,k-1$;  $\mathbf{1}_2$ appears by unit $2\times 2$ matrix.
\vspace{1ex}

For massive case this theorem is proved for unconstrained BRST
Lagrangian  formulation in \cite{Reshetnyak2} by explicit
resolution of BRST $Q$-complex within first order Lagrangian
formulation. Here we  reached the equivalence in question  for
massless case without above procedure but with using the results
of study the structure of the physical states for topological
dynamical system subject to the first and second-class constraints
on a base of BRST--BFV approach.

There are some consequences from suggested construction. First, the constrained  BRST Lagrangian formulation due to nilpotency of  $Q_C$  on $\mathcal{H}_C$ without spin operator imposing may be used to determine Lagrangian formulation for so called fermionic HS fields with \emph{continuous spin} (i.e. with not restricted by the spin constraints set of spin-tensor fields), suggested for totally-symmetric integer HS fields in $d=4$ in  \cite{Bentssoncont}.  Second, without off-shell constraints (\ref{constralg}) imposing, but with generalized spin condition (given by the middle set in (\ref{Qchic})--(\ref{Qchiic}))  we will have so-called \emph{generalized triplet formulation} for half-integer HS fields on Minkovski space-time with many auxiliary spin-tensors with different  generalized spin values. Third, including of the part BRST-extended constraints, corresponding to the gamma-trace   constraints: $\widehat{T}_i|\chi^{l (m)}_c\rangle_{(n)_k} =0$, (with obvious resolution the mixed-symmetric constraints $\widehat{T}_{rs}$ for whole set of field and gauge parameters $|\chi^{l (m)}_c\rangle_{(n)_k}$) into Lagrangian dynamics with help of  additional gauge transformations and Lagrangian multipliers permits to construct so-called \emph{generalized quartet formulation}  for half-integer HS field  with spin $(s)_k=(n)_k+(1/2)_k$, suggested for totally-symmetric integer and half-integer  HS fields in
\cite{quartmixbosemas}.

 \subsection{Constrained  Lagrangian formulation for integer HS fields}\label{constr-Lagrint}

  Another  direct consequence from the procedure of constrained  Lagrangian formulation for  half-integer HS fields concerns,  the equivalence among the  unconstrained BRST Lagrangian formulation developed in \cite{BuchbinderRmix}, \cite{BurdikReshetnyak}  and constrained BRST Lagrangian formulations for the integer HS irreducible representations of the Poincare group in Minkowski space-time with fixed generalized spin $(s)_k=(s_1,s_2,...,s_k)$, $s_i\geq s_j, i>j$. and constrained BRST Lagrangian formulation, in fact, suggested here in full details (considered without spin operator for totally-symmetric case in \cite{Barnich} and for mixed-symmetric case in  \cite{BRST-BV2}).

 Indeed, the HS symmetry algebra $\mathcal{A}(Y(k),\mathbb{R}^{1,d-1})$ for integer HS fields in $\mathbb{R}^{1,d-1}$ \cite{BuchbinderRmix} (described by tensor fields, $\Phi_{(\mu^1)_{s_1},...,(\mu^k)_{s_k}}(x)$ subject to the d'Alamber, traceless and  divergentless equations:
  \begin{equation}\label{eqbose}
    \Big(\partial^\mu\partial_\mu,\, \partial_\mu^i,\,  \eta^{\mu^i_{m_i} \mu^j_{m_j}}\Big)\Phi_{(\mu^1)_{s_1},...,(\mu^k)_{s_k}}(x) =0,\  1 \leq m_i\leq s_i;\  i,j=1,...,k,
  \end{equation}
  instead of (\ref{Eq-0}), (\ref{Eq-1}) in addtion to the mixed-symmetry conditions (\ref{Eq-2})) may be  obtained from the superalgebra $\mathcal{A}^f(Y(k),\mathbb{R}^{1,d-1})$ by extracting Grassmann-odd generators $t_0,t_i,t_i^+$ with ignoring the matrix-valued representation for the rest operators  $o^B_I$: $\{o^B_I\} = \{o_I\}\setminus \{o^F_p\}$, for $\{o^F_p\}=\{t_0,t_i,t_i^+\}$, as well as for the respective BRST, $Q^B$, generalized  spin operator $\sigma^{i}_B$ not depending on the ghost: $q_0,q^{(+)}_i,p^{(+)}_i,p_0$,   and auxiliary (for conversion) oscillators:  $f_i, f_i^+$, for $\{o^F_p\}$.

  The unconstrained  Lagrangian formulation , in fact, is given  by the second-order Lagrangian action (\ref{secorder}) adapted for integer spin case:
\begin{equation}\label{SunB}
  {\cal{}S}^{B}_{(s)_k}= \int d \eta_0 {}_{(s)_k}\langle\chi^0_B|K^B_{(s)_k}Q^B_{(s)_k} |\chi^0_B\rangle_{(s)_k}, \ \ K^B=1\otimes K^{B\prime}\otimes 1_{gh}:\ K^B_{(s)_k}Q^B_{(s)_k}=(Q^{B})^+_{(s)_k}K_{(s)_k},
\end{equation}
reproducing the Lagrangian equations of motion: $Q^B_{(s)_k}|\chi^0_B\rangle_{(s)_k}=0$, being invariant with respect to reducible gauge transformations
\begin{eqnarray}
 \delta|\chi^l_B \rangle_{(s)_k}
=Q^B_{(s)_k}|\chi^{l+1}_B\rangle_{(s)_k} \,, l=0,1,..., k(k+1)-1...\,,
\delta|\chi^{k(k+1)}_B \rangle_{(s)_k}=0\label{Q12Bose}
\end{eqnarray}
and, thus, determining the gauge theory of $[k(k+1)-1]$-stage of reducibility.
Here, the standard Grassman-even scalar product: $\langle\chi_B|\phi_B\rangle $, for Lorentz- scalar vectors  is used, instead of Grassman-odd scalar product and Lorentz-spinor vectors for the fermionic HS fields. The operator $K^B$ is reduced from $K$ and explicitly given in \cite{BuchbinderRmix} (see for $K^{B\prime}$ denoted in \cite{BuchbinderRmix} as ${K}'$ Eq. (3.14), as well as for the BRST operator  $Q^B$ Eq. (5.2)).
The form of the field ($l=0$) and gauge parameters ($l=1,...,k(k+1)$) $|\chi^l_B\rangle_{(s)_k}$, $gh(|\chi^l_B\rangle)=-l$ is determined accordingly to (\ref{chif}) for $n^0_{c}=n_{b{}0}=n_{a{}e}=n_{b{}g}=0$ with component vector $|\Phi(a^+_i)^{0_{b{}0} 0_{f 0};  (0)_{a{}e} (0)_{b{}g} (n)_{f i}(n)_{p j}(n)_{f lm}
(n)_{pno}(n)_{f rs}(n)_{\lambda
tu}}_{(n^0)_{c};(n)_{ij}(p)_{rs}}\rangle$ (instead of $|\Psi(a^+_i)^{...}_{...}\rangle$) being by the proper eigen-states  for generalized spin operator $\sigma^{i}_B(G) $, $\sigma^{i}_B(G) =  \sigma^{i}(G)\big|_{(q^{(+)}_i=p^{(+)}_i= f^{(+)}_i=0)} $ (\ref{sigmai}),  (\ref{g'0iFf})  for integer spin with value for the  $h^i_B$ in
 \begin{equation}\label{genspineqB}
   {\sigma}^{i}_B (G)|\chi_B\rangle_{(s)_k} = ({\sigma}^{i}_B +h^i_B) |\chi_B\rangle_{(s)_k} = 0 \Longrightarrow- h^i_B = s^i+\frac{d-2-4i}{2} \;, \quad
i=1,..,k.
 \end{equation}
 so that the vector $|\chi^0_B\rangle_{(n)_k}$ contains the physical field $|\Phi\rangle$.
In the corresponding spin (\ref{nidecomposf}) and ghost number (\ref{ghnumf}) distributions for  $|\chi^l_B\rangle_{(s)_k}$ one should put $n^0_{c}=n_{b{}0}=n_{a{}e}=n_{b{}g}=0$ and $n_i=s_i$.

Note, the  Statements 1, 2, 3  are valid for the integer HS field case. The constrained BRST Lagrangian formulation for integer HS field in Minkowski space with generalized spin $(s)_k$ may be derived equivalently both in the self-consistent way and from the unconstrained  BRST Lagrangian formulation above as it was done respectively in the Subsection~\ref{constr-self} and Subsection~\ref{constr-unconstr}.
The final result for  constrained BRST Lagrangian formulation for integer HS field under consideration can be determined by the \vspace{1ex}

\noindent
 \textbf{Statement 5:} The  constrained gauge-invariant Lagrangian formulation for integer HS field, $\Phi_{(\mu^1)_{s_1},...,(\mu^k)_{s_k}}(x)$ with generalized spin $(s)_k$, is determined by the action and sequence of reducible gauge transformation,
  \begin{eqnarray}
 &&  {\cal{}S}^{B}_{C|(s)_k}= \int d \eta_0 {}_{(s)_k}\langle\chi^0_{C|B}|Q^B_{C} |\chi^0_{c|B}\rangle_{(s)_k},\ \  \delta|\chi^l_{c|B} \rangle_{(s)_k}
=Q^B_{C}|\chi^{l+1}_{c|B}\rangle_{(s)_k} \,, \
\delta|\chi^{k}_{c|B} \rangle_{(s)_k}=0, \label{LconB}\\
&& Q^B_C = Q_C\vert_{q_0=0} = \eta_0l_0+\eta_i^+l^i+l^{i+}\eta_i
 + {\imath}\sum_l\eta_l^+\eta^l{\cal{}P}_0, \label{QconB}
\end{eqnarray}
for $l=0,1,..., k-1$, $|\chi^l_{c|B} \rangle \in \mathcal{H}{}_{C|tot}^{B|l}$, with $\mathcal{H}{}_{C|tot}^{B}= \mathcal{H}\otimes H^{o^B_A}_{gh}$ = $ \oplus_l \mathcal{H}{}_{C|tot}^{B|l}  $ , $gh(|\chi^l_{c|B} \rangle)=-l$,  describing the gauge theory of  $(k-1)$-stage of reducibility with off-shell  BRST-extended constraints   imposed on the whole set of field and gauge parameters:
\begin{equation}
\Big(\widehat{L}_{ij}, \widehat{T}_{rs}\Big)|\chi^{l }_{c|B}\rangle_{(s)_k} =0, \ \  l=0,1,...,k; \ \Big(\widehat{L}_{ij}, \widehat{T}_{rs}\Big) =\Big(l_{ij} +\frac{1}{2}\eta_{\{i} \mathcal{P}_{j\}}, t_{rs}-\eta^+_r\mathcal{P}_s -\mathcal{P}^+_r  \eta_s \Big)  \label{constralgB}.
\end{equation}
     The set of $|\chi^{l }_{c|B}\rangle_{(s)_k}$ is determined by the decomposition(\ref{chifconst}) for ${n_{b{}0}}\equiv 0$  and appears by the set of proper eigen-states for the constrained spin operator $\sigma^i_{c|B}\equiv \sigma^i_{c}$ with proper eigen-values:
 \begin{equation}\label{spindiffB}
  {\sigma}^i_C |\chi^l_{c|B}\rangle_{(s)_k} = \left( s^i+\frac{d-2}{2}\right)|\chi^l_{c|B}\rangle_{(s)_k},
 \end{equation}
which determine the same spin and ghost number grading (\ref{nidecomposfconst}), (\ref{ghnumconnstr})  [but for ${n_{b{}0}}\equiv 0$]  as for the constrained half-integer Lagrangian formulation. \vspace{1ex}

The constrained  and unconstrained  Lagrangian formulations for integer HS field  in Minkowski space-time  with generalized spin $(s)_k$ are equivalent, but former one contains less auxiliary HS fields as compared to latter formulation.
Concluding, the subsection, note the same comments in the end of the previous subsection, concerning  bosonic HS fields with \emph{continuous spin}, {generalized triplet formulation} and generalized quartet formulation  are valid as well.

Let us now shortly consider  the Lagrangian formulations for the massive HS fields in Minkowski space-time   subject to Young diagram with $k$ rows.

\subsection{On Constrained Lagrangian Formulations for  Massive  Fields}\label{constrmasex}

The unconstrained BRST Lagrangian formulations for massive  half-integer and massive integer HS fields in $d$-dimensional Minkowski space-time with generalized respective spins, $(n)_k+(\frac{1}{2})_k$,  $(s)_k$ were elaborated in \cite{Reshetnyak2}, \cite{BuchbinderRmix} (for $k=1$ and example, see \cite{0505092}, \cite{ReshMosh}). It was done  on a base  of the derivation of the HS symmetry superalgebra $\mathcal{A}_m^f(Y(k),\mathbb{R}^{1,d-1})$ for massive half-integer  HS fields in $\mathbb{R}^{1,d-1}$ and HS symmetry algebra $\mathcal{A}_m(Y(k),\mathbb{R}^{1,d-1})$ for massive integer  HS fields from respective HS symmetry superalgebra  $\mathcal{A}^f(Y(k),\mathbb{R}^{1,d})$ and algebra $\mathcal{A}(Y(k),\mathbb{R}^{1,d})$ for massless   HS fields in $\mathbb{R}^{1,d}$ with help of the dimensional reduction (see Subsections~3.3 in \cite{Reshetnyak2}, \cite{BuchbinderRmix}).

In the case of a massive half-integer HS field
$\Psi_{(\mu^1)_{n_1},...,(\mu^k)_{n_k}}(x)$ of spin
$(s)_k=(n)_k+(\frac{1}{2})_k$, the Dirac equation
\begin{equation}\label{Eq-om}
  \Big(\imath\gamma^{\mu}\partial_{\mu}-m\Big)\Psi_{(\mu^1)_{n_1},(\mu^2)_{n_2},...,(\mu^k)_{n_k}}
=0  \Longleftrightarrow (\imath \tilde{\gamma}^\mu\partial_\mu- \tilde{\gamma}
m)\Psi_{(\mu^1)_{n_1},(\mu^2)_{n_2},...,(\mu^k)_{n_k}}
 =0
\end{equation}
contains a massive term in both even and odd space-time dimensions, but equivalent description in terms of Clifford algebra elements $\tilde{\gamma}^\mu$, $\tilde{\gamma}$
is possible only for $d=2N$ explicitly given by (\ref{tgammas}),
 (\ref{evengamm}),  with unaltered gamma-traceless and mixed-symmetry equations
(\ref{Eq-1}), (\ref{Eq-2}).
The unconstrained Lagrangian formulation in this case is determined
by the same relations as those in the massless case
with some modifications, first of all, for the initial operators $t_0, l_0$,
\begin{equation}\label{changeferm}
  \big(t_0,\, l_0\big) \to \big(\check{t}_0,\, \check{l}_0\big) = \big({t}_0 + \tilde{\gamma} m ,\, {l}_0 + m^2\big),
\end{equation}
which, along with the remaining unaltered elements from $o_I$,
obey the same HS symmetry superalgebra
$\mathcal{A}^f(Y(k),\mathbb{R}^{1,d-1})$, except for the additional
commutators
\begin{equation}\label{ll+}
  [t^+_i, l_j] = \delta_{ij}(\check{t}_0 - \tilde{\gamma}m),\qquad  [t_i, l^+_j] = -\delta_{ij}(\check{t}_0 - \tilde{\gamma}m),\qquad   [l_i, l^+_j] = \delta_{ij}(\check{l}_0 -  m^2).
\end{equation}
Secondly, the additional parts $o'_I(B,B^+)$ coincide with those
of the massless case, whereas the converted set of constraints has
the form $O^m_I=\check{o}_I+o'_I$, no longer with a central charge
$\tilde{\gamma}m$, for massive HS fields with $\check{o}_I =
{o}_I+{\hat{o}}_I(b_i, b_i^+)$ (for additional bosonic
$2k$-oscillators $[b_i,
b_j^+]=\delta_{ij}$ acting in the Fock space $H_m$),
determined by adding the terms induced by dimensional reduction:
\begin{eqnarray}
&& \left(\hat{t}_0,\, \hat{l}_0\,, \hat{l}_i,\, \hat{l}_i^+, \hat{g}_0^i\right)  = \left(0,\, 0,\,  mb_i, \,   mb_i^+,\,b^+_ib_{i} + \frac{1}{2}\right) , \label{hat0}\\
 && \left(\hat{t}_i,\, \hat{t}_i^+\,, \hat{l}_{ij},\, \hat{l}_{ij}^+,\,\hat{t}_{ij},\,\hat{t}_{ij}^+\right) = -\left(\tilde{\gamma}b_i,\, \tilde{\gamma}b_i^+,\,
\frac{1}{2}b_ib_{j}, \,
 \frac{1}{2}b^+_ib^+_{j},\,b^+_{i}b_j\theta^{ji},\,b_{i}b^+_j\theta^{ji}\right) . \label{hat1}
\end{eqnarray}
The set of $O^m_I$ satisfies the same HS symmetry superalgebra $\mathcal{A}^f(Y(k),\mathbb{R}^{1,d-1})$  as for massless half-integer Poincare group irreducible representations, but for even-valued $d$. Third, the generalized spin, $\sigma^i(G_m)$, BRST, $Q(O^m)$, operators   as well as the arbitrary vector $|\chi_{m}\rangle \in \mathcal{H}^m_{c|tot}$, $\mathcal{H}^m_{c|tot} = \mathcal{H}\otimes {H}'_m\otimes H_{gh}$ for ${H}'_m=H'\otimes H_m$ coincide by the form respectively with ones for massless case (\ref{sigmai}), (\ref{Q}) with change $(g, o)\to (G_m, O^m)$, whereas   $|\chi_{m}\rangle$ has the vector $|\chi\rangle$ (\ref{chif}) as the massless limit for $b^+_i=0$:
\begin{eqnarray}
|\chi_m \rangle &=& \sum_{n'_l\geq (0)_l} \prod_{l=1}^k(b_l^+)^{n'_l}|\chi_{(n')_l} (a^+,B^+,q^+,p^+,\eta^+,\mathcal{P}^+, \vartheta^+, \lambda^+ ) \rangle  \ \mathrm{ for} \ |\chi_{(n')_l}\rangle \in \mathcal{H}_{tot}\,. \label{chifm}
\end{eqnarray}
Note, the $b_l^+$-independent  vectors $|\chi_{(n')_l}\rangle$ have the decomposition in powers of oscillators  presented by  (\ref{chif}).

The unconstrained Lagrangian formulation for massive half-integer HS field of spin $(n)_k+(\frac{1}{2})_k$  is determined for $d=2N$, $N\in \mathbb{N}$ almost the same relations as for massless case (\ref{Lun}), (\ref{GT1}) with use of (\ref{strQ})--(\ref{deltaQ}) for $Q(O^m)$:
\begin{eqnarray}
\hspace{-0.5em}&\hspace{-0.5em}&\hspace{-0.5em} {\cal{}S}^m_{(n)_k}  = \left({}_{(n)_k}\langle\tilde{\chi}^{0}_{m|0}|\  {}_{(n)_k}\langle\tilde{\chi}^{1}_{m|0}|\right) K_{m|(n)_k} \left(\begin{array}{lr} \overline{t}_0 & \Delta{}Q \\
\Delta{}Q  &  \frac{1}{2}\bigl\{\overline{t}_0,\eta_i^+\eta_i\bigr\} \end{array}
\right) \left(\begin{array}{c}   | \chi^{0}_{m|0}\rangle_{(n)_k} \\
|\chi^{1}_{m|0}\rangle_{(n)_k}  \end{array}
\right),\label{Lunm}\\
\hspace{-0.5em}&\hspace{-0.5em}&\hspace{-0.5em}\delta\left(\begin{array}{c}   | \chi^{s(0)}_{m|0}\rangle_{(n)_k} \\
|\chi^{s(1)}_{m|0}\rangle_{(n)_k}  \end{array}
\right) =\ \left(\begin{array}{lr}
\Delta{}Q  &  \frac{1}{2}\bigl\{\overline{t}_0,\eta_i^+\eta_i\bigr\} \\
\overline{t}_0 & \Delta{}Q \end{array}
\right)\hspace{-0.5em} \left(\begin{array}{c}   | \chi^{s+1(0)}_{m|0}\rangle_{(n)_k} \\
|\chi^{s+1(1)}_{m|0}\rangle_{(n)_k}  \end{array}
\right) ,  \  \delta\left(\hspace{-0.3em}\begin{array}{c}   | \chi^{s_{\mathrm{max}}(0)}_{m|0}\rangle_{(n)_k} \\
|\chi^{s_{\mathrm{max}}(1)}_{m|0}\rangle_{(n)_k}  \end{array}
\hspace{-0.3em}\right) =0,
\label{GT1m}
\end{eqnarray}
where the operator $K_{m|(n)_k}$, which realizes the hermitian conjugation in $\mathcal{H}^m_{c|tot}$:  $K_m = 1\otimes K'\otimes 1_m\otimes 1_{gh}$ [with  $1_m$ being by unit operator in $H_m$] as well as $\Delta{}Q = \Delta{}Q(O^m)$ are  obtained  after substitution for  of the proper eigen-values for real constants $h^i_m$  from the spectrum problem for
generalized spin equations, for $e=0,1$; $s=0,...,s_{\mathrm{max}}$:
\begin{eqnarray}
\sigma^i(G_m)  | \chi^{s(e)}_{m|0} \rangle_{(n)_k} = (\sigma^i_m+h^i_m)| \chi^{s(e)}_{m|0} \rangle_{(n)_k}=0 \Longrightarrow  h^i_m(n) =- \left( n^i+\frac{d+1-4i}{2}
\right)\,,  \label{mstate}
\end{eqnarray}
as follows, $K_{m|(n)_k}=K_m\big|_{h^i_m\to h^i_m(n)}$.
From the spin and ghost number distributions (\ref{nidecomposf}), (\ref{ghnumf}) for massless HS field the only first one is modified as:
 \begin{eqnarray}\label{nidecomposfm}
n_i &= & p_i+n'_i+ n_{a{}i}+ n^0_{i}+
\sum_{j=1}(1+\delta_{ij})(n_{ij}+n_{f{}ij}+n_{p{}ij} )
+n_{f{}i}+n_{p{}i} \nonumber \\
&{}& +\sum_{r<i} (p_{ri} + n_{f{}ri}+ n_{\lambda{}ri}) -\sum_{r>i}
(p_{ir}+ n_{f{}ir}+ n_{\lambda{}ir} )\,,\  i=1,\ldots,k.
\end{eqnarray}

In turn, omitting the details of the derivation of the constrained BRST Lagrangian formulation for massive half-integer HS field, $\Psi_{(\mu^1)_{n_1},,...,(\mu^k)_{n_k}}$, of spin $(n)_k+(\frac{1}{2})_k$,    the final expressions for the gauge-invariant action, sequence of reducible (of the same stage reducibility as for massless case) gauge transformations and independent off-shell BRST-extended constraints  are given by the expressions almost coinciding with (\ref{Lcon}-- (\ref{constralg}), but in  terms of the constrained operators and vectors on $\mathcal{H}^m_C$, $\mathcal{H}^m_C=\mathcal{H}_C\otimes {H}_m$:
\begin{eqnarray}
\hspace{-0.5em}&\hspace{-0.5em}&\hspace{-0.5em} {\cal{}S}^m_{C|(n)_k} \ = \ \left({}_{(n)_k}\langle\tilde{\chi}^{0}_{m|0|c}|\  {}_{(n)_k}\langle\tilde{\chi}^{1}_{m|0|c}|\right)  \left(\begin{array}{lr} \check{t}_0 & \Delta{}Q_C \\
\Delta{}Q_C  & \check{t}_0 \eta_i^+\eta_i \end{array}
\right) \left(\begin{array}{c}   | \chi^{0}_{m|0|c}\rangle_{(n)_k} \\
|\chi^{1}_{m|0|c}\rangle_{(n)_k}  \end{array}
\right), \label{Lconm}\\
\hspace{-0.5em}&\hspace{-0.5em}&\hspace{-0.5em} \phantom{{\cal{}S}^m_{c|(n)_k}}
\ \mathrm{for} \  \Delta{}Q_C= \eta_i^+\check{l}_i+ \check{l}_i^+\eta_i, \nonumber \\
\hspace{-0.5em}&\hspace{-0.5em}&\hspace{-0.5em} \delta\left(\begin{array}{c}   | \chi^{l-1(0)}_{m|0|c}\rangle_{(n)_k} \\
|\chi^{l-1(1)}_{m|0|c}\rangle_{(n)_k}  \end{array}
\right)  =  \left(\begin{array}{lr}
\Delta{}Q_C  &  \check{t}_0 \eta_i^+\eta_i \\
\check{t}_0 & \Delta{}Q_C
\end{array}
\right) \left(\begin{array}{c}   | \chi^{l(0)}_{m|0|c}\rangle_{(n)_k} \\
|\chi^{l(1)}_{m|0|c}\rangle_{(n)_k}  \end{array}
\right) ,  \  \delta\left(\begin{array}{c}   | \chi^{k(0)}_{m|0|c}\rangle_{(n)_k} \\
|\chi^{k(1)}_{m|0|c}\rangle_{(n)_k}  \end{array}
\right) =0
\label{GTconst1m},\\
\hspace{-0.5em}&\hspace{-0.5em}&\hspace{-0.5em} \Big(\widehat{T}{}^m_i, \widehat{T}{}^m_{rs}\Big)\sum_{e=0}^1q_0^e|\chi^{l (e)}_{m|0|c}\rangle_{(n)_k} =0,  \ \mathrm{for} \  \Big(\widehat{T}{}^m_i, \widehat{T}{}^m_{rs}\Big) =  \Big(\widehat{T}_i+\hat{t}_{i},\, \widehat{T}{}_{rs}+\hat{t}_{rs}^+\Big), \;  l=0,1,...,k. \label{constralgm}
\end{eqnarray}
Here, the constrained field and gauge parameters are by the proper eigen-functions for the constrained massive spin operator, $\sigma^i_{m|C}$  determined as in  (\ref{sigmaidec}):
 \begin{equation}\label{spindiffm}
  {\sigma}^i_{m|C}|\chi^{l (e)}_{m|0|c}\rangle_{(n)_k} =  \left( n^i+\frac{d-1}{2}\right)|\chi^{l (e)}_{m|0|c}\rangle_{(n)_k},\ \texttt{for}\  d=2N, N\in \mathbb{N},
 \end{equation}
which together with constrained BRST operator, $Q_C^m$:   $Q_C^m =Q_C(\check{o}_A)$ (\ref{Qc}), constraints $\widehat{T}{}^m_i, \widehat{T}{}^m_{rs}$ and algebraically dependent constraints $\widehat{L}{}^m_{ij}$: $\widehat{L}{}^m_{ij}=\widehat{L}_{ij}+\hat{l}_{ij}$, forms the same closed superalgebra
with respect to $[\ ,\ \}$- multiplication as theirs analogs from massless case, namely, nilpotent  $Q_C^m$ supercommutes with any from  $\Big\{\widehat{T}^m_i$, $\widehat{T}^m_{rs}$, $\widehat{L}{}^m_{ij}$, $\sigma^i_{m|C}\Big\}$, which satisfy to the relations in the Table~\ref{table in} for $t_i, l_{ij}, t_{rs}, g_0^i$.

Note, it is easy to establish  again that both approaches: first, reduced from unconstrained BRST Lagrangian formulation and, second, obtained in   self-consistent way; to derive  the constrained BRST Lagrangian formulation for massive HS fields developed in the Subsections~\ref{constr-unconstr},~\ref{constr-self} are equivalent.
The equivalence among the unconstrained and constrained BRST Lagrangian formulations for the same massive half-integer HS field of spin $(n)_k+(\frac{1}{2})_k$   follows, in fact, from the validity of the  Statement 4 and Corollary 3 because of the  constraints subsystem, $\check{o}_A = \{\check{t}_0,\,\check{l}_0, \check{l}_i,\,\check{l}_i^+\}$ appears by the first-class ones in $\mathcal{H}_m \equiv \mathcal{H} \otimes {H}_m$ (not in $\mathcal{H}$!).

The problem of derivation of the (un)constrained  Lagrangian formulation for massive half-integer HS fields in odd-valued dimensions, $d=2N+1$, may be solved within BRST-BFV approach by means of calculation of whole pairings for ghost $\mathcal{C},\overline{\mathcal{P}}$ and auxiliary $B, B^+$ oscillators in (\ref{Lunm}) and (\ref{Lconm}) with use of the partial gauge-fixing procedure up to the representation of the respective Lagrangian formulations where the only initial $\gamma^\mu$-matrices will survive, without $\tilde{\gamma}$ object due to the property described in the footnote~2. The last Lagrangian formulations obtained for $d=2N$ can be extrapolated to be by Lagrangian formulations for odd-valued dimensions, $d=2N+1$ if there will not appear another restrictions.

To be complete, we remind the form of unconstrained BRST Lagrangian formulation for massive integer HS field of spin $(s)_k$  in $\mathbb{R}^{1,d-1}$ \cite{BuchbinderRmix} and present new results for respective constrained BRST Lagrangian formulation.
For the former case we  ignore, first, spinor-matrix-like
$2^{\left[\frac{d}{2}\right]}\times 2^{\left[\frac{d}{2}\right]}$
structure for the operators and spinor structure for the states, and extracting from the massive HS symmetry superalgebra  $\mathcal{A}_m^f(Y(k),\mathbb{R}^{1,d-1})$  the $(2k+1)$ Grassman-odd elements: $\check{t}_0,\check{t}_i,\check{t}_i^+$, to get massive HS symmetry algebra  $\mathcal{A}_m(Y(k),\mathbb{R}^{1,d-1})$ because of having instead of the wave equation in (\ref{eqbose}) the Klein-Gordon equation
\begin{eqnarray}
\label{Eq-0bm} && (\partial^\mu\partial_\mu+
m^2)\Phi_{(\mu^1)_{s_1},(\mu^2)_{s_2},...,(\mu^k)_{s_k}}
 =0.
\end{eqnarray}
Second, we adapt the results of the previous subsection for the massive integer HS field  for any dimension $d$, and representation for the Grassmann-even constraints, $O^{B|m}_I$,  to present  the unconstrained Lagrangian formulation on a base of one for massless case (\ref{SunB}), (\ref{Q12Bose}) with BRST operator $Q^B_m=Q^B(O^B_m)$ for:
 \begin{eqnarray}\label{SunBm}
  \hspace{-0.5em}&\hspace{-0.5em}&\hspace{-0.5em} {\cal{}S}^{B}_{m|(s)_k}= \int d \eta_0 {}_{(s)_k}\langle\chi^0_{m|B}|K^B_{m|(s)_k}Q^B_{m|(s)_k} |\chi^0_{m|B}\rangle_{(s)_k}, \ \ K^B_m=1\otimes {K^{B\prime}_m}\otimes 1_{gh}: \\
 &&  K^B_{m|(s)_k}Q^B_{m|(s)_k}=(Q^{B})^+_{m|(s)_k}K^B_{(s)_k}, \nonumber\\
\hspace{-0.5em}&\hspace{-0.5em}&\hspace{-0.5em} \delta|\chi^lp_{m|B} \rangle_{(s)_k}
=Q^B_{m|(s)_k}|\chi^{l+1}_{m|B}\rangle_{(s)_k} \,, l=0,1,..., k(k+1)-1...\,,
\delta|\chi^{k(k+1)}_{m|B} \rangle_{(s)_k}=0\label{Q12Bosem}
\end{eqnarray}
and, thus, determining the gauge theory of $[k(k+1)-1]$-stage of reducibility.
The form of the field ($l=0$) and gauge parameters ($l=1,...,k(k+1)$) $|\chi^l_{m|B}\rangle_{(s)_k}$, with the same Grassman and ghost number grading  as ones for massless case is determined accordingly to (\ref{chifm}) for $n^0_{c}=n_{b{}0}=n_{a{}e}=n_{b{}g}=0$ with component vector $|\Phi(a^+_i)^{0_{b{}0} 0_{f 0};  (0)_{a{}e} (0)_{b{}g} (n)_{f i}(n)_{p j}(n)_{f lm}
(n)_{pno}(n)_{f rs}(n)_{\lambda
tu}}_{{(n')_l}(n^0)_{c};(n)_{ij}(p)_{rs}}\rangle$  being by the proper eigen-states  for massive generalized spin operator $\sigma^{i}_{m|B}(G) $, $\sigma^{i}_{m|B}(G) =  \sigma^{i}_B(G_m)$  for integer spin with value for the  $h^i_{m|B}$ in
 \begin{equation}\label{genspineqBm}
   {\sigma}^{i}_B (G_m)|\chi_{m|B}\rangle_{(s)_k} = ({\sigma}^{i}_{m|B} +h^i_{m|B}) |\chi_{m|B}\rangle_{(s)_k} = 0 \Longrightarrow- h^i_B = s^i+\frac{d-1-4i}{2} \;, \quad
i=1,..,k.
 \end{equation}
According to the Statement 5,   the  constrained gauge-invariant Lagrangian formulation for massive integer HS field, $\Phi_{(\mu^1)_{s_1},...,(\mu^k)_{s_k}}(x)$ with generalized spin $(s)_k$, is determined by the action and sequence of reducible gauge transformation,
  \begin{eqnarray}
 && {\cal{}S}^{B}_{m|C|(s)_k}= \int d \eta_0 {}_{(s)_k}\langle\chi^0_{m|c|B}|Q^B_{m|C} |\chi^0_{m|c|B}\rangle_{(s)_k}, \label{LconBm}\\
   && \delta|\chi^l_{m|c|B} \rangle_{(s)_k}
=Q^B_{m|C}|\chi^{l+1}_{m|c|B}\rangle_{(s)_k} \,, l=0,1,..., k-1...\,,
\delta|\chi^{k}_{m|c|B} \rangle_{(s)_k}=0 \label{LconBm1}
\end{eqnarray}
describing the gauge theory of  $(k-1)$-stage of reducibility with off-shell  BRST-extended constraints, $\widehat{L}{}^m_{ij}$, $\widehat{T}{}^m_{rs}$   imposed on the whole set of massive field and gauge parameters:
\begin{equation}
\Big(\widehat{L}{}^m_{ij}, \widehat{T}{}^m_{rs}\Big)|\chi^{l }_{m|c|B}\rangle_{(s)_k} =0, \ \  l=0,1,...,k; \ \Big(\widehat{L}{}^m_{ij}, \widehat{T}{}^m_{rs}\Big) =\Big(\widehat{L}_{ij} +\hat{l}_{ij} , \widehat{T}_{rs}+ \hat{t}_{rs}\Big) \label{constralgBm}.
\end{equation}
     The set of $|\chi^{l }_{m|c|B}\rangle_{(s)_k}$ is determined by (\ref{chifconst}) and  (\ref{chifm}) for ${n_{b{}0}}\equiv 0$  and appears by the set of proper eigen-states for the massive constrained spin operator $\sigma^i_{m|C|B}\equiv \sigma^i_{m|C}$ with proper eigen-values:
 \begin{equation}\label{spindiffBm}
  {\sigma}^i_{m|C} |\chi^l_{m|c|B}\rangle_{(s)_k} = \left( s^i+\frac{d-1}{2}\right)|\chi^l_{m|c|B}\rangle_{(s)_k}.
 \end{equation}
  Again, as well as for the  massive half-integer  HS fields  from the dimensional reduction procedure it follows  that both approaches: first, reduced from unconstrained BRST Lagrangian formulation and, second, obtained in   self-consistent way; to get the constrained BRST Lagrangian formulation for massive HS fields developed in the Subsections~\ref{constr-unconstr},~\ref{constr-self} are equivalent.
The equivalence among the unconstrained and constrained BRST Lagrangian formulations for the same massive integer HS field of spin $(s)_k$   follows, in fact, from the validity of the  Statement 4 and Corollary 3 adapted to integer spin case because of the  constraints subsystem, $\check{o}^m_A = \{\check{l}_0, \check{l}_i,\,\check{l}_i^+\}$ appears by the first-class ones in $\mathcal{H}_m$.

\section{Example: Spin $(n+\frac{1}{2})$ totally-symmetric field}\label{examples}
\setcounter{equation}{0}

Here, we, firstly, realize the general prescriptions of our
constrained BRST--BFV   Lagrangian formulations in the known case
of totally-symmetric  fermionic fields in the metric-like
formulation.

\subsection{Spin $(n+\frac{1}{2})$ field in BRST constrained  approach}\label{totasymconbfv}

The equations expressing the belonging of the spin-tensor, $\Psi_{\mu_1...\mu_n}(x)$ to the Poincare group irreducible representation space   of spin, $n+\frac{1}{2}$ contains only two equations from (\ref{Eq-0}), (\ref{Eq-1})
\begin{align}
\label{Eq-01tot}
\imath\gamma^{\mu}\partial_{\mu}\Psi_{(\mu)_{n}}(x)
=0\,, & \quad
 \gamma^{\mu_{l}}\Psi_{\mu_1...\mu_n}(x) =0\,
\end{align}
which are described by two Grassmann-odd operators ${t}_0 = -\imath\tilde{\gamma}^\mu \partial_\mu$ \,,
$t_1 =  \tilde{\gamma}^\mu a_\mu$ ($a_\mu \equiv  a^1_\mu$) acting on  the basic vector
\begin{eqnarray}
\label{PhysStatetot} |\Psi\rangle &=&
\sum_{n=0}^{\infty}
\frac{\imath^{n}}{n!}\Psi_{(\mu)_{n}}\,
a^{+\mu_{1}}\ldots  a^{+\mu_{n}}|0\rangle.
\end{eqnarray}
The closed algebra of all constraints: $\{{t}_0, l_0,  t_1, t^+_1, l_1, l^+_1, l_{11}, l^+_{11} ;  g_0 = - \frac{1}{2}\{a^{\mu},a^+_\mu\}\}$ determined by (\ref{lilijpr}),  (\ref{lilijt+})  forms the half-integer HS symmetry algebra in Minkowski space for totally-symmetric fields,
  $\mathcal{A}^f(Y(1),
\mathbb{R}^{1,d-1})$,    with the differential first-class $\{{t}_0, l_0,   l_1, l^+_1\}$, holonomic second-class  $\{  t_1, t^+_1$, $l_{11}, l^+_{11}\}$ constraints and number particle operator $g_0$. Following to the results of the section~\ref{constrBRST} the nilpotent constrained BRST operator for the first-class system, off-shell BRST extended constraints $\{ \widehat{T}_1, \widehat{L}_{11} \}$ and constrained spin operator $\widehat{\sigma}_C(g)$ in the Fock space $\mathcal{H}_C $  have the respective form:
\begin{eqnarray}
   Q_C & = &  q_0t_0+\eta_0l_0+\eta_1^+l_1+l_1^{+}\eta_1
 + {\imath}\bigl(\eta_1^+\eta_1-q_0^2\bigr){\cal{}P}_0, \label{Qctotsym}
  \\
   \big\{\widehat{T}_1, \widehat{L}_{11}\big\}& = &
  \big\{ t_1-\imath \eta_1p_0 -2q_0 \mathcal{P}_1, \  l_{11}+\eta_{1} \mathcal{P}_{1}   \big\}, \label{brstext0tot}\\
  \widehat{\sigma}_C(g) & = &  g_0+ \eta_1^+\mathcal{P}_{1} -\eta_1\mathcal{P}_{1}^+ \label{spinctotsym}
\end{eqnarray}
whose closed algebra  satisfy to the  relations (\ref{commOsig}) for $K_a^i\equiv \big( K_1^1,\,  K_{11}^1)=(1,\,2)$  and  to (\ref{eqQctot}).

 According to the Statement 4 we have the first-order constrained irreducible  gauge-invariant Lagrangian formulation for half-integer HS field, $\Psi_{(\mu)_{n}}(x)$, described by the action
  \begin{eqnarray}
{\cal{}S}_{C|(n)} &=& \left({}_{n}\langle\tilde{\chi}^{0}_{0|c}|\  {}_{n}\langle\tilde{\chi}^{1}_{0|c}|\right)  \left(\begin{array}{lr} {t}_0 & \eta_1^+l_1+ l_1^+\eta_1 \\
\eta_1^+l_1+ l_1^+\eta_1  &  {t}_0\eta^+_1\eta_1 \end{array}
\right) \left(\begin{array}{c}   | \chi^{0}_{0|c}\rangle_{n} \\
|\chi^{1}_{0|c}\rangle_{n}  \end{array}
\right) ,\label{Lcontot}
\end{eqnarray}
to be invariant with respect to the gauge transformations (for $n\geq 1$)
\begin{eqnarray}
\delta\left(\begin{array}{c}   | \chi^{0}_{0|c}\rangle_{n} \\
|\chi^{1}_{0|c}\rangle_{n}  \end{array}
\right) &=& \left(\begin{array}{lr}
\eta_1^+l_1+ l_1^+\eta_1   &  {t}_0\eta_1^+\eta_1 \\
{t}_0 & \eta_1^+l_1+ l_1^+\eta_1
\end{array}
\right) \left(\begin{array}{c}   | \chi^{1(0)}_{0|c}\rangle_{n} \\
0  \end{array}
\right) ,  \  \delta  | \chi^{1(0)}_{0|c}\rangle_{n} =0
\label{GTconst1totsym}
\end{eqnarray}
for  $| \chi^{1(1)}_{0|c}\rangle_{n} \equiv 0$ (because of $gh(| \chi^{1(m)}_{0|c}\rangle_{n}) = -m-1$ and $(\mathcal{P}^+_i)^2 = 0$),   with off-shell algebraically independent BRST-extended constraint   imposed on the  fields, $| \chi^{m}_{0|c}\rangle_{n}$, $m=0,1$  and gauge parameter $ | \chi^{1(0)}_{0|c}\rangle_{n}$:
\begin{equation}
\widehat{T}_1 \Big(|\chi^{0}_{0|c}\rangle_{n} + q_0|\chi^{1}_{0|c}\rangle_{n}\Big)=0, \ \  \ \widehat{T}_1 |\chi^{1(0)}_{0|c}\rangle_{n} =0. \label{constralgtotsym}
\end{equation}
The field vectors and gauge parameter being proper for the spin operator $\widehat{\sigma}_C(g)$  have the decomposition in ghosts $\eta_1^+, \mathcal{P}_1^+$:
\begin{eqnarray}
   && |\chi^{0}_{0|c}\rangle_{n}= |\Psi \rangle_{n} + \eta_1^+\mathcal{P}_1^+ |\chi \rangle_{n-2} = |\Psi \rangle_{n}+ \frac{\imath^{n-2}}{(n-2)! } \eta_1^+\mathcal{P}_1^+ \chi^{(\mu)_{n-2}}\prod_{k=1}^{n-2}a^+_{\mu_k}|0\rangle ,\label{chi0ts}\\
   && |\chi^{1}_{0|c}\rangle_{n} =  \mathcal{P}_1^+ \tilde{\gamma} |\chi_{1} \rangle_{n-1}=\frac{\imath^{n-1}}{(n-1)! }\mathcal{P}_1^+ \tilde{\gamma} \chi_1^{(\mu)_{n-1}}\prod_{k=1}^{n-1}a^+_{\mu_k}|0\rangle ,\label{chi01ts}\\
  && |\chi^{1(0)}_{0|c}\rangle_{n} = \mathcal{P}_1^+ |\xi\rangle_{n-1}=\frac{\imath^{n-1}}{(n-1)! }\mathcal{P}_1^+ \xi^{(\mu)_{n-1}}\prod_{k=1}^{n-1}a^+_{\mu_k}|0\rangle  ,  \label{chi1ts}
\end{eqnarray}
with $|\Psi \rangle$ given by (\ref{PhysStatetot}).

The constraints (\ref{constralgtotsym}) are easily resolved as the gamma-traceless constraint for the gauge parameter spin-tensor  $\xi^{(\mu)_{n-1}}$ and   triple gamma-traceless constraint for the field spin-tensor $\Psi^{(\mu)}$ and with expressing $\chi_1^{(\mu)_{n-1}}$, $\chi^{(\mu)_{n-2}}$ in terms of  $\Psi^{(\mu)}$
\begin{eqnarray}
\hspace{-0.8em}  &\hspace{-0.8em}& \hspace{-0.8em}\widehat{T}_1|\chi^{1(0)}_{0|c}\rangle_{n} = t_1|\chi^{1(0)}_{0|c}\rangle_{n}  \Longleftrightarrow \ \gamma^\mu  \xi_{(\mu)_{n-1}}=0,  \label{gammaxi}\\
\hspace{-0.8em}  &\hspace{-0.8em}&\hspace{-0.8em} \widehat{T}_1\Big(\hspace{-0.1em}|\chi^{0}_{0|c}\rangle_{n}\hspace{-0.3em}+q_0|\chi^{1}_{0|c}\rangle_{n}\hspace{-0.3em}\Big)\hspace{-0.1em}=0 \hspace{-0.2em} \Longleftrightarrow  \hspace{-0.1em}\Big\{\hspace{-0.1em}t_1|\Psi \rangle_{n}\hspace{-0.1em}= \tilde{\gamma}|\chi_{1} \rangle_{n-1},\,   |\chi \rangle_{n-2}\hspace{-0.1em}=\hspace{-0.1em}-\frac{1}{2}t_1\tilde{\gamma}|\chi_{1} \rangle_{n-1}\hspace{-0.1em} , \, \hspace{-0.2em}t_1 |\chi \rangle_{n-2}\hspace{-0.1em} = 0\hspace{-0.1em}\Big\}.\label{gammafield0}
\end{eqnarray}
Indeed, from (\ref{gammafield0}) it follows that
\begin{eqnarray}
&& (t_1)^3|\Psi \rangle_{n}= 0, \quad   \tilde{\gamma}|\chi_{1} \rangle_{n-1} = t_1|\Psi \rangle_{n} , \quad   |\chi \rangle_{n-2}= - \frac{1}{2}(t_1)^2|\Psi \rangle_{n}\label{gammafield}
\end{eqnarray}
and therefore, $\prod_{i=1}^3\gamma^{\mu_i}\Psi_{(\mu)_{n}} = 0 $. The resolution of the constraint (\ref{gammafield}) for the fields $|\chi^{m}_{0|c}\rangle_{n}$  means the validity of the representation for them denoted as $|\chi^{m}_{r|c}\rangle_{n}$, $ m=0,1$:
\begin{eqnarray}
   &\hspace{-0.5em}&\hspace{-0.5em} |\chi^{0}_{r|c}\rangle_{n}=\Big(1-\frac{1}{2}\eta_1^+\mathcal{P}_1^+(t_1)^2\Big) |\Psi \rangle_{n} =   \frac{\imath^{n}}{n! } \big(a^+_{\mu_1}a^+_{\mu_2}+ n(n-1)\eta_1^+\mathcal{P}_1^+ \eta_{\mu_1\mu_2}\big)  \Psi^{(\mu)_{n}}\prod_{k=3}^{n}a^+_{\mu_k}|0\rangle ,\label{chi0tsres}\\
   &\hspace{-0.5em}&\hspace{-0.5em} |\chi^{1}_{r|c}\rangle_{n} =  \mathcal{P}_1^+ t_1|\Psi \rangle_{n}=-\frac{\imath^{n-1}}{(n-1)! }\mathcal{P}_1^+ \tilde{\gamma}\gamma_{\mu_1} \Psi^{(\mu)_{n}}\prod_{k=2}^{n}a^+_{\mu_k}|0\rangle.\label{chi01tsres}
\end{eqnarray}
Dual (bra-vectors) vectors ${}_n\langle\widetilde{\chi}{}^{m}_{0|c}| $, $m=0,1$ and ${}_n\langle\widetilde{\chi}{}^{1(0)}_{0|c}| $ have the form
\begin{eqnarray}
\hspace{-1.0em}&\hspace{-1.0em}& \hspace{-1.0em}  {}_{n}\langle \widetilde{\chi}{}^{0}_{r|c}\big|\hspace{-0.1em}  =\hspace{-0.1em}  {}_{n}\langle\widetilde{\Psi}\big|\Big(\hspace{-0.15em}1   -   \hspace{-0.15em} \frac{1}{2}(t_1^+)^2 \mathcal{P}_1\eta_1\hspace{-0.15em}
 \Big) \hspace{-0.15em}=\hspace{-0.1em} -\frac{(\hspace{-0.15em}-\hspace{-0.15em}\imath)^{n}}{n! } \langle0|\hspace{-0.15em}\prod_{k=3}^{n}\hspace{-0.15em}a_{\nu_k}\hspace{-0.15em}\overline{\Psi}{}^{(\nu)_{n}}\big(a_{\nu_1}a_{\nu_2}\hspace{-0.1em}+ n(n\hspace{-0.15em}-\hspace{-0.15em}1)\mathcal{P}_1\eta_1 \eta_{\nu_1\nu_2}\big) \tilde{\gamma}(\hspace{-0.1em}-1\hspace{-0.1em})^d\hspace{-0.15em}, \label{brachi00}\\
   \hspace{-1.0em}&\hspace{-1.0em}& \hspace{-1.0em} {}_{n}\langle \widetilde{\chi}{}^{1}_{r|c}\big| \hspace{-0.15em} = \hspace{-0.15em} {}_{n}\langle\widetilde{\Psi}\big|t_1^+ \mathcal{P}_1 \hspace{-0.15em}= \hspace{-0.1em}-\frac{(\hspace{-0.15em}-\imath)^{n\hspace{-0.15em}-\hspace{-0.15em}1}}{(n-1)! } \langle0|\hspace{-0.15em}\prod_{k=2}^{n}\hspace{-0.15em}a_{\nu_k}\hspace{-0.1em}\Psi^+_{(\nu)_{n}}\hspace{-0.3em}\gamma_0\hspace{-0.15em}\gamma_{\nu_1}
   \hspace{-0.15em}\tilde{\gamma}_0\hspace{-0.1em}\mathcal{P}_1\hspace{-0.15em}\tilde{\gamma}_0 \hspace{-0.1em}= \hspace{-0.1em}
   \frac{(\hspace{-0.15em}-\imath)^{n-1}}{(n\hspace{-0.15em}-\hspace{-0.15em}1)! } \langle0|\hspace{-0.25em}\prod_{k=2}^{n}\hspace{-0.15em}a_{\nu_k}\hspace{-0.15em}\overline{\Psi}{}^{(\nu)_{n}}\hspace{-0.1em}\gamma_{\nu_1}\hspace{-0.1em}\mathcal{P}_1\hspace{-0.1em} (\hspace{-0.1em}-1\hspace{-0.1em})^d\hspace{-0.15em} ,\label{brachi10}\\
   \hspace{-0.9em}&\hspace{-0.9em}& \hspace{-0.9em} {}_{n}\langle \widetilde{\chi}{}^{1(0)}_{r|c}\big| \hspace{-0.15em} = \hspace{-0.15em}  {}_{n}\langle\widetilde{\xi}\big| \mathcal{P}_1 = \frac{(-\imath)^{n-1}}{(n-1)! } \langle0|\prod_{k=1}^{n-1}a_{\nu_k}\xi^{+(\nu)_{n-1}}\mathcal{P}_1\tilde{\gamma}_0=-\frac{(-\imath)^{n-1}}{(n-1)! } \langle0|\prod_{k=1}^{n-1}a_{\nu_k}\overline{\xi}{}^{(\nu)_{n-1}}\mathcal{P}_1\tilde{\gamma}(\hspace{-0.1em}-1\hspace{-0.1em})^d\hspace{-0.15em} ,\label{brachi01}
\end{eqnarray}
where use has been made of the definition
\cite{totfermiMin,totfermiAdS,Reshetnyak2}
for the Hermitian conjugation of the $\tilde{\gamma}_\nu$-matrix,
adapted to odd and even dimensions $d$,
$(\tilde{\gamma}_\nu)^+ = (-1)^d\tilde{\gamma}_0\tilde{\gamma}_\nu\tilde{\gamma}_0$;
we have also used the properties (\ref{tgammas}), (\ref{tgammasodd}),
the fact that $\{\tilde{\gamma}_0, \,\mathcal{P}_1\} =\{\tilde{\gamma}, \,\mathcal{P}_1\}=0$
and $(\tilde{\gamma}_0)^2=(-1)^d$, as well as the definition of a Dirac-conjugated
spin-tensor, $\overline{\Psi}{}^{(\nu)_{n}} = {\Psi}{}^{+(\nu)_{n}}\gamma_0 $.

The action (\ref{Lcontot})  in terms of independent  field vector $|\Psi\rangle_n$ in the ghost-free form (not depending on value of $d\geq 4$) looks as
\begin{eqnarray}
{\cal{}S}_{C|(n)}\hspace{-0.5em} &=&\hspace{-0.5em} {}_{n}\langle\tilde{\Psi}| \hspace{-0.2em} \left(\hspace{-0.2em}{t}_0  - \frac{1}{4}(t_1^+)^2{t}_0t_1^2- t_1^+{t}_0t_1 + l^+_1t_1+ t_1^+l_1+\frac{1}{2}t_1^+l_1^+t_1^2 +\frac{1}{2}(t_1^+)^2l_1t_1
\hspace{-0.2em}\right) \hspace{-0.2em} |\Psi \rangle_{n} ,\label{Lcontotgh}
\end{eqnarray}
to be invariant with respect to the gauge transformations\footnote{In terms of only $\gamma^\mu$-matrices the action  (\ref{Lcontotgh}) takes the same form but with
operators $(\tilde{t}_1^+, {\tilde{t}}_0, \tilde{t}_1)$ (\ref{totit12}).}
\begin{eqnarray}
\delta | \Psi\rangle_{n} = l_1^+ | \xi\rangle_{n-1}.
\label{GTtotsymgh}
\end{eqnarray}
The gauge invariance for the action ${\cal{}S}_{C|(n)}$ is easily checked from the Noether  identity:
\begin{eqnarray}
\hspace{-0.6em}\delta{\cal{}S}_{C|(n)} \frac{\overleftarrow{\delta}}{\delta |\xi\rangle}\hspace{-0.6em}  &=&\hspace{-0.6em} {}_{n}\langle\tilde{\Psi}|\hspace{-0.2em}  \left(\hspace{-0.2em}{t}_0 \hspace{-0.15em} - \frac{1}{4}(t_1^+)^2{t}_0t_1^2\hspace{-0.1em}- t_1^+{t}_0t_1\hspace{-0.1em}+ l^+_1t_1\hspace{-0.1em}+ t_1^+l_1\hspace{-0.1em}+\frac{1}{2}t_1^+l_1^+t_1^2\hspace{-0.1em} +\frac{1}{2}(t_1^+)^2l_1t_1
\hspace{-0.3em}\right) \hspace{-0.2em}l_1^+\hspace{-0.1em} =\hspace{-0.1em}0 ,\label{Noether_identcontotgh}
\end{eqnarray}
modulo the operators $\mathcal{L}(t_1^+,t_0,l_1, l_1^+)t_1$ vanishing  when acting on the  gamma-traceless vectors like   the gauge parameter $| \xi\rangle_{n-1}$ (\ref{gammaxi}). Here, the variational derivative of  the functional $\delta {\cal{}S}_{C|(n)} = \langle\tilde{\Psi}| L(t_0,t_1,...)|\xi \rangle + \langle\tilde{\xi}| L^+(t_0,t_1,...)|{\Psi} \rangle $ (with the kernel $L(t_0,t_1,...)$ written in (\ref{Noether_identcontotgh})) with respect to the vector $|\xi \rangle$ was introduced.

Indeed, the sum of the  first and the fifth terms  vanishes, due to $[t_1,l_1^+\}=-t_0$.  The sum of the fourth term, transformed into $-\frac{1}{2}t_1^+l_1^+t_1t_0=- t_1^+l_1^+l_1$ (due to the half-integer HS symmetry algebra,
  $\mathcal{A}^f(Y(1),\mathbb{R}^{1,d-1})$) with the third term gives $t_1^+l_0$. The same expression, but with opposite sign: $-t_1^+l_0$  follows from the sixth term. Finally, from the second and the last terms we get respectively, $\frac{1}{4}(t_1^+)^2{t}_0t_1t_0$ = $\frac{1}{2}(t_1^+)^2{t}_0l_1$  and  $-\frac{1}{2}(t_1^+)^2l_1t_0$ which justify  the validity of   (\ref{Noether_identcontotgh}) with accuracy up to the terms $\mathcal{L}(t_1^+,t_0,l_1, l_1^+)t_1$.

 In the spin-tensor  form the  action and the gauge transformations take the familiar form  \cite{Fronsdalhalfint} not depending on the values of the dimension $d \geq 4$  and representation for $\tilde{\gamma}_{\mu}, \tilde{\gamma}$ (\ref{evengamm}), (\ref{tgammasodd}),
  with accuracy up  to the common coefficient $(n!)^{-1}$, (which may be absorbed into definition of $| \Psi\rangle_{n}$ with writing $(n!)^{-(1/2)}$ instead of $(n!)^{-1}$ in (\ref{PhysStatetot})):
\begin{eqnarray}
{\cal{}S}_{C|(n)}\hspace{-0.1em}(\Psi)\hspace{-0.6em} &=&\hspace{-0.6em} (-1)^n \hspace{-0.3em}\int \hspace{-0.2em}d^dx \hspace{-0.1em} \overline{\Psi}{}^{(\nu)_{n}}\hspace{-0.1em}\Big\{\hspace{-0.2em} - \imath \gamma^\mu \partial_\mu  {\Psi}{}_{(\nu)_{n}} \hspace{-0.15em}+ \frac{1}{4}n(n\hspace{-0.1em}-1)\hspace{-0.1em}\eta_{\nu_{n-1}\nu_{n}} (\imath \gamma^\mu \partial_\mu )\eta^{\mu_{n-1}\mu_{n}} {\Psi}{}_{(\nu)_{n-2}\mu_{n-1}\mu_{n}}   \label{LcontotghFronsdal} \\
 &-&  n
\gamma_{\nu_{n}}( \imath \gamma^\mu \partial_\mu )\gamma^{\mu_{n}} {\Psi}{}_{(\nu)_{n-1}\mu_{n}}+ n (\imath \partial_{\nu_n})\gamma^{\mu_{n}} {\Psi}{}_{(\nu)_{n-1}\mu_{n}} +  n ( \imath \partial^{\mu_n})\gamma_{\nu_{n}} {\Psi}{}_{(\nu)_{n-1}\mu_{n}} \nonumber \\
 &-& \hspace{-0.5em}\frac{1}{2}n(n-1) \Big(\gamma_{\nu_{n-1}}(\imath \partial_{\nu_n}) \eta^{\mu_{n-1}\mu_{n}} {\Psi}{}_{(\nu)_{n-2}\mu_{n-1}\mu_{n}} + \eta_{\nu_{n-1}\nu_{n}}\gamma^{\mu_{n-1}}(\imath \partial^{\mu_n})  {\Psi}{}_{(\nu)_{n-2}\mu_{n-1}\mu_{n}}\Big)
\hspace{-0.2em}\Big\} ,\nonumber\\
&=&\hspace{-0.5em} (-1)^n \int d^dx  \Big\{ \overline{\Psi}p\hspace{-0.4em}/{\Psi}- \frac{1}{4} n(n-1)\overline{\Psi}{}^{\prime\prime}p\hspace{-0.4em}/{\Psi}{}^{\prime\prime}+ n\overline{\Psi}{}^{\prime}p\hspace{-0.4em}/{\Psi}{}^{\prime} - n\overline{\Psi}{}\cdot  p {\Psi}{}^{\prime}- n \overline{\Psi}{}^{\prime} p \cdot {\Psi}{} \label{LcontotghFronsdal1}\\
&+& \frac{1}{2} n(n-1)\Big(\overline{\Psi}{}^{\prime} \cdot  p  {\Psi}{}^{\prime\prime} +\overline{\Psi}{}^{\prime\prime} p \cdot  {\Psi}{}^{\prime}\Big)\Big\}, \nonumber\\
\delta {\Psi}^{(\mu)_{n}} &=& - \sum_{i=1}^n\partial^{\mu_i}{\xi}^{\mu_{1}...\mu_{i-1}\mu_{i+1}...\mu_{n}},  \label{gtrFronsdal}
\end{eqnarray}
where each term in (\ref{LcontotghFronsdal}) and
(\ref{LcontotghFronsdal1}) corresponds to the respective summand
in (\ref{Lcontotgh}), whereas for the last expression we have used
Fang--Fronsdal  notations \cite{Fronsdalhalfint} with
identifications, $p_\mu=-\imath\partial_{\mu}$, $ - \imath
\gamma^\mu \partial_\mu= p\hspace{-0.4em}/$ and  $p \cdot {\Psi} =
p_{\mu_1} {\Psi}^{(\mu)_{n}}$.

This result was obtained earlier from the unconstrained BRST approach  \cite{totfermiMin}  by means  of the
tedious  gauge-fixing procedure eliminating additional reducible set of gauge parameters and additional to $|\Psi\rangle_n$ field vectors  from the unconstrained Lagrangian formulation.
We stress that the actions ${\cal{}S}_{C|(n)}$ (\ref{Lcontotgh})
and  ${\cal{}S}_{C|(n)}(\Psi)$ (\ref{LcontotghFronsdal1}) when identifying, $\imath{\Psi}^{(\mu)_{n}}=h^{(\mu)_{n}}$,
coincide with those given by the respective Eqs. (122) and (128)  of \cite{totfermiMin}.

In turn, the triplet formulation to describe  Lagrangian dynamic of massless totally-symmetric field ${\Psi}{}_{(\nu)_{n}}$ with help of the
triplet of spin-tensors ${\Psi}{}_{(\nu)_{n}},  \chi_1^{(\mu)_{n-1}},  \chi^{(\mu)_{n-2}}$ and gauge parameter $ \xi^{(\mu)_{n-1}}$ subject to the off-shell 3 constraints on the fields vectors, $|\Psi \rangle_{n}$, $|\chi_{1} \rangle_{n-1}$, $|\chi \rangle_{n-2}$ (\ref{gammafield})  and 1 gamma-traceless constraint on $|\xi\rangle_{n-1}$ (\ref{gammaxi}) in the ghost-independent  vector form
\begin{eqnarray}
\label{Lcontotghtrip} && {\cal{}S}_{C|(n)}(\Psi,\chi_{1} ,\chi )\ =\ {}_{n}\langle\tilde{\Psi}|  {t}_0|\Psi \rangle_{n}  - {}_{n-2}\langle\tilde{\chi}|  {t}_0|\chi \rangle_{n-2}+{}_{n-1}\langle\tilde{\chi}_1| \tilde{\gamma}{t}_0\tilde{\gamma}|\chi_1 \rangle_{n-1}\\
&& \phantom{{\cal{}S}_{c|(n)}(\Psi,\chi_{1} ,\chi )}\  -\Big({}_{n-1}\langle\tilde{\chi}_1|  \tilde{\gamma}\big\{ l_1|\Psi \rangle_{n}  -l^+_1|\chi \rangle_{n}
 \big\} +h.c.\Big) \hspace{-0.2em}  , \nonumber \\
&& \delta \Big(| \Psi\rangle_{n}, |\chi\rangle_{n-2},|\chi_{1} \rangle_{n-1}\Big)\ =\  \Big(l_1^+, l_1, \tilde{\gamma} t_0\Big) | \xi\rangle_{n-1}
\label{GTtotsymghtrip}
\end{eqnarray}
coincides with one suggested in \cite{FranciaSagnottitrip}. Again, in terms of only $\gamma^\mu$-matrices the relations (\ref{Lcontotghtrip}), (\ref{GTtotsymghtrip}) take the same form with
operators $(\tilde{t}_1^+, {\tilde{t}}_0, \tilde{t}_1)$ without $\tilde{\gamma}$  according to footnote~6. With absence of the off-shell constraints the triplet formulation describes the  free propagation of couple of massless particles with respective spins $(n+\frac{1}{2})$, $(n-\frac{1}{2})$,...,$\frac{1}{2}$.

 \subsection{Unconstrained quartet Lagrangians for spin $(n+\frac{1}{2})$ field and massive case}\label{2totasymconbfv}

 It was shown in \cite{quartmixbosemas} that this formulation  maybe  described within \emph{unconstrained quartet formulation} with additional to the triplet,  compensator field vector $|\varsigma\rangle_{n-2}$, whose gauge transformation  is proportional to  the constraint on $|\xi\rangle_{n-1}$: $\delta|\varsigma\rangle_{n-2}=\tilde{\gamma}t_1|\xi\rangle_{n-1}$ and the whole off-shell constraints (\ref{gammafield}) are augmented by the terms proportional to $|\varsigma\rangle$ to provide theirs total gauge invariance with respect to (\ref{GTtotsymghtrip}) and above gauge transformations for $|\varsigma\rangle$ as follows
\begin{eqnarray}
\hspace{-0.9em}  &\hspace{-0.9em}&\hspace{-0.9em}\big\{ t_1|\Psi \rangle-\tilde{\gamma}|\chi_{1} \rangle+l_1^+\tilde{\gamma}|\varsigma\rangle ,\,   |\chi \rangle+\textstyle\frac{1}{2}t_1\tilde{\gamma}|\chi_{1} \rangle+\textstyle\frac{1}{2}t_0\tilde{\gamma} |\varsigma\rangle ,  t_1 |\chi \rangle + l_1\tilde{\gamma}|\varsigma\rangle\big\} = \big\{0,0,0\big\}\hspace{-0.1em}.\label{gammafieldquart}
\end{eqnarray}
Introducing the respective Lagrangian multipliers: fermionic  ${}_{n-1}\langle\tilde{\lambda}_1 |$, bosonic  ${}_{n-2}\langle\tilde{\lambda}_2 |$,  fermionic  ${}_{n-3}\langle\tilde{\lambda}_3 |$ with trivial gauge transformations,  the equations (\ref{gammafieldquart}) and theirs hermitian conjugated may be derived from the action functional
\begin{eqnarray}\label{Saddquart}
  && {\cal{}S}_{\mathrm{add}|(n)}( \lambda)= {}_{n-1}\langle\tilde{\lambda}_1 |\Big(t_1|\Psi \rangle_{n}\hspace{-0.1em}-\tilde{\gamma}|\chi_{1} \rangle_{n-1}+l_1^+\tilde{\gamma}|\varsigma\rangle_{n-2}\Big) + {}_{n-2}\langle\tilde{\lambda}_2 |\Big(|\chi \rangle_{n-2}\hspace{-0.1em}\\
  && \ \ +\frac{1}{2}t_1\tilde{\gamma}|\chi_{1} \rangle_{n-1}+\frac{1}{2}t_0\tilde{\gamma} |\varsigma\rangle_{n-2}\Big)+{}_{n-3}\langle\tilde{\lambda}_3 |\Big(t_1 |\chi \rangle_{n-2} + l_1\tilde{\gamma}|\varsigma\rangle_{n-2} \Big) + h.c., \nonumber
\end{eqnarray}
so that, the gauge-invariant functional\footnote{As for the Fang-Fronsdal (see footnote~6) and triplet formulations the representation in terms of $\gamma^\mu$-matrices turn the Lagrangian formulation (\ref{Saddquart}) into the same form, but with
operators $(\tilde{t}_1^+, {\tilde{t}}_0, \tilde{t}_1)$,  fermionic $\langle\tilde{\lambda}'_2 |= \langle\tilde{\lambda}_2 |\tilde{\gamma}$, property $\big(t_1, t_0\big)\tilde{\gamma}$ = $-\big(\tilde{t}_1, \tilde{t}_0\big) $ and $\delta|\varsigma\rangle =\tilde{t}_1|\xi\rangle$ without $\tilde{\gamma}$.},
\begin{equation}\label{unconstrStot}
  {\cal{}S}_{(n)}\ = \ {\cal{}S}_{C|(n)}(\Psi,\chi_{1} ,\chi )+ {\cal{}S}_{\mathrm{add}|(n)}( \lambda)
\end{equation}
determines the unconstrained Lagrangian formulation for massless spin-tensor of spin $(n+\frac{1}{2})$ in terms of quartet of the spin-tensor fields ${\Psi}{}_{(\nu)_{n}},  \chi_1^{(\mu)_{n-1}},  \chi^{(\mu)_{n-2}}, \varsigma^{(\mu)_{n-2}}$ with help of three Lagrangian multipliers $\lambda_i^{(\mu)_{n-i}}$, $i=1,2,3$ with trivial for the latters, as it was shown in \cite{quartmixbosemas}, dynamics.

The constrained Lagrangian formulation for the massive
totally-symmetric spin-tensor may be obtained according to the
recipe (\ref{Lconm})--(\ref{constralgm}) of
Subsection~\ref{constrmasex}. Note that the constrained Lagrangian
formulation for such HS fields in constant-curvature spaces was
considered using the metric-like formalism in Ref.~\cite{0609029}
beyond the BRST--BFV approach.

Once again, the constrained first-order irreducible
gauge-invariant Lagrangian formulation for a massive half-integer
HS field $\Psi_m^{(\mu)_{n}}(x)$ satisfying,
instead of (\ref{Eq-01tot}), the Dirac equation (\ref{Eq-om})
for $k=1$ and any even $d =2N  \geq 4$,
\begin{equation}\label{Eq-omtot}
  \Big(\imath\gamma^{\mu}\partial_{\mu}-m\Big)\Psi_m^{(\mu)_{n}}
=0  \Longleftrightarrow (\imath \tilde{\gamma}^\mu\partial_\mu-\tilde{\gamma}
m)\Psi_m^{(\mu)_{n}}
 =0,
\end{equation}
is described by the action
\begin{eqnarray}
\hspace{-0.5em}&\hspace{-0.5em}&\hspace{-0.5em} {\cal{}S}^m_{C|n} \ = \ \left({}_{n}\langle\tilde{\chi}^{0}_{m|0|c}|\  {}_{n}\langle\tilde{\chi}^{1}_{m|0|c}|\right)  \left(\begin{array}{lr} \check{t}_0 & \eta_i^+\check{l}_i+ \check{l}_i^+\eta_i \\
\eta_i^+\check{l}_i+ \check{l}_i^+\eta_i  &  \check{t}_0\eta_1^+\eta_1\ \end{array}
\right) \left(\begin{array}{c}   | \chi^{0}_{m|0|c}\rangle_{n} \\
|\chi^{1}_{m|0|c}\rangle_{n}  \end{array}
\right), \label{Lconmtot}\\
\hspace{-0.5em}&\hspace{-0.5em}&\hspace{-0.5em} \delta\left(\begin{array}{c}   | \chi^{0}_{m|0|c}\rangle_{n} \\
|\chi^{1}_{m|0|c}\rangle_{n}  \end{array}
\right) \ = \ \left(\begin{array}{lr}
\eta_1^+\check{l}_1+ \check{l}_1^+\eta_1   &  {\check{t}}_0\eta_1^+\eta_1 \\
{\check{t}}_0 & \eta_1^+\check{l}_1+ \check{l}_1^+\eta_1
\end{array}
\right) \left(\begin{array}{c}   | \chi^{1(0)}_{m|0|c}\rangle_{n} \\
0  \end{array}
\right) ,  \  \delta  | \chi^{1(0)}_{m|0|c}\rangle_{n} =0
\label{GTconst1mtotsym}
\end{eqnarray}
with an off-shell algebraically independent BRST-extended constraint
imposed on the fields $| \chi^{e}_{m|0|c}\rangle_{n}$, $e=0,1$,
and gauge parameter $ | \chi^{1(0)}_{m|0|c}\rangle_{n}$:
\begin{equation}
\widehat{T}{}^m_1\Big(|\chi^{0}_{m|0|c}\rangle_{n} + q_0|\chi^{1}_{m|0|c}\rangle_{n}\Big)=0, \ \  \ \widehat{T}{}^m_1 |\chi^{1(0)}_{m|0|c}\rangle_{n} =0. \label{constralgtotsymm}
\end{equation}
Here, the constrained field and gauge parameters are by the proper eigen-functions for the constrained massive spin operator, $\sigma_{m|C}$  determined as in  (\ref{sigmaidec}) for $G_0=g_0+\hat{g}_0 $:
 \begin{equation}\label{spindiffmtot}
  {\sigma}_{m|C}|\chi^{l (e)}_{m|0|c}\rangle_{n} =  \left( n+\frac{d-1}{2}\right)|\chi^{l (e)}_{m|0|c}\rangle_{n}.
 \end{equation}
The above operators $\check{o}_I = {o}_I+\hat{o}_I(b,b^+)$
are determined according to (\ref{changeferm}), (\ref{hat0}), (\ref{hat1}):
\begin{eqnarray}
  &&  \big(\check{t}_0,\, \check{l}_0\big) = \big({t}_0 + \tilde{\gamma} m ,\, {l}_0 + m^2\big), \nonumber\\
&& \big(\hat{l}_1,\, \hat{l}_1^+, \hat{g}_0,\hat{t}_1,\, \hat{t}_1^+\big)\ = \  \big(mb, \,mb^+,\,b^+b + \frac{1}{2}, -\tilde{\gamma}b,\, -\tilde{\gamma}b^+\big)\label{changefermtot}
\end{eqnarray}
for $2$ additional bosonic oscillators,
$[b, b^+]=1$, acting in the Fock space $\mathcal{H}_m$.
The massive field vectors and gauge
parameter, being eigenvectors of the spin operator $\sigma_{m|C}$,
have the following decompositions in the ghosts $\eta_1^+, \mathcal{P}_1^+$ and $b^+$:
\begin{eqnarray}
\hspace{-0.5em}   &\hspace{-0.9em}&\hspace{-0.9em} |\chi^{0}_{m|0|c}\rangle_{n}= |\Psi_m \rangle_{n} + \eta_1^+\mathcal{P}_1^+ |\chi_m \rangle_{n-2} =\sum_{k=0}^n \frac{(b^+)^k|}{k!}\Psi_{m|k} \rangle_{n-k} + \eta_1^+\mathcal{P}_1^+\sum_{k=0}^{n-2} \frac{(b^+)^k}{k!} |\chi_{m|k} \rangle_{n-k-2},   \label{chi0tsm}\\
 \hspace{-0.5em}  &\hspace{-0.7em}&\hspace{-0.9em} |\chi^{1}_{m|0|c}\rangle_{n} =  \mathcal{P}_1^+ \tilde{\gamma} |\chi_{m|1} \rangle_{n-1}=\mathcal{P}_1^+ \tilde{\gamma} \sum_{k=0}^{n-1}\frac{(b^+)^k}{k!}   |\chi_{m|1|k} \rangle_{n-k-1} ,\label{chi01tsm}\\
\hspace{-0.4em}  &\hspace{-0.7em}& \hspace{-0.9em} |\chi^{1(0)}_{m|0|c}\rangle_{n} = \mathcal{P}_1^+ |\xi_{m}\rangle_{n-1} =  \mathcal{P}_1^+ \sum_{k=0}^{n-1}\frac{\imath^{n-k-1}}{k!(n-k-1)!}(b^+)^k |\xi_{m|k}\rangle_{n-k-1}   ,  \label{chi1tsm}
\end{eqnarray}
 for $|\Psi_{m|k} (a^+) \rangle_{n-k}$,  $ |\chi_{m|k} \rangle_{n-k-2}$, $|\chi_{m|1|k} \rangle_{n-k-1}$, $|\xi_{m|k}\rangle_{n-k-1}$ having the decomposition in $a^+$,
   \begin{eqnarray}\label{psink}
   \hspace{-1.0em}  &\hspace{-1.0em}& \hspace{-1.0em} |\Psi_{m|k} \rangle_{n-k}= \frac{\imath^{n-k}}{(n-k)! }\Psi^{(\mu)_{n-k}}_{m|k}\prod_{e=1}^{n-k}\hspace{-0.2em}a^+_{\mu_e}|0\rangle, \ \ |\chi_{m|k} \rangle_{n-k-2} = \frac{\imath^{n-k-2}}{(n-k-2)! } \chi^{(\mu)_{n-k-2}}_{m|k}\hspace{-0.2em}\prod_{e=1}^{n-k-2}\hspace{-0.3em}a^+_{\mu_e}|0\rangle,\\
   \hspace{-0.5em}  &\hspace{-0.7em}& \hspace{-0.7em} \label{psink1} \Big(|\chi_{m|1|k} \rangle,\, |\xi_{m|k}\rangle\Big)_{n-k-1} =  \frac{\imath^{n-k-1}}{(n-k-1)! }\Big( \chi_{m|1|k}^{(\mu)_{n-k-1}},\,  \xi_{m|k}^{(\mu)_{n-k-1}} \Big)\prod_{e=1}^{n-k-1}a^+_{\mu_e}|0\rangle .
   \end{eqnarray}
The constraints (\ref{constralgtotsymm}) transform the similar
massless case (\ref{gammaxi}), (\ref{gammafield0}) to a ghost independent  form,
\begin{eqnarray}
\hspace{-0.8em}  &\hspace{-0.8em}& \hspace{-0.8em}\widehat{T}{}^m_1|\chi^{1(0)}_{m|0|c}\rangle_{n} = {\check{t}}_1|\chi^{1(0)}_{m|0|c}\rangle_{n}  \Longleftrightarrow \ t_1|\xi_{m}\rangle_{n-1} = \tilde{\gamma} b |\xi_{m}\rangle_{n-1},  \label{gammaxim}\\
\hspace{-0.8em}  &\hspace{-0.8em}&\hspace{-0.8em}
{\check{t}}_1|\Psi_m \rangle_{n}\hspace{-0.1em}= \tilde{\gamma}|\chi_{m|1} \rangle_{n-1},\ \ \   |\chi_m \rangle_{n-2}\hspace{-0.1em}=\hspace{-0.1em}-\frac{1}{2}{\check{t}}_1\tilde{\gamma}|\chi_{m|1} \rangle_{n-1} , \ \  {\check{t}}_1 |\chi_m \rangle_{n-2} = 0.\label{gammafield0m}
\end{eqnarray}
The analogue of the constraints (\ref{gammafield}) reads as follows:
\begin{eqnarray}
&& (\check{t}_1)^3|\Psi_m \rangle_{n}= 0, \quad   \tilde{\gamma}|\chi_{m|1} \rangle_{n-1} = \check{t}_1|\Psi_m \rangle_{n} , \quad   |\chi_m \rangle_{n-2}= - \frac{1}{2}(\check{t}_1)^2|\Psi_m \rangle_{n}.\label{gammafieldm}
\end{eqnarray}
The ghost-independent Lagrangian formulation
for any even $d$ in terms of  field vector $|\hspace{-0.1em}\Psi_m(a^+\hspace{-0.1em},b^+) \hspace{-0.1em}\rangle_{n}$
which  contains $(n-1)$ spin-tensor  fields
$|\Psi_{m|k} (a^+) \rangle_{n-k}$, $k=1,...,n,$
in addition to the initial field
$|\Psi_{m|0} (a^+) \rangle_{n}\equiv |\Psi_{m} (a^+) \rangle_{n}$
can be obtained from (\ref{Lconmtot}) and has the form
\begin{eqnarray}
\hspace{-0.4em}&\hspace{-0.4em}&\hspace{-0.4em} {\cal{}S}^m_{C|n}\  =\   {}_{n}\langle\tilde{\Psi}_m| \hspace{-0.2em} \left(\hspace{-0.2em}\check{t}_0  - \frac{1}{4}(\check{t}_1^+)^2\check{t}_0\check{t}_1^2- \check{t}_1^+\check{t}_0\check{t}_1 +\check{l}^+_1\check{t}_1+ \check{t}_1^+\check{l}_1+\frac{1}{2}\check{t}_1^+\check{l}_1^+\check{t}_1^2 +\frac{1}{2}(\check{t}_1^+)^2\check{l}_1\check{t}_1
\hspace{-0.2em}\right) \hspace{-0.2em} |\Psi_m \rangle_{n} ,\label{Lcontotghm}\\
\hspace{-0.4em}&\hspace{-0.4em}&\hspace{-0.4em}\delta |\Psi_m (a^+,b^+) \rangle_{n} \ = \  \check{l}_1^+|\xi_m(a^+,b^+)\rangle_{n-1}, \label{gaugetransm}
\end{eqnarray}
with $(n-2)$ spin-tensor gauge parameters (additional to
$|\xi_{m|0}(a^+)\rangle_{n-1}$) $|\xi_{m|k}(a^+)\rangle_{n-k-1}$ for
$k=1,...,n-1$ and off-shell constraints
\begin{eqnarray}
 (\check{t}_1)^3|\Psi_m \rangle_{n}= 0, \quad   \check{t}_1|\xi_{m}\rangle_{n-1} =0, \ n>2.\label{gammafieldmas}
\end{eqnarray}
Note that the constrained Lagrangian formulation
(\ref{Lcontotghm})--(\ref{gammafieldmas}) is valid for any $d=2N$ and
was suggested in \cite{Klishevich} as a product of dimensional
reduction (not being the constrained BRST--BFV procedure) for even
dimensions only. The triplet and quartet Lagrangian formulations
for massive spin-tensor fields have the same form as those of the
massless case, albeit with the change
\begin{equation}\label{changemastrq}
\hspace{-0.5em}  \left({t}_0, {t}_i^{(+)}, {l}_i^{(+)}; |\Psi \rangle , |\chi \rangle, |\chi_1 \rangle,  |\zeta \rangle,  |{\lambda}_p\rangle, |\xi \rangle \right) \rightarrow  \left(\check{t}_0, \check{t}_i^{(+)}, \check{l}_i^{(+)}; |\Psi_m \rangle , |\chi_m \rangle,|\chi_{m|1} \rangle,  |\zeta_m \rangle,  |{\lambda}_{m|p}\rangle, |\xi_m \rangle\right)\hspace{-0.2em},
\end{equation}
(for $ p=1,2,3$) in (\ref{Lcontotghtrip}), (\ref{GTtotsymghtrip})
and (\ref{gammafieldquart}), (\ref{Saddquart}), (\ref{unconstrStot}),
respectively, with a representation
for the massive vectors as in (\ref{psink}), (\ref{psink1}).
Resolving the constraint (\ref{gammaxim}) in a $b^+$-independent
form leads to the representation (for $k=1,...,n-1$):
\begin{eqnarray}
&&     |\xi_{m|k}\rangle_{n-k-1} \ = \  (t_1\tilde{\gamma})^k|\xi_{m|0}\rangle_{n-1}\ =\ \tilde{t}_1^k|\xi_{m|0}\rangle_{n-1}. \label{resgammaxim}
\end{eqnarray}
with the only independent gauge parameter
$|\xi_{m|0}\rangle_{n-1}$.
In turn,  the $(\check{t}_1)^3$-constraint  (\ref{gammafieldmas})
at a fixed degree in $(b^+)^k$ for $k=0,...,n$ is rewritten
in an unfolded $b^+$-independent form, with $k=0,..., n-3$
and the spin index omitted in $|\Psi_{m|k}\rangle_{n-k}$,
\begin{eqnarray}
&&    t_1^3 |\Psi_{m|k}\rangle -3 \tilde{\gamma}t_1^2 |\Psi_{m|k+1}\rangle - 3t_1|\Psi_{m|k+2}\rangle +  \tilde{\gamma}|\Psi_{m|k+3}\rangle = 0\label{gammafieldmasu}\\
&& \Longleftrightarrow  \  \tilde{t}_1^3 |\Psi_{m|k}\rangle -3 \tilde{t}_1^2 |\Psi_{m|k+1}\rangle -\hspace{-0.1em}3\tilde{t}_1|\Psi_{m|k+2}\rangle + |\Psi_{m|k+3}\rangle\Big) \  =\  0 . \label{gammafieldmasu1}
\end{eqnarray}
From the gauge transformations for the field vectors $|\Psi_k \rangle_{n-k}$
at a fixed degree in $(b^+)^k$ for $k=0,...,n$, with allowance for (\ref{resgammaxim}),
\begin{eqnarray}\label{gtrpsink}
&&     (b^+)^k: \quad \delta |\Psi_{m|k} (a^+) \rangle_{n-k} = {l}_1^+|\xi_{m|k}(a^+)\rangle_{n-k-1}\theta_{nk}+ k\, m|\xi_{m|k-1}(a^+)\rangle_{n-k} \\
 && \phantom{(b^+)^k: \quad \delta |\Psi_{m|k} (a^+) \rangle_{n-k}}= \Big({l}_1^+t_1\tilde{\gamma}\theta_{nk}+  k\, m\Big)(t_1\tilde{\gamma})^{k-1}|\xi_{m|0}(a^+)\rangle_{n-1},\label{gtrpsink1}
   \end{eqnarray}
one can remove the field $|\Psi_n  \rangle_{0}$
by using the gauge parameter $|\xi_{m|n-1}\rangle_{0}$,
so that a constraint appears for the independent gauge
parameter  $|\xi_{m|0}\rangle_{n-1}$:
\begin{equation}\label{constrgn-1}
    t_1^{n-1}|\xi_{m|0}\rangle_{n-1}=0 \Longleftrightarrow \prod_{i=1}^{n-1}\gamma_{\mu_i} \xi_{m|0}^{(\mu)_{n-1}}=0.
  \end{equation}
In the spin-tensor notation, the off-shell constraints
and gauge transformations read
\begin{eqnarray}
\hspace{-0.7em}&\hspace{-0.7em}& \hspace{-0.7em}     \xi_{m|k}^{(\mu)_{n-k-1}}= (-\imath)^k\prod_{i=n-k}^{n-1}\gamma_{\mu_i}\xi_{m|0}^{(\mu)_{n-1}}, \label{resgammaximten}\\
\hspace{-1.1em}&\hspace{-1.1em}& \hspace{-1.1em}    -\imath \hspace{-0.35em}\prod_{j=n-k-2}^{n-k}\hspace{-0.35em}\gamma_{\mu_j}\Psi_{m|k}^{(\mu)_{n-k}}  +3  \hspace{-0.35em} \prod_{j=n-k-2}^{n-k-1}\hspace{-0.35em}\gamma_{\mu_j}\Psi_{m|k+1}^{(\mu)_{n-k-1}}  +3\imath\gamma_{\mu_{n-k-2}}\Psi_{m|k+2}^{(\mu)_{n-k-2}} + |\Psi_{m|k+3}^{(\mu)_{n-k-3}} = 0,  \label{gammafieldmasuten}\\
\hspace{-1.1em}&\hspace{-1.1em}& \hspace{-1.1em} \delta \Psi_{m|e}^{(\mu)_{n-e}}= (-\imath)^e\Big(-\hspace{-0.2em}\sum_{i=1}^{n-e}\hspace{-0.15em}\partial^{\mu_i}\hspace{-0.15em}\prod_{j=1}^{e}\hspace{-0.15em}\gamma_{\nu_j}\xi_{m|0}^{\mu_1...\mu_{i-1}\mu_{i+1}...\mu_{n-e}(\nu)_{e}}\theta_{ne}+ \imath e\,m \hspace{-0.15em}\prod_{j=1}^{e-1}\hspace{-0.15em}\gamma_{\nu_j}\xi_{m|0}^{(\mu)_{n-e}\nu_1...\nu_{e-1}}\hspace{-0.15em}\Big) \label{gtrpsink1ten}
\end{eqnarray}
for $k=0,...,n-3$, $ e=0,...,n$. Without removing the spinor $\Psi_{m|n}(x)$,
the gauge parameter $\xi_{m|0}^{(\mu)_{n-1}}(x)$ is an unconstrained spin-tensor
of rank $(n-1)$. In the opposite case, the constraint (\ref{constrgn-1}) holds.
In terms of the initial constrained massive spin-tensor
${\Psi}{}_{m|0}^{(\nu)_{n}}$ and the auxiliary  spin-tensors
${\Psi}{}_{m|k}^{(\nu)_{n-k}}$, $k=1,...,n$ with lower ranks
and allowance for the representation of the dual vector ${}_{n}\langle\widetilde{\Psi}_m\big|$
\begin{eqnarray}
\hspace{-0.8em}&\hspace{-0.8em}& \hspace{-0.7em}  {}_{n}\langle\widetilde{\Psi}_m\big|  = -\sum_{k=0}^n \frac{(-\imath)^{n-k}}{k!(n-k)!} \langle0| b^k\prod_{e=1}^{n-k}a_{\nu_e}\overline{\Psi}{}_{m|k}^{(\nu)_{n-k}} \tilde{\gamma}, \label{brachi00m}
\end{eqnarray}
the action takes the equivalent forms
\begin{eqnarray}
\hspace{-0.5em}&\hspace{-0.5em}&\hspace{-0.5em} {\cal{}S}^m_{C|(n)}(\Psi_{m|k}) \ =\  (-1)^n \hspace{-0.3em}\int \hspace{-0.2em}d^dx \hspace{-0.1em} \sum_{k=0}^n(-1)^{k }{}C^k_n\Bigg[-\overline{\Psi}{}_{m|k}^{(\nu)_{n-k}}\big( \imath \gamma^\mu \partial_\mu -m\big) {\Psi}{}_{m|k\,(\nu)_{n-k}} \nonumber \\
&& \ - (n-k)\Big\{\overline{\Psi}{}_{m|k}^{(\nu)_{n-k}}\gamma_{\nu_{n-k}}\big( \imath \gamma^\mu \partial_\mu +m\big) \gamma_{\nu_{n-k}} {\Psi}_{m|k\,(\nu)_{n-k-1}}{}^{\nu_{n-k}} \nonumber\\
&& \ \   -\Big(\imath \overline{\Psi}{}_{m|k}^{(\nu)_{n-k}}\gamma_{\nu_{n-k}}\big( \imath \gamma^\mu \partial_\mu +m\big) {\Psi}{}_{m|k+1\,(\nu)_{n-k-1}}+h.c.\Big)  \nonumber\\
&& \ \ +  \overline{\Psi}{}_{m|k+1}^{(\nu)_{n-k-1}}\big( \imath \gamma^\mu \partial_\mu +m\big)  {\Psi}_{m|k+1\,(\nu)_{n-k-1}}\Big\}\theta_{nk}\nonumber\\
& &\ +    \frac{1}{4}\prod_{j=0}^1(n-k-j)\Big\{\overline{\Psi}{}_{m|k}^{(\nu)_{n-k}}\eta_{\nu_{n-k-1}\nu_{n-k}} \big(\imath \gamma^\mu \partial_\mu -m\big)\eta^{\rho_1\rho_2}{\Psi}{}_{m|k (\nu)_{n-k-2}\rho_1\rho_2}\nonumber \\
  && \ \ + \Big(\overline{\Psi}{}_{m|k}^{(\nu)_{n-k}}\eta_{\nu_{n-k-1}\nu_{n-k}} \big(\imath \gamma^\mu \partial_\mu -m\big){\Psi}{}_{m|k+2 (\nu)_{n-k-2}}+h.c.\Big) \nonumber \\
 &&\ \ +  \overline{\Psi}{}_{m|k+2}^{(\nu)_{n-k-2}}\big( \imath \gamma^\mu \partial_\mu -m\big)  {\Psi}_{m|k+2\,(\nu)_{n-k-2}}\Big\}\theta_{n-1,k} \nonumber\\
 &&\ -  (n-k)\left[\Big\{\overline{\Psi}{}_{m|k}^{(\nu)_{n-k}}\big(-\imath \partial_{\nu_{n-k}}\big)\Big(  \gamma^{\nu}
 {\Psi}_{m|k(\nu)_{n-k-1}\nu}\,  -\, \imath {\Psi}_{m|k+1(\nu)_{n-k-1}}\Big) \right.\nonumber \\
 &&\ \ - \left.  m  \overline{\Psi}{}_{m|k+1}^{(\nu)_{n-k-1}} \Big( \imath \gamma^{\nu}
 {\Psi}_{m|k(\nu)_{n-k-1}\nu}\,  +\,  {\Psi}_{m|k+1(\nu)_{n-k-1}}\Big)\Big\}\,+\, h.c.\right]\theta_{n k} \nonumber \\
 &&\ +  \frac{1}{2}\prod_{j=0}^1(n-k-j)\hspace{-0.15em}\left[ \hspace{-0.15em} \Big\{\overline{\Psi}{}_{m|k}^{(\nu)_{n-k-2}\mu_1\mu_2}\gamma_{\mu_1}\hspace{-0.15em}\big(\hspace{-0.1em}-\imath \partial_{\mu_2}\hspace{-0.15em}\big) \hspace{-0.1em} \Big( \hspace{-0.1em}\eta^{\rho_1\rho_2} {\Psi}{}_{m|k(\nu)_{n-k-2}\rho_1\rho_2} \hspace{-0.1em}+   \hspace{-0.1em} {\Psi}{}_{m|k+2(\nu)_{n-k-2}} \hspace{-0.15em} \Big)
 \right. \nonumber\\
&&\ \  -  \imath m\overline{\Psi}{}_{m|k+1}^{(\nu)_{n-k-1}}\gamma_{\nu_{n-k-1}} \Big( \eta^{\rho_1\rho_2} {\Psi}{}_{m|k(\nu)_{n-k-2}\rho_1\rho_2}  +  {\Psi}{}_{m|k+2(\nu)_{n-k-2}}\Big) \nonumber \\
&& \ \ +  \imath \overline{\Psi}{}_{m|k+1}^{(\nu)_{n-k-1}}\big(-\imath \partial_{\nu_{n-k-1}}\big) \Big( \eta^{\rho_1\rho_2} {\Psi}{}_{m|k(\nu)_{n-k-2}\rho_1\rho_2}  +   {\Psi}{}_{m|k+2(\nu)_{n-k-2}}\Big) \nonumber \\
&& \ \ +  \left.m\overline{\Psi}{}_{m|k+2}^{(\nu)_{n-k-2}}\Big( \eta^{\rho_1\rho_2} {\Psi}{}_{m|k(\nu)_{n-k-2}\rho_1\rho_2}  +   {\Psi}{}_{m|k+2(\nu)_{n-k-2}}\Big) \Big\} +\ h.c\ \right]\theta_{n-1,k}\Bigg]\label{LcontotghFronsdalm}   \\
\hspace{-0.5em}&\hspace{-0.5em}&\hspace{-0.5em}  \ =\   (-1)^n \int d^dx  \sum_{k=0}^n(-1)^{k }{}C^k_n\Bigg[\overline{\Psi}{}_{m|k}(p\hspace{-0.4em}/+m){\Psi}{}_{m|k} +   (n-k)\Big\{\overline{\Psi}{}^{\prime}_{m|k}\big(p\hspace{-0.4em}/ -m\big) {\Psi}_{m|k}^{\prime} \nonumber\\
&& \ \  - \Big(\imath \overline{\Psi}{}_{m|k}^{\prime}\big(p\hspace{-0.4em}/ -m\big) {\Psi}{}_{m|k+1}+h.c.\Big)+  \overline{\Psi}{}_{m|k+1}\big(p\hspace{-0.4em}/ -m\big) {\Psi}_{m|k+1}\Big\}\theta_{nk}  \nonumber\\
&& \ - \frac{1}{4} \prod_{j=0}^1(n-k-j)\Big\{\overline{\Psi}{}_{m|k}^{\prime\prime}(p\hspace{-0.4em}/+m){\Psi}{}_{m|k}^{\prime\prime}+  \Big(\overline{\Psi}{}_{m|k}^{\prime\prime}(p\hspace{-0.4em}/+m){\Psi}{}_{m|k+2}+h.c.\Big) \nonumber\\
&& \ +  \overline{\Psi}{}_{m|k+2}(p\hspace{-0.4em}/+m)  {\Psi}_{m|k+2}\Big\}\theta_{n-1,k}  -  (n-k)\left[\Big\{\Big(\overline{\Psi}{}_{m|k} \cdot p -\imath m  \overline{\Psi}{}_{m|k+1}\Big)\right.\nonumber\\
&&\  \times  \left. \Big(  {\Psi}{}_{m|k}^{\prime} \,  -\, \imath {\Psi}_{m|k+1}\Big)\Big\}\,+\, h.c.\right]\theta_{n k} +  \frac{1}{2}\prod_{j=0}^1(n-k-j)\hspace{-0.15em}\left[ \hspace{-0.15em} \Big\{\Big(\overline{\Psi}{}_{m|k}^{\prime} \cdot p -\imath m\overline{\Psi}{}_{m|k+1}^{\prime}
 \right. \nonumber\\
&&\ \    \left.   +  \imath \overline{\Psi}{}_{m|k+1}\cdot p  +m\overline{\Psi}{}_{m|k+2}\Big)\Big(  {\Psi}{}_{m|k}^{\prime\prime}  +   {\Psi}{}_{m|k+2}\Big) \Big\} +\ h.c\ \right]\theta_{n-1,k}\Bigg] \label{LcontotghFronsdal1f}
\end{eqnarray}
where $C^k_n=\frac{n!}{k!(n-k)!}$. Thereby, we have obtained a
constrained gauge-invariant Lagrangian formulation for the massive
spin-tensor field ${\Psi}{}_{m|0}^{(\nu)_{n}}\equiv
\Psi_m^{(\nu)_{n}}$ of spin $s= n+1/2$ in Minkowski space-time of
any even dimension. In a similar way, the triplet and quartet
Lagrangian formulations can be derived in spin-tensor forms for
the massive spin-tensor field following the recipes for the above
massless case with any $d=2N$. In the massless case determined by the
vanishing of all ${\Psi}{}_{m|k}^{(\nu)_{n-k}}$, $k\geq 1$ with
$m=0$ in (\ref{gammafieldmasuten})--(\ref{LcontotghFronsdal1f}),
we derive a Fang--Fronsdal Lagrangian formulation containing no
dependence (!) on either odd or even values of $d$. Secondly, in
view of the spin-tensor representation for the constrained
Lagrangian and its gauge symmetries in the massive case not
depending on the Grassmann-odd gamma-matrices $\tilde{\gamma}^\mu$, we
may use  Lagrangian formulation (\ref{resgammaximten})--(\ref{gtrpsink1ten}), (\ref{LcontotghFronsdalm}), (\ref{LcontotghFronsdal1f}) for $d=2N+1$ dimensions as well, thus providing   its applicability for any $d\geq 4$.\footnote{The author highly appreciates
the referee's critical remarks on issues concerning an explicit
realization of the odd matrix $\tilde{\gamma}$ for different
values of $d$ and also concerning a Lagrangian formulation for
massive half-integer HS fields. These remarks have proved to be
instrumental in solving the mentioned problems.} Let us consider
a massive field of spin $s=5/2$, first examined in \cite{Berends}
for $d=4$. In this case, the off-shell constraints for the
independent gauge parameters and the spin-tensor fields
$\Psi_{m\mu\nu}$ parameterizing the configuration space,
\begin{equation}\label{conf52}
  \big(\Psi_{m|0}^{\mu\nu},\, \Psi_{m|1}^{\mu},\,\Psi_{m|2}\big), \ \ \  \big(\xi_{m|0}^{\mu},\,\xi_{m|1}\big)
\end{equation}
look trivial for the fields, which is implied by (\ref{gammafieldmasuten}),
and have the representation (\ref{resgammaximten}) for $\xi$, albeit for arbitrary $d$ (for $d=2N$ by construction and for $d=2N+1$ by extrapolation of the Lagrangian formulation (\ref{resgammaximten})--(\ref{gtrpsink1ten}), (\ref{LcontotghFronsdalm}) from even to odd dimensions),
\begin{eqnarray}\label{resgammaximten52}&& \xi_{m|1}= (-\imath)\gamma_{\mu}\xi_{m|0}^{\mu}.
\end{eqnarray}
Using the gauge transformations (\ref{gtrpsink1ten}), we find
\begin{eqnarray}&& \delta \Psi_{m|0}^{\mu\nu}= -\partial^{\mu}\xi_{m|0}^{\nu}\ - \  \partial^{\nu}\xi_{m|0}^{\mu},  \label{gtrpsink1ten5/20}\\
&&  \delta \Psi_{m|1}^{\mu}= \imath\Big(\partial^{\mu}\gamma_{\nu}\xi_{m|0}^{\nu}- \imath \,m \hspace{-0.15em}\xi_{m|0}^{\mu}\Big) , \label{gtrpsink1ten5/21}\\
&& \delta \Psi_{m|2}= 2  \,m \gamma_{\mu}\xi_{m|0}^{\mu}, \label{gtrpsink1ten5/22}
\end{eqnarray}
where the constraint (\ref{resgammaximten52}) has been resolved, with
the gauge parameter $\xi_{m|1}$ expressed in terms of the $\gamma$-trace
of $\xi_{m|0}^{\mu}$. Using (\ref{gtrpsink1ten5/22}), we resolve
the Stueckelberg gauge symmetries, thereby removing the field $\Psi_{m|2}$
with a $\gamma$-traceless residual gauge parameter $\xi_{m|0}^{\mu}$,
$\gamma_{\mu}\xi_{m|0}^{\mu}=0$. We then decompose the field $\Psi_{m|1}^{\mu}$
as follows:
\begin{equation}\label{represf52}
  \Psi_{m|1}^{\mu} = \Psi_{m|1|\bot}^{\mu}+\gamma^\mu {\Psi}_{m|1|L}: \quad \gamma_{\mu}\Psi_{m|1|\bot}^{\mu}=0,
\end{equation}
so that (\ref{gtrpsink1ten5/21}) implies the gauge transformations
for $\Psi_{m|1|\bot}^{\mu}$ and $ \Psi_{m|1|L}$
\begin{eqnarray}
&& \delta \Psi_{m|1|\bot}^{\mu}=  \,m \hspace{-0.15em}\xi_{m|0}^{\mu}, \quad \delta \Psi_{m|1|L} \equiv \delta (\gamma_\mu \Psi_{m|1}^{\mu})= 0 \label{gtrpsink1ten5/211}
\end{eqnarray}
whence the spin-tensor $\Psi_{m|1|\bot}^{\mu}$ is gauged away
entirely by using the remaining degrees of freedom in
$\xi_{m|0}^{\mu}$. Thus the theory of a massive field of spin
$5/2$ is described, in accordance with \cite{0609029}, by the
non-gauge unconstrained fields $\Psi_{m|0}^{\mu\nu}$,
$\Psi_{m|1|L}$ whose dynamics is governed by an action which
follows from (\ref{LcontotghFronsdal1f}) for $n=2$ in the
Fang--Fronsdal-like notation
\begin{eqnarray}\hspace{-0.5em}&\hspace{-0.5em}&\hspace{-0.5em} {\cal{}S}^m_{C|(2)}(\Psi_{m|0}, \Psi^\prime_{m|1|L})  \ =\    \int d^dx  \Bigg[\overline{\Psi}{}_{m|0}(p\hspace{-0.4em}/+m){\Psi}{}_{m|0}+   2\overline{\Psi}{}^{\prime}_{m|0}\big(p\hspace{-0.4em}/ -m\big) {\Psi}_{m|0}^{\prime} \nonumber \\
&& -\frac{1}{2}\overline{\Psi}{}^{\prime\prime}_{m|0}\big(p\hspace{-0.4em}/ + m\big) {\Psi}_{m|0}^{\prime\prime} -\Big\{\Big(2 \overline{\Psi}{}_{m|0}\cdot p {\Psi}{}^\prime_{m|0}- \overline{\Psi}{}_{m|0}^\prime \cdot p {\Psi}{}^{\prime\prime}_{m|0}\Big) +h.c.\Big\}\nonumber \\
&&
  +2d\Big\{\big[2-d\big] \overline{\Psi}{}_{m|1|L}p\hspace{-0.4em}/ {\Psi}{}_{m|1|L} +d\,m \overline{\Psi}{}_{m|1|L}{\Psi}{}_{m|1|L}\Big\}\nonumber\\
&&  - 2 \left\{ \imath \Big(\overline{\Psi}{}_{m|0}^{\prime\prime}\big(p\hspace{-0.4em}/- m\big){\Psi}{}_{m|1|L} - \overline{\Psi}{}_{m|0}^{\prime}\cdot p {\Psi}{}_{m|1|L} +m  \overline{\Psi}{}_{m|1|L}{\Psi}{}_{m|0}^{\prime\prime}\phantom{\frac{1}{2}} \right.\nonumber\\
&& \qquad \left. - \frac{1}{2}\overline{\Psi}{}_{m|1|L}\big(p\hspace{-0.4em}/- d\,m\big){\Psi}{}_{m|0}^{\prime\prime}\Big)  + h.c.\right\}  \Bigg]. \label{LcontotghFronsdal1f52}
\end{eqnarray}
For an arbitrary massive spin-tensor field of spin $n+1/2$,
we can derive from (\ref{LcontotghFronsdalm}) or (\ref{LcontotghFronsdal1f})
a Lagrangian non-gauge description (equivalent to the gauge-invariant one)
whose configuration space contains only the constrained fields  (due to (\ref{gammafieldmasuten}))
\begin{eqnarray}\label{confngn}
 \hspace{-1.1em}&\hspace{-1.1em}& \hspace{-1.1em}    \Psi_{m|0}^{(\mu)_n},\,  \Psi^{(\mu)_{n-2}}_{m|1|L},\, \ldots ,   \Psi^{\mu}_{m|n-2|L}, \Psi_{m|n-1|L},\  \mathrm{for} \ \Psi^{(\mu)_{n-j-1}}_{m|j|L}=\gamma_{(\mu)_{n-j}}\Psi^{(\mu)_{n-j}}_{m|j}, j=1,...,n-1.   \label{gammafieldmasutenfin}
\end{eqnarray}
The derivations of the constrained Lagrangain formulations for
massless and massive spin-tensor fields of spin $s=n+1/2$
according to general prescriptions of the suggested constrained
BRST--BFV approach presents the basic results of the subsection.

\section{Conclusion }\label{con}
\setcounter{equation}{0}

In this paper, we have developed a constrained BRST--BFV method to
construct gauge-invariant Lagrangian formulations for free
massless and massive half-integer spin-tensor fields with an
arbitrary fixed generalized spin
$\mathbf{s}=(n_{1}+1/2,n_{2}+1/2,...,n_{k}+1/2)$, in Minkowski
space-time $\mathbb{R}^{1,d-1}$ of any dimension in the
\textquotedblleft metric-like\textquotedblright\ formulation. This
result is presented in Statement 4 and explicitly includes, for
the field $\Psi_{(\mu^{1})_{n_{1}}, ...,(\mu^{k})_{n_{k}}}$, a
first-order Lagrangian action ${S}_{C|(n)_{k}}$ (\ref{Lcon})
invariant with respect to reducible gauge transformations
(\ref{GTconst1}) (which determine a gauge theory of $(k-1)$-th
stage of reducibility), and off-shell independent BRST-extended
constraints $\widehat{T}_{i},\widehat{T}_{rs}$ (\ref{extconstr}),
(\ref{solextconstr}) imposed on the whole set of field
(incorporating the initial spin-tensor
$\Psi_{(\mu^{1})_{n_{1}},...,(\mu^{k})_{n_{k}}}$) and gauge
parameter vectors \ref{constralg}) from the resultant Hilbert
space $\mathcal{H}\otimes H_{gh}^{o_{A}}$, whose vectors have the
representation (\ref{chifconst}).  The crucial point is that the
superalgebra formed by a constrained BRST operator (only for the
first-class constraint system $o_{A}$ with a subsuperalgebra of
Minkowski space $\mathbb{R}^{1,d-1}$ isometries (\ref{TkTk}) in
the HS symmetry superalgebra
$\mathcal{A}^{f}(Y(k),\mathbb{R}^{1,d-1})$), a generalized spin
operator, and BRST extended second-class constraints
$\{Q_{C},\,\widehat{\sigma}{}_{C}^{i}(g),\widehat{T}_{i},\widehat{T}_{rs},
\widehat{L}_{ij}\}$ is closed with respect to the $[\ ,\
\}$-multiplication in $\mathcal{H}\otimes H_{gh}^{o_{A}}$ and
forms, with the exception of $Q_{C}$ and with an addition of
$\widehat{T}{}_{i}^{+},
\widehat{T}{}_{rs}^{+},\widehat{L}{}_{ij}^{+}$, an $osp(1|2k)$
orthosymplectic superalgebra.  This fact guarantees a common set of
eigenstates in $\mathcal{H}\otimes H^{o_A}_{gh}$, which depends
on less ghost coordinates and momenta (hence, for a smaller set of
auxiliary fields and gauge parameters) than the ones in the
unconstrained Lagrangian formulation (\ref{Lun}), (\ref{GT1})
\cite{Reshetnyak2} for the same spin-tensor field, and therefore
also ensures the consistency of dynamics in the constrained
formulation with holonomic off-shell constraints.  It is shown in the Theorem (\ref{equivconstinit}), (\ref{equivconstinit2}),  on the
basis of general results in operator quantization for dynamical
systems with first- and second-class constraints, that the
Lagrangian dynamics for the same element of an irreducible
half-integer HS representation of Poincare group in
$\mathbb{R}^{1,d-1}$ subject to $Y(s_{1},...,s_{k})$ in the
constrained and unconstrained BRST--BFV approaches are equivalent,
i.e., for both dynamics equivalent to irreducibility conditions
with a given spin (\ref{Eq-0})--(\ref{Eq-2}).
The equivalence of
the constrained and unconstrained BRST--BFV methods in question is
two-fold.  First, it is based on derivation starting from the
unconstrained HS symmetry superalgebra $\mathcal{A}^{f}(Y(k),
\mathbb{R}^{1,d-1})$ and respective BRST operators $Q^{\prime}$
(\ref{Q'k}) and $Q$ (\ref{Q}), as one disregards the additional
(due to conversion) oscillators $B^{a},B^{a+}$ given by
(\ref{sigmaidec})--(\ref{brstextended1}), and resulting in
Statement 3 and Corollary 3 in (\ref{equivsec11}),
(\ref{equivsec21}), or, equivalently, in terms of the $Q$- and
$Q_{C}$-complexes in (\ref{equivsec1f}), (\ref{equivsec2f}.  Second, in a self-consistent form, the nilpotent constrained BRST
operator $Q_{C}$, the spin operators
$\widehat{\sigma}{}_{C}^{i}(g)$, and the off-shell BRST-extended
constraints $\widehat{T}_{i},
\widehat{T}_{rs}$ are derived explicitly in Section~\ref{constr-self}
from the irreducibility conditions (\ref{Eq-0})--(\ref{Eq-2}), on
a basis of solving the generating equations (\ref{eqQctot}). The
constrained Lagrangian formulation is then obtained from the
second-order one (\ref{Q12c})--(\ref{2ordercon}) by partial
gauge-fixing, thereby eliminating the zero-mode ghost operators
$q_{0},\eta_{0}$, $p_{0},\mathcal{P}_{0}$ for the Dirac and
D'Alembert operators from the whole set of gauge parameters and
field vectors, in a way compatible with off-shell BRST-extended
constraints having the form (\ref{compoffshelc}).

As a byproduct, we have derived the constrained BRST--BFV
Lagrangian formulation (\ref{LconB})-(\ref{constralgB}), for the
integer mixed-symmetric HS field
$\Phi_{(\mu^{1})_{s_{1}},...,(\mu^{k})_{s_{k}}}$ in
$\mathbb{R}^{1,d-1}$ subject to the irreducibility conditions
(\ref{eqbose}), from the unconstrained formulation, albeit with
BRST-extended traceless off-shell constraints, $\widehat{L}_{ij}$,
instead of gamma-traceless ones, $\widehat {T}_{i}$. The stages of
reducibility for both integer and half-integer massless HS fields
coincide and can be used as a starting point for constrained
BRST--BFV Lagrangian formulations to accommodate SUSY models of HS
fields.

It should be noted that the constrained BRST--BFV approach, as well
as the unconstrained one, implies automatically a gauge-invariant
Lagrangian description, reflecting the general fact of BV--BFV
duality \cite{AKSZ}, \cite{BV-BFV}, \cite{GMR}, which reproduces a
Lagrangian action for the initial non-Lagrangian equations
(reflecting the fact that the (spin)-tensor belongs to an
irreducible representation space of the Poincare group) by means
of a Hamiltonian object.

It is shown in Section~\ref{constrmasex} that the case of massive
half-integer and integer HS fields with a corresponding arbitrary
Young tableaux $Y(s_{1},...,s_{k})$ allows one to obtain
constrained gauge-invariant Lagrangian formulations, for fermionic
(\ref{Lconm})--(\ref{constralgm}), initially for even dimensions $d$, and bosonic
(\ref{LconBm})--(\ref{constralgBm}) fields for any $d$, derived from the
respective unconstrained formulations (\ref{Lunm}), (\ref{GT1m})
and (\ref{SunB}), (\ref{Q12Bose}). This was achieved by
dimensional reduction procedure for the respective massless HS
symmetry (super)algebra,
$\mathcal{A}^{(f)}(Y(k),\mathbb{R}^{1,d})$ in $\mathbb{R}^{1,d}$,
to the massive one,
$\mathcal{A}_{m}^{(f)}(Y(k),\mathbb{R}^{1,d-1})$ in
$\mathbb{R}^{1,d-1}$, which means conversion for the sets of
differential constraints, being this time second-class
constraints. In both cases, the constrained BRST operator for
differential constraints, the generalized spin operators, and the
BRST-extended off-shell constraints are modified by $k$ pairs of
conversion oscillators, $b_{i},b_{i}^{+}$, $i=1,...,k$, thereby
preserving their superalgebra, albeit in a larger Hilbert space.
Both resulting gauge theories possess the same reducibility stage
as the ones for massless fields. We differentiate the cases of odd
and even values of space-time dimension $d$ for half-integer HS
fields when realizing the explicit Grassmann-odd gamma-matrix-like
objects suggested in (\ref{evengamm}) and (\ref{tgammasodd}) which
do not influence either the form of HS symmetry superalgebra or
the resulting BRST--BFV Lagrangian formulation. Besides, for massless  half-integer HS fields we  shown in (\ref{grassmapargam}), (\ref{grassmapargam1}) the realization of the (un)constrained Lagrangian formulations with only standard $\gamma^\mu$ matrices as Grassmann-odd quantities,  following to totally-symmetric HS fields \cite{totfermiMin}.

As an example demonstrating the applicability of the suggested
scheme, it is shown that for the particular case of
totally-symmetric massless half-integer HS field
$\Psi_{(\mu)_{n}}$ of spin $s=n+\frac{1}{2}$ the constrained
BRST--BFV method permits one to immediately reproduce the
Fang--Fronsdal \cite{Fronsdalhalfint} gauge-invariant Lagrangian
action (\ref{LcontotghFronsdal})--(\ref{gtrFronsdal}) in terms of
a triple-gamma-traceless field $\Psi_{(\mu)_{n}}$ and a gamma-traceless
gauge parameter $\xi _{(\mu)_{n-1}}$ for any value of $d\geq 4$. Note that this formulation
was reproduced in \cite{totfermiMin} from an unconstrained
Lagrangian with the help of a rather tedious gauge-fixing
procedure. The same action ${S}_{C|(n)}$ in terms of a
ghost-independent field vector has the form (\ref{Lcontotgh}),
with independent gauge transformations (\ref{GTtotsymgh}) and
off-shell holonomic constraints $t_{1}^{3}|\Psi\rangle_{n}=0$,
$t_{1}|\xi\rangle_{n-1}=0$. The constrained Lagrangian formulation
in terms of the triplet of fields $\Psi^{(\mu)_{n}}$,
$\chi_{1}^{(\mu)_{n-1}}$, $\chi^{(\mu)_{n-2}}$ with the action
(\ref{Lcontotghtrip}), invariant with respect to the gauge
transformations (\ref{GTtotsymghtrip}) and subject to the
off-shell constraints (\ref{gammaxi}), (\ref{gammafield}),
coincides with those of \cite{FranciaSagnottitrip}. This
Lagrangian served as a starting point to construct an
unconstrained Lagrangian formulation in a quartet form
\cite{quartmixbosemas}, with the addition of a \textquotedblleft
fourth\textquotedblright\ compensating spin-tensor,
$\varsigma^{(\mu)_{n-2}}$, and $3$ Lagrangian multipliers to the
rest of the augmented gauge-invariant constraints
(\ref{gammafieldquart}), the resulting gauge-invariant action
being of the form (\ref{Saddquart}), (\ref{unconstrStot}).
A constrained gauge-invariant Lagrangian formulation for a massive
spin-tensor for even  space-time dimension $d$ has been obtained in the form of ghost-independent Fock
space (\ref{Lcontotghm}) and in the Fang--Fronsdal-like spin-tensor
form (\ref{LcontotghFronsdalm}), (\ref{LcontotghFronsdal1f}),
with a set of $(n-1)$ auxiliary spin-tensors and a single unconstrained
gauge parameter $\xi_{m|0}^{(\mu)_{n-1}}$, in accordance with
\cite{0609029,Berends},  with a different analogue (as for $m=0$)
of the off-shell constraints (\ref{gammafieldmasuten})
in the total set of fields. {Because of the final Lagrangian formulation for massive half-integer field does not depend on the Grassmann-odd matrices  $\tilde{\gamma}^\mu,  \tilde{\gamma}$, we suggested to use it as an ansatz  for the Lagrangian formulation for massive spin-tensor in  odd space-time dimension $d$.
}

The above construction of the constrained BRST--BFV approach for Lagrangian
formulations was considered from the general viewpoint in
Appendix~\ref{mixgen} for a finite-dimensional dynamical system with
Hamiltonian $H_{0}(\Gamma)$ subject to first-class $T_{A}(\Gamma)=0$ and
second-class $\Theta_{\alpha}(\Gamma)=0$ constraints satisfying special
commutation relations only in terms of the Poisson superbrackets
(\ref{firstclassconstrp}), (\ref{hamconstrp}). The crucial point here is a
construction on the basis of solving the \emph{generating equations}
(\ref{eqQctotgen}) for a superalgebra of BRST-extended (in $\mathcal{M}_{\min
}$)\ second-class constraints $\widehat{\Theta}_{\alpha}$, from the
requirement of commutation with the BRST charge and Hamiltonian, respecting
only the first-class constraints $T_{A}$ in the minimal sector of ghost
coordinates and momenta, and also (trivially) in the total phase-space. The
explicit form of $\widehat{\Theta}_{\alpha}$ was found in
(\ref{extBRSTconstr}), with accuracy up to the second order in $C^{A}$, and satisfying the same
Poisson bracket relations (\ref{extPB2}) as in (\ref{firstclassconstrp}) for
vanishing new structure functions $f_{\alpha\beta A}^{CD}(\Gamma)$ resolving
the Jacobi identity for $T_{A},\Theta_{\alpha},\Theta_{\beta}$, albeit in
$\mathcal{M}_{\min}$ with a BRST-invariant extension of the invertible
supermatrix \ref{extPB2Delta}. The supercommutative algebra of a BRST charge,
a unitarizing Hamiltonian, and BRST-invariant second-class constraints lead to
a new representation for the generating functional of Green's functions
(\ref{ZPIr}), (\ref{ZPI1r}), in terms of $\widehat{\Theta}_{\alpha}$ for the
dynamical system in question. At the operator level, the operators of the
latter quantities allow one to describe equivalently, with some prescriptions
for a choice of ordering in (\ref{equivsec11fin})--(\ref{equivsec1fin}), a set
of physical states by imposing half of the constraints $\widehat{\Theta
}_{\alpha}$ on the Hilbert space vectors.

Concluding, we present some ways of extending the results obtained
in this paper. First, the development of a Lagrangian construction
for tensor and spin-tensor fields with an arbitrary index symmetry
in AdS space. Second, the derivation of constrained BRST--BFV
Lagrangian formulations for irreducible representations of the
SUSY Poincare supergroup along the lines of
\cite{BuchControkicos}. Third, the development of the constrained
and unconstrained BRST--BV method to construct respective minimal
field-antifield BV actions for half-integer HS fields in terms of
Fock-space vectors. Fourth, the construction of a quantum action
for HS fields within an $N=1$ BRST approach,\footnote{For a
generalization of quantization rules intended to accommodate
theories with a gauge group on the basis of an $N$-parametric
BRST symmetry, see \cite{reshetnyakN34}.}
where the space-time variables $x^{\mu}$ are to be considered
on equal footing with the total Fock space variables.
Fifth, a consistent deformation of the (un)constrained
BRST--BFV and BRST--BV approaches applied to bosonic and fermionic
mixed-symmetric HS fields will make it possible to construct an
interacting theory with mixed-symmetry fermionic HS fields,
including the case of curved (AdS) backgrounds. We intend to carry
out a study of these problems in our forthcoming works.

\section*{Acknowledgements}
The author is grateful to J.L. Buchbinder for helpful advices  and
numerous discussions. He also appreciates discussions with the
participants of the International Workshop ``Supersymmetries and
Quantum Symmetries'' SQS'17, where some of the results of this
paper were presented. The author thanks the referee for criticism
which sufficiently improved the paper's contents. He is grateful
to A.A.~Sharapov, V.A.~Krykhtin and P.Yu.~Moshin for advices and
important comments, as well as to Yu.M. Zinoviev, V.N. Tolstoy, D.~Francia, A.~Campoleoni
and E.D.~Skvortsov for correspondence and valuable remarks on massive half-integer fields, structure of HS symmetry algebra,
the terminology concerning Lagrangian formulations. The work was
supported by the Program of Fundamental Research sponsored by the
Russian Academy of Sciences, 2013-2020.

\appendix
\section*{Appendix}
\section{On  quantization of the dynamical system with mixed-class constraints}\label{mixgen}
\renewcommand{\theequation}{\Alph{section}.\arabic{equation}}
\setcounter{equation}{0}

 Let us consider the dynamical system with Hamiltonian,  $H_0(\Gamma)$, subject to the finite set of the

 \noindent
  first-class $T_A(\Gamma)=0$ and second-class $\Theta_\alpha(\Gamma)=0$,
constraints given on the $2n$-dimensional phase-
space $(\mathcal{M}, \omega)$, $\varepsilon(\omega)=0$ satisfying to the  particular case  of the relations (\ref{firstclassconstr})--(\ref{hamconstr}) with respect to the Poisson superbracket, for vanishing functions $V^A_{\alpha}(\Gamma) =V^\beta_{\alpha}(\Gamma)$ = $f^{\alpha\beta}_{AB}(\Gamma) $=  $V^A_{\alpha}(\Gamma)=0$:
\begin{align}\label{firstclassconstrp}
 & \{T_A,\, T_B\} = f^C_{AB}(\Gamma)T_C ,  &  \{T_A, \Theta_\alpha\} =  f^C_{A\alpha}(\Gamma)T_C, &&  \{\Theta_\alpha,\, \Theta_\beta \} = \Delta_{\alpha\beta}(\Gamma)+f_{\alpha\beta}^\gamma(\Gamma)\Theta_\gamma,\\
     &\label{hamconstrp}  \{H_0,\,\Theta_\alpha\} = 0   ,  &\{H_0,\,T_A\} =V^B_A(\Gamma)T_B, &&
  \end{align}
where structure functions $\Delta_{\alpha\beta}(\Gamma)$, compose invertible (on the surface, $T_A=\Theta_\alpha = 0$ in $\mathcal{M}$)  supermatrix.
The dynamics and gauge transformations   are described by the equations
(\ref{Eqham}).

Because of, the algebra of the functions  $(T_A,\,H_0)$ is closed with respect to $\{\ ,\ \}$-multiplication, we may restrict ourselves  by choice of
 the total phase space $\mathcal{M}_{\mathrm{tot}}$, ($\mathcal{M} \subset \mathcal{M}_{\mathrm{tot}}$)  underlying the generalized
canonical  quantization \cite{BFV}, which is parameterized (for linearly independent  constraints $T_A, \Theta_\alpha$) by the canonical
phase-space variables, $\Gamma^{\mathbf{P}}_{T}= \left(  \Gamma^{p}, \Gamma_{\mathrm{gh}
}\right)$, $\varepsilon(\Gamma^{\mathbf{P}}_{T})=\varepsilon_{\mathbf{P}}%
$, (\ref{tpsv}), with Grassmann and ghost number distributions (\ref{tpsv1}). subject to (\ref{ghPbr}). In this case the generating functional of Green's functions (and respective partition function for $I_{\mathbf{P}}(t)=0$) has the form (\ref{ZPI}), (\ref{ZPI1})%
$I_{\mathbf{P}}(t)$ to ${\Gamma}_T^{\mathbf{P}}$, but with another
unitarizing Hamiltonian $H_{r|\Psi_r}(t)=H_{r|\Psi_r}({\Gamma}_T(t))$ to be determined by
three $t$-local functions:  $\mathcal{H}_r(t)$ with $(\varepsilon,{gh})(\mathcal{H}_r)=(0,0)$,
  $\Omega_r(t)$, with $(\varepsilon,{gh})(\Omega_r)=(1,1)$, and  $\Psi_r(t)$, with $(\varepsilon,{gh})(\Psi)=(1,-1)$ given by the equations in terms of Poisson (instead of Dirac in (\ref{Hphi})--(\ref{DiracBracket})) superbracket: %
\begin{align}
&  H_{r|\Psi}(t)=\mathcal{H}_r(t)+ \left\{
\Omega_r(t), \Psi(t)\right\} _{t}%
 ,
& \mathrm{for} \ \   \left\{  \Omega_r,\Omega_r\right\} =  0\ ,\ \ \left\{  \mathcal{H}_r%
,\Omega_r\right\}  = 0\ , \label{HOmegar}
\end{align}
with simplest choice for the gauge Fermion $\Psi$ determined by (\ref{gaugefer}).
The solutions for the generating  equations (\ref{HOmegar}) are given by the expressions (\ref{Hrep}), (\ref{Orep}) but with strong equality:
in the form of series in powers of \emph{ minimal} ghost coordinates and momenta $C^A, \overline{\mathcal{P}}_A$ with use of $C \overline{\mathcal{P}}$-ordering  up to the second order in $\Gamma_{\mathrm{gh}}$:
з\begin{eqnarray}\label{Hrepr}
 \mathcal{H}_r  &= & H_0 + (-1)^{\varepsilon_C} C^AV^C_{A}(\Gamma)\overline{\mathcal{P}}_C + O(C^2),   \\
   \Omega_r &= & \Omega_{r|\min}+ \pi_A \mathcal{P}^A\ = \  C^A\Big(T_A+ \frac{1}{2}(-1)^{\varepsilon_C+\varepsilon_A} C^Bf^C_{BA}(\Gamma)\overline{\mathcal{P}}_C + O(C^2)\Big) + \pi_A \mathcal{P}^A ,\label{Orepr}
\end{eqnarray}
which encode in $\mathcal{H}_r$ and $\Omega_{r|\min}=\Omega_{r|\min}(\Gamma, \Gamma_{gh|m})$ the structure functions of the first-class constraints system $T_A$.

Let us introduce  new (additional) equations in $\mathcal{M}_{\mathrm{tot}}$ without using  Dirac superbracket due to the presentation (\ref{firstclassconstrp}), (\ref{hamconstrp}):
\begin{align}\label{eqQctotgen}
  &  \{\Omega_{r|\min},\,  \widehat{\Theta}_\alpha\}    = 0, & \{\mathcal{H}_r,\, \widehat{\Theta}_\alpha\}    = 0 ,   && \ (\varepsilon, gh)\widehat{\Theta}_\alpha = (\varepsilon_\alpha, 0),
\end{align}
which we call by the \emph{generating equations} for superalgebra of the  BRST extended in $\mathcal{M}_{\mathrm{tot}}$  second-class constraints $\widehat{\Theta}_\alpha = \widehat{\Theta}_\alpha(\Gamma, \Gamma_{gh|m})$ with boundary condition
\begin{equation}\label{boundext}
  \widehat{\Theta}_\alpha(\Gamma, 0) = {\Theta}_\alpha(\Gamma).
\end{equation}
The solution of the first equations in (\ref{eqQctotgen}), being analogous as one for the latter equation in (\ref{HOmegar}) exists in the form of series in powers of \emph{ minimal} ghost coordinates and momenta $C^A, \overline{\mathcal{P}}_A$ with use of $C \overline{\mathcal{P}}$-ordering  up to the second order in $\Gamma_{\mathrm{gh}}$ because of any $p$-times applied  Poisson bracket from:
\begin{eqnarray}\label{t1nth}
 && \left\{T_{A_1},\left\{T_{A_2}.\, \left\{,,,,\, , \left\{T_{A_p}.\,{\Theta}_\alpha\right\}...\right\}\right\}\right\} =   f^A_{A_1...A_p \alpha}(\Gamma) T_{A}, \\
\label{t1nHth}&& \left\{T_{A_1},\left\{T_{A_2}.\, \left\{,,,,\, , \left\{T_{A_p-1},\,\left\{H_0.\,{\Theta}_\alpha\right\}\right\}...\right\}\right\}\right\} =   f^A_{A_1...A_{p-1} \alpha}(\Gamma) T_{A},
\end{eqnarray}
for $p=2,...$ is proportional to the constraints $T_A$ with some regular $f's$. Hence, we have
\begin{equation}\label{extBRSTconstr}
  \widehat{\Theta}_\alpha(\Gamma, \Gamma_{gh|m}) = {\Theta}_\alpha(\Gamma)+ (-1)^{\varepsilon_\alpha+\varepsilon_C}  C^Af^C_{A\alpha}(\Gamma) \overline{\mathcal{P}}_C + O(C^2).
\end{equation}
The validity of the second equations in (\ref{eqQctotgen})  for commutativity of $\widehat{\Theta}_\alpha$ with   $\mathcal{H}_r$ is considered as the restrictions on the form of $ \widehat{\Theta}_\alpha$ and
will follow up to the second order in $C^A, \overline{\mathcal{P}}_A$, first, from (\ref{firstclassconstrp}), (\ref{hamconstrp}), second,  from  the Jacobi  identity for Poisson brackets of $H_0, T_A, \Theta_\alpha$:
\begin{eqnarray}
  \label{jcident3} && \left\{\left\{H_0,\,\Theta_\alpha\right\},\,T_A\right\}+(-1)^{\varepsilon_A\varepsilon_\alpha}
  \left\{\left\{T_A,\,H_0\right\},\,\Theta_\alpha\right\} +\left\{\left\{\Theta_\alpha,\,T_A\right\},\,H_0\right\}=0, \\
&&  \Longleftrightarrow  (-1)^{\varepsilon_A\varepsilon_\alpha} \Big( \left\{H_0,\,f^{C}_{A\alpha}\right\}+ V_A^Bf^{C}_{B\alpha} +f^{B}_{A\alpha}V_B^C +(-1)^{\varepsilon_C\varepsilon_\alpha}\left\{V_A^C,\,\Theta_{\alpha}\right\} \Big)T_C \ =\  0, \label{jcident31}
\end{eqnarray}
 whose resolution (as it was shown for the first-class constraints in \cite{bf})  means the presence of new structural functions $f^{CD}_{A\alpha}(\Gamma)$, $f^{C\beta}_{A\alpha}(\Gamma)$  given on $\mathcal{M}$ as follows
\begin{eqnarray} \label{resjcident31}
  &&  \left\{H_0,\,f^{C}_{A\alpha}\right\}+ V_A^Bf^{C}_{B\alpha} +f^{B}_{A\alpha}V_B^C +(-1)^{\varepsilon_C\varepsilon_\alpha}\left\{V_A^C,\,\Theta_{\alpha}\right\} = -f^{CD}_{A\alpha}(\Gamma)T_D,\\
  && \ \ \mathrm{with} \ f^{CD}_{A\alpha}(\Gamma) \ = \  -(-1)^{\varepsilon_C\varepsilon_D}f^{DC}_{A\alpha}  \ = \  0, \label{resjcident311}
\end{eqnarray}
for the dynamical system in question.
The additional terms (proportional to the second order in $(C^A)^2$ in the searched-for  BRST extended constraints $\widehat{\Theta}_\alpha$ (\ref{extBRSTconstr}) should correspond to new  structural functions, analogous to one in the right  (\ref{resjcident31}) which follows from  the resolution of the Jacobi identity with Poisson brackets for $p=2$ in (\ref{t1nth}) of the form
\begin{eqnarray}
  \label{jcident5}\hspace{-0.9em} &\hspace{-0.7em}&\hspace{-0.7em} (-1)^{\varepsilon_A\varepsilon_B}\left\{\left\{T_B,\Theta_\alpha\right\},T_A\right\}+(-1)^{\varepsilon_A\varepsilon_\alpha}
  \left\{\left\{T_A,T_B\right\},\Theta_\alpha\right\} +(-1)^{\varepsilon_B\varepsilon_\alpha}\left\{\left\{\Theta_\alpha,T_A\right\}, T_B\right\}=0,
\end{eqnarray}
with two upper indices $f_{AB\alpha}^{CD}$. If  $f_{AB\alpha}^{CD}=0$ then the presentation in (\ref{extBRSTconstr}) is final without $O(C^2)$, but with possible $O(C^3)$ terms, whose presence or absence  should be described analogously.
To establish  the algebra of $\widehat{\Theta}_\alpha$ with respect to the Poisson superbracket  one should to study the Jacobi identity of the form (\ref{jcident5}) but for the quantities  $T_A,\, {\Theta}_\alpha,\, , {\Theta}_\beta$
\begin{eqnarray}
  \label{jcident6} \hspace{-1.0em}&\hspace{-0.9em}&\hspace{-0.9em} (-1)^{\varepsilon_A\varepsilon_\beta}\left\{\left\{T_A,\Theta_\alpha\right\},{\Theta}_\beta\right\}+(-1)^{\varepsilon_\beta\varepsilon_\alpha}
  \left\{\left\{{\Theta}_\beta,T_A\right\},\Theta_\alpha\right\} +(-1)^{\varepsilon_A\varepsilon_\alpha}\left\{\left\{\Theta_\alpha,{\Theta}_\beta\right\},T_A\right\}=0\\
\label{jcident61}   &\hspace{-0.7em}&\hspace{-0.7em} \Rightarrow (-1)^{\varepsilon_A\varepsilon_\beta}\Bigg( \Big[\big(f_{A\alpha}^D f_{D\beta}^C+ (-1)^{\varepsilon_C\varepsilon_\beta}\left\{f_{A\alpha}^C,\,\Theta_\beta\right\} \big)  - (-1)^{\varepsilon_\alpha\varepsilon_\beta}(\alpha\longleftrightarrow \beta) \Big]  \\
\hspace{-0.5em} &\hspace{-0.8em}&\hspace{-0.8em} \phantom{\Rightarrow} - (-1)^{\varepsilon_A(\varepsilon_\alpha+\varepsilon_\beta+\varepsilon_\gamma)} f^\gamma_{\alpha\beta}f_{A\gamma}^C\Bigg)T_C  = (-1)^{\varepsilon_A\varepsilon_\beta}\left\{ T_A,\, \Delta_{\alpha\beta}\right\}+ (-1)^{\varepsilon_A\varepsilon_\beta} \left\{T_A,\, f^\gamma_{\alpha\beta}\right\}\Theta_\gamma. \nonumber
\end{eqnarray}
Due to the linear independence of $T_C$ and $\Theta_\gamma$ on the surface $T_C=\Theta_\gamma=0$ it follows from the right-hand side of (\ref{jcident61}) it follows that there exist
structure functions, $\Delta^C_{\alpha\beta A}(\Gamma)$, $\Delta^\gamma_{\alpha\beta A}(\Gamma)$, regular (i.e. smooth) in $\mathcal{M}$
\begin{equation}\label{resjcident61}
  \left\{ T_A,\, \Delta_{\alpha\beta}\right\} = - \Delta^C_{\alpha\beta A}(\Gamma)T_C -\Delta^\gamma_{\alpha\beta A}(\Gamma)\Theta_\gamma,\ \ \big(\Delta^C_{\alpha\beta A}, \Delta^\gamma_{\alpha\beta A}\big)\ =\ -(-1)^{\varepsilon_\alpha\varepsilon_\beta}\big(\Delta^C_{\beta\alpha A}, \Delta^\gamma_{\beta\alpha A}\big)
\end{equation}
with $\varepsilon\big(\Delta^C_{\alpha\beta A}, \Delta^\gamma_{\alpha\beta A}\big) = \varepsilon_\alpha+\varepsilon_\beta +\varepsilon_A + (\varepsilon_C,\,\varepsilon_\gamma)$ so that the resolution of the equations  (\ref{jcident61}) looks as
\begin{eqnarray}\label{resjcident62}   \hspace{-0.5em}&\hspace{-0.9em}& \hspace{-0.9em} \left(\hspace{-0.2em} \Big[\big(f_{A\alpha}^D f_{D\beta}^C+ (-1)^{\varepsilon_C\varepsilon_\beta}\left\{f_{A\alpha}^C,\,\Theta_\beta\right\} \big)  - (-1)^{\varepsilon_\alpha\varepsilon_\beta}(\alpha\longleftrightarrow \beta) \Big]- (-1)^{\varepsilon_A(\varepsilon_\alpha+\varepsilon_\beta+\varepsilon_\gamma)} f^\gamma_{\alpha\beta}f_{A\gamma}^C\hspace{-0.2em}\right)  \\
\hspace{-0.5em} &\hspace{-0.7em}&\hspace{-0.7em} \phantom{\Rightarrow}  + \Delta^C_{\alpha\beta A}(\Gamma) = f^{CD}_{\alpha\beta A}(\Gamma)T_D, \nonumber \\
 \hspace{-0.5em}&\hspace{-0.5em}& \hspace{-0.5em}\label{resjcident63}   \left\{T_A,\, f^\gamma_{\alpha\beta}\right\}- \Delta^\gamma_{\alpha\beta A}(\Gamma) \ =\ f^{\gamma\delta}_{\alpha\beta A}(\Gamma)\Theta_\delta .
\end{eqnarray}
In (\ref{resjcident62}), (\ref{resjcident63}) the new structure regular functions, $f^{CD}_{\alpha\beta A},f^{\gamma\delta}_{\alpha\beta A}$,  of the algebra of mixed-class constraints are introduced, with the properties:
 \begin{eqnarray}\label{resjcident64}
 \hspace{-1.0em}&\hspace{-1.0em}&\hspace{-1.0em} \big(f^{CD}_{\alpha\beta A}, f^{\gamma\delta}_{\alpha\beta A}\big) =\hspace{-0.1em} -(-1)^{\varepsilon_\alpha\varepsilon_\beta}\big(f^{CD}_{\beta\alpha A}, f^{\gamma\delta}_{\beta\alpha A}\big),\, \big(f^{CD}_{\alpha\beta A},f^{\gamma\delta}_{\beta\alpha A}\big)=\hspace{-0.1em}-\big((-1)^{\varepsilon_C\varepsilon_D}f^{DC}_{\alpha\beta A},(-1)^{\varepsilon_\gamma\varepsilon_\delta}f^{\delta\gamma}_{\beta\alpha A}\big),\\
 \hspace{-0.8em} &\hspace{-0.5em}&\hspace{-0.8em} \quad  \varepsilon\big(f^{CD}_{\alpha\beta A}, f^{\gamma\delta}_{\alpha\beta A}\big) = \varepsilon_\alpha+\varepsilon_\beta+ \varepsilon_A + \big(\varepsilon_C+\varepsilon_D,\,\varepsilon_\gamma+\varepsilon_\delta\big). \nonumber
\end{eqnarray}
Therefore, we have from the representation (\ref{extBRSTconstr}) with account for (\ref{firstclassconstrp}),(\ref{secclassconstr}), (\ref{resjcident62}):
\begin{eqnarray} \label{extPB2}
 &&  \left\{\widehat{\Theta}_\alpha,\,\widehat{\Theta}_\beta\right\} = \widehat{\Delta}_{\alpha\beta}(\Gamma,\Gamma_{gh|m}) + f^{\gamma}_{\alpha\beta }\widehat{\Theta}_\gamma + (-1)^{\varepsilon_\alpha+\varepsilon_\beta+ \varepsilon_C}  C^A f^{CD}_{\alpha\beta A}(\Gamma) T_D \overline{\mathcal{P}}_C + O(C^2), \\
 && \ \mathrm{with} \ \ \widehat{\Delta}_{\alpha\beta}(\Gamma,\Gamma_{gh|m}) = {\Delta}_{\alpha\beta}(\Gamma) - (-1)^{\varepsilon_\alpha+\varepsilon_\beta+ \varepsilon_C}  C^A \Delta^{C}_{\alpha\beta A}  \overline{\mathcal{P}}_C + O(C^2),\label{extPB2Delta}
\end{eqnarray}
that means the BRST invariant extension of the second-class constraints ${\Theta}_\alpha$ satisfies to the same Poisson bracket relations as ones in (\ref{firstclassconstrp}) for $f^{CD}_{\alpha\beta A}(\Gamma) =0$, but in the phase-space $\mathcal{M}_{\min} \subset \mathcal{M}_{\mathrm{tot}}$ with accuracy up to second order in ghost coordinates $C^A$. The BRST-invariant extension of the invertible supermatrix $\|{\Delta}_{\alpha\beta}(\Gamma)\|$ (\ref{secclassconstr}): supermatrix $\|{\widehat{\Delta}}_{\alpha\beta}(\Gamma,\Gamma_{gh|m})\|$, obeys to the same
  generalized antisymmetry property and is invertible as well on the surface $\widehat{\Theta}=0$ in $\mathcal{M}_{\min}$ due to the representation:
 \begin{eqnarray}
    && \mathrm{sdet} \|\widehat{\Delta}_{\alpha\beta}(\Gamma,\Gamma_{gh|m})\|\big|_{\widehat{\Theta}_\alpha = 0} =  \mathrm{sdet} \|{\Delta}_{\alpha\beta}(\Gamma)\|\times \nonumber\\
    &&  \times \mathrm{sdet} \|(\delta^{\alpha}_{\beta}- (-1)^{\varepsilon_\sigma+\varepsilon_\beta+ \varepsilon_C}({\Delta}^{-1})^{\alpha\sigma}    C^A \Delta^{C}_{\sigma\beta A}  \overline{\mathcal{P}}_C + O(C^2) \|\big|_{\widehat{\Theta}_\alpha = 0} \ne 0.
   \label{extsecclassconstr}\end{eqnarray}
 If the superalgebra of the constraints $T_A, \Theta_\alpha, H_0$ with respect to the Poisson superbracket  appears by Poisson-Lie one, i.e. all the structure functions in (\ref{firstclassconstrp}), (\ref{hamconstrp}) are constants, the form of BRST-invariant constraints  $\widehat{\Theta}_\alpha$ in (\ref{extBRSTconstr}) is exact without terms $O(C^2)$ with unchanged elements, ${\widehat{\Delta}}_{\alpha\beta}={\Delta}_{\alpha\beta}$.

 In general case, the algebra of the second-class constraints $\widehat{\Theta}_\alpha$ is open by first-class constraints term $f^{CD}_{\alpha\beta A}(\Gamma) T_DA$.
 From the Jacobi identities for the Poisson superbrackets of $\Omega_{r|\min}, \widehat{\Theta}_\alpha, \widehat{\Theta}_\beta$ and of $\mathcal{H}_{r}, \widehat{\Theta}_\alpha, \widehat{\Theta}_\beta$ it follows that $\big\{\widehat{\Theta}_\alpha,\,\widehat{\Theta}_\beta\big\}$ satisfy the same generating equations (\ref{eqQctot}) as ones for $\widehat{\Theta}_\alpha$.

Note, that considered in the Section~\ref{constr-self} case of
BRST-extended second-class constraints for the constrained
BRST--BFV Lagrangian formulation for half-integer HS fields with
generating equations (\ref{eqQctot}) for the superalgebra   of
$Q_C$, spin operators $\widehat{\sigma}{}^i_C(g)$ and extended in
$\mathcal{H}_C$  off-shell constraints
$\widehat{T}_i,\,\widehat{T}_{rs}$ and theirs hermitian
conjugated, $\widehat{T}{}^+_i,\,\widehat{T}{}^+_{rs}$:
    \begin{equation}\label{solextconstr+}
 \Big(\widehat{T}{}^+_i,\,\widehat{T}{}^+_{rs}\Big)=  \Big(t_i^+ +\imath \eta^+_ip_0 -2q_0 \mathcal{P}^+_i ,\, t^+_{rs} -\mathcal{P}^+_s\eta_r -  \eta^+_s\mathcal{P}_r  \Big),
\end{equation}
in terms of supercommutators, in fact for vanishing hamiltonian $H_0=\mathcal{H}_r =0$, repeats the construction developed in the Appendix. Remind, that $\widehat{\sigma}{}^i_C(g)$ composes the invertible supermatrix $\|\widehat{\Delta}_{ab}(\widehat{\sigma}{}^i_C(g))\|$  which may be explicitly constructed with respect to one $\|\Delta_{ab}(g_0^i)\|$ in (\ref{inconstraintsd}) by change of $g_0^i$ (\ref{gocond}) on $\widehat{\sigma}{}^i_C(g)$ (\ref{extconstr}), (\ref{solextconstr}).

Thus, instead of the functional measure,  $d \mu({\Gamma}_T)$ (\ref{ZPI1}), in the definition of the generating functional $Z_\Psi\left(  I\right)$  (\ref{ZPI}) one should use  measure $d \widehat{\mu}({\Gamma}_T)$ in
\begin{eqnarray}
&& Z_{\Psi}\left(  I\right)  =\int d \widehat{\mu}({\Gamma}_{T})\exp\left\{  \frac{i}{\hbar}\int
dt\left[  \frac{1}{2}\Gamma^{\mathbf{P}}_{T}(t){\omega}_{T|\mathbf{P}\mathbf{Q}}\dot{\Gamma}{}_{T}^{\mathbf{Q}}(t)-H_{r|\Psi
}(t)+I(t){\Gamma}_{T}(t)\right]  \right\}, \label{ZPIr}\\
 && \  d \widehat{\mu}({\Gamma}_T)  = d {\Gamma}_T  \delta(\widehat{\Theta}) \mathrm{sdet}^{\frac{1}{2}}\| \{\widehat{\Theta}_\alpha, \widehat{\Theta}_\beta\}\|, \ d \widehat{\mu}({\Gamma}_T) = \prod_t d \widehat{\mu}({\Gamma}_T(t)),  \label{ZPI1r}%
\end{eqnarray}
being invariant with respect to another [than in (\ref{Binf})]  BRST transformations with generators $\overleftarrow{s}_r =\left\{\bullet , \Omega_r \right\}$.

On the operator level,  we  suppose, first,  that the set of $\Theta_\alpha$ admits the splitting  (and therefore for  BRST-extended constraints $\widehat{\Theta}_\alpha$) as well,  at least, locally, on two subsystems, which of them appears by the first-class constraints:
\begin{equation}\label{splittingapp}
\Theta_\alpha(\Gamma) \rightarrow \Theta^{\prime}_\alpha(\Gamma) = \Lambda^\beta_\alpha (\Gamma) \Theta_\beta(\Gamma)= \big(\theta_{\bar{\alpha}}, \theta_{\underline{\alpha}}\big), \ \  \mathrm{sdet}\|\Lambda^\beta_\alpha \|\vert_{(\Theta_\alpha = T_A=0)}\ne 0 ,
\end{equation}
   for the index $\alpha$ division:  $\alpha = (\bar{\alpha}, \underline{\alpha})$    with $\bar{\alpha} = 1,...,\frac{1}{2}m$ and $\underline{\alpha}=\frac{1}{2}m+1,...,m$ such that  each subsystems $\theta_{\bar{\alpha}}$, $\theta_{\underline{\alpha}}$.  subject to the relations (\ref{split}), (\ref{split1}) between themselves and with specific Poisson brackets with $H_0$ and $T_A$ following from (\ref{firstclassconstrp}), (\ref{hamconstrp}):
  \begin{eqnarray}\label{splitapp1}
    && \{T_A, \theta_{\bar{\alpha}}\} =  \bar{f}{}^C_{A\bar{\alpha}}(\Gamma)T_C,\quad \{T_A, \theta_{\underline{\alpha}}\} =  \underline{f}{}^C_{A\underline{\alpha}}(\Gamma)T_C,\\
    && \label{splitapp2} \{\theta_{\bar{\alpha}},\, \theta_{\bar{\beta}}\} = \bar{f}{}^{\bar{\gamma}}_{{\bar{\alpha}}{\bar{\beta}}}(\Gamma) \theta_{\bar{\gamma}},\quad \{\theta_{\underline{\alpha}},\, \theta_{\underline{\beta}}\} = \underline{f}{}^{\underline{\gamma}}_{{\underline{\alpha}}{\underline{\beta}}}(\Gamma) \theta_{\underline{\gamma}}, \\
    && \label{splitapp3}\{\theta_{\bar{\alpha}},\, \theta_{\underline{\beta}}\} = \check{\Delta}_{\bar{\alpha}\underline{\beta}}(\Gamma)+\check{f}_{\bar{\alpha}\underline{\beta}}^{\bar{\gamma}}(\Gamma)\theta_{\bar{\gamma}} + \check{f}_{\bar{\alpha}\underline{\beta}}^{\underline{\gamma}}(\Gamma)\theta_{\underline{\gamma}}, \quad \{H_0, \, \theta_{\bar{\alpha}}\} =  \{H_0, \, \theta_{\underline{\alpha}}\} = 0,
  \end{eqnarray}
  with invertible
$\|\check{\Delta}_{\alpha\beta}(\Gamma)\|\big|$ satisfying to the relations (\ref{split2}).
\vspace{1ex}

 In the  Hilbert space  $H({Q_r})=H_{\Gamma}\otimes H_{gh|m}$ for the minimal sector,    [with the correspondence $\big(\Gamma^p$, $\hat{C}^A, \hat{\overline{P}}_A\big) \to$ $\big(\hat{\Gamma}{}^p, \hat{C}{}^A, \hat{\overline{P}}_{A}\big)$: $[\hat{\Gamma}{}^p, \hat{\Gamma}{}^q \} = \imath\hbar \omega^{pq} $,  and $[\hat{C}{}^A, \hat{\overline{P}}_{B}\}=\imath \hbar \delta^A_B$,   $\omega^{pq}=\mathrm{const}$]   the nilpotency for the  operator
 \begin{equation}\label{Qrpmega}
   Q_r(\hat{\Gamma},\hat{\Gamma}_{gh|m}) = \Omega_{r|\min}({\Gamma},{\Gamma}_{gh|m})\vert_{({\Gamma},{\Gamma}_{gh|m})\to (\hat{\Gamma},\hat{\Gamma}_{gh|m})}:  \quad Q_r^2 = (1/2)[Q_r,\,Q_r\}=0
 \end{equation}
holds,   with representation respecting the division for $\Theta_\alpha$,  and  the choice of some $qp-$, $\hat{C}\hat{\overline{P}}$-orderings for $\Theta_\alpha(\hat{\Gamma})$, $ Q_r$ and rest operator functions. The operator form for the  \emph{generating equations} (\ref{eqQctotgen}), in addition to commutativity of $Q_r$ with operator of the Hamiltonian, $H_r=\mathcal{H}_r(\hat{\Gamma},\hat{\Gamma}_{gh|m})$: $\left[Q_r,\,{H}_r\right\}=0$ looks as:
    \begin{align}\label{eqQctotgenop}
  &  \left[Q_r,\,  \widehat{\Theta}_\alpha(\hat{\Gamma},\hat{\Gamma}_{gh|m})\right\}    = 0, & \left[{H}_r,\, \widehat{\Theta}_\alpha(\hat{\Gamma},\hat{\Gamma}_{gh|m})\right\}    = 0 ,   && \ (\varepsilon, gh)\widehat{\Theta}_\alpha = (\varepsilon_\alpha, 0),
\end{align}
    for superalgebra of $Q_r, H_r, \widehat{\Theta}_\alpha(\hat{\Gamma},\hat{\Gamma}_{gh|m}) $ which determines the  BRST extended operator  second-class constraints $\widehat{\Theta}_\alpha(\hat{\Gamma},\hat{\Gamma}_{gh|m}) $ and, in particular, the first-class subsystem $\widehat{\theta}_{\bar{\alpha}}(\hat{\Gamma},\hat{\Gamma}_{gh|m})$. The latter, according to (\ref{splittingapp}), (being true for  $\widehat{\Theta}{}^{\prime}_\alpha$) have the form
\begin{equation}\label{extBRSTconstrop}
  \widehat{\theta}_{\bar{\alpha}}(\hat{\Gamma},\hat{\Gamma}_{gh|m}) = {\theta}_{\bar{\alpha}}(\hat{\Gamma}) + (-1)^{\varepsilon_{\bar{\alpha}}+\varepsilon_C}  \hat{C}{}^A\overline{f}{}^C_{A\bar{\alpha}}(\hat{\Gamma}) \hat{\overline{\mathcal{P}}}_C + O(\hat{C}{}^2),
\end{equation}
where for $\Lambda^\beta_\alpha (\Gamma) = \delta^\beta_\alpha$ the equality: ${f}^C_{A\bar{\alpha}}=\overline{f}{}^C_{A\bar{\alpha}}$ is true. As it follows from the involution relations for  ${\theta}_{\bar{\alpha}}(\Gamma)$  (\ref{splitapp2}) and  the $[\ ,\ \}$-analog of the Poisson brackets for $ \big(\widehat{\Theta}_{{\alpha}}, \widehat{\Theta}_{{\beta}}\big)({\Gamma},{\Gamma}_{gh|m})$ (\ref{extPB2}) the superalgebra of the operators $Q_r, H_r,  \widehat{\theta}_{\bar{\alpha}}(\hat{\Gamma},\hat{\Gamma}_{gh|m})$ is closed with respect to $[\ ,\ \}$-multiplication:
    \begin{align}\label{eqQctotgenop2}
  &  \left[Q_r,\,  \widehat{\theta}_{\bar{\alpha}}(\hat{\Gamma},\hat{\Gamma}_{gh|m})\right\}    = 0, & \left[{H}_r,\, \widehat{\theta}_{\bar{\alpha}}(\hat{\Gamma},\hat{\Gamma}_{gh|m})\right\}    = 0 ,   &&   \left[\widehat{\theta}_{\bar{\alpha}},\,\widehat{\theta}_{\bar{\beta}}\right\} = \bar{f}{}^{\bar{\gamma}}_{{\bar{\alpha}}{\bar{\beta}}}(\hat{\Gamma})\widehat{\theta}_{\bar{\gamma}},
\end{align}
  for vanishing the functions  $f^{CD}_{\bar{\alpha}\bar{\beta} A}(\hat{\Gamma})$ in the operator analog of (\ref{resjcident62}) which follows from the Jacobi identity  (\ref{jcident6}), but with respect to supercommutator, $[\ , \}$, and for $T_A(\hat{\Gamma}), {\theta}_{\bar{\alpha}}(\hat{\Gamma}), {\theta}_{\bar{\beta}}(\hat{\Gamma})$.

 The same superalgebra (\ref{eqQctotgenop2}) will be obviously true for the set of operators $Q$, $H_{r|\Psi}(\hat{\Gamma}_{T})$ = $H_r + (\imath\hbar)^{-1}[Q,\,\Psi\}$,  $\widehat{\theta}_{\bar{\alpha}}(\hat{\Gamma},\hat{\Gamma}_{gh|m})$  which corresponds to the total BRST operator, $Q\hspace{-0.1em} = \hspace{-0.1em}\Omega_r(\Gamma_T)\vert_{\Gamma_T\to\hat{\Gamma}_T}$, and to unitarizing Hamiltonian $H_{r|\Psi}(\hat{\Gamma}_{T})$ with gauge Fermion, $\Psi(\hat{\Gamma}_T)$ in the total, for BFV method \cite{bf}, Hilbert space, $H(Q)=H({Q_r})\otimes H_{gh|nm}$. Here the Hilbert subspace $H_{gh|nm}$
 generated by the rest fictitious operator pairs, $\hat{\overline{C}}_A, \hat{\mathcal{P}}{}^A$, $\hat{\pi}_{A}, \hat{\lambda}^{A}\ $  with non-vanishing supercommutators:
\begin{equation}\label{ghcuperc}
  \big[\hat{\overline{C}}{}_{A},\, \hat{\mathcal{P}}{}^{B
}\big\}= \imath\hbar\delta_A^B, \ \ \big[\hat{\pi}_{A},\, \hat{\lambda}{}^{B}\big\}=\imath\hbar\delta^B_A.
\end{equation}

  Following to the results of the Sections~\ref{BRST--BFVorig}, \ref{constr-unconstr} with Corollaries~2, 3 and in agreement  with Dirac approach \cite{Dirac}, \cite{Dirac1}  the physical states $|\psi\rangle \in  H^{phys}_{\Gamma}$ for constrained dynamical system subject to the operator analogs of the  relations (\ref{firstclassconstrp}),(\ref{hamconstrp}) from Hilbert subspace  of physical vectors  $H^{phys}_{\Gamma}\subset H_{\Gamma}$
  may be derived according to
\vspace{1ex}

  \noindent
 \textbf{Statement 6}: A set of the states  $H_{\Theta_\alpha, T_A}$: $H_{\Theta_\alpha, T_A} \subset H({Q_r})$  with vanishing ghost number from the Hilbert  subspace: $\ker Q_r \diagup {Im }\, Q_r$,  with nilpotent BRST--BFV operator $Q_r$ (\ref{Orepr}), (\ref{Qrpmega})  for the subsuperalgebra of constraints $T_A$ acting in   $H({Q_r})$ both  being  annihilated by  the half of the BRST invariant (extended) second-class constraints $\widehat{\theta}_{\bar{\alpha}}$ and satisfying to the Schrodinger equation with Hamiltonian $H_r$  is equivalent to the set of the states from the Hilbert  subspace $H^{phys}_{\Gamma}\subset H_{\Gamma}$ annihilated by the constraints $T_A, {\theta}_{\bar{\alpha}}$ and satisfying the Schrodinger equation with Hamiltonian $H_0$:
\begin{eqnarray}\label{equivsec11fin}
       H_{\Theta_\alpha, T_A}& =& \big\{|\psi\rangle| \  \big(T_A, {\theta}_{\bar{\alpha}},\imath\hbar \partial_t -H_0(\hat{\Gamma})\vert_{\theta_{\underline{\alpha}}=0} \big) |\psi\rangle =(0, 0, 0),  \  |\psi
       \rangle \in   H_{\Gamma} \big\} \\
       &=&   \big\{|\psi_r\rangle|\ \big(\widehat{\theta}_{\bar{\alpha}},\, \imath\hbar \partial_t -H_r,\,   gh\big) |\psi_r\rangle =(0, 0, 0),  \        |\psi_r\rangle \in \ker Q_r \diagup {Im }\, Q_r \big\}. \label{equivsec21fin}
    \end{eqnarray}

Equivalently, in terms of the respective   $Q_r$-complex   the equivalence  above means that  the found set of states,   $H_{\Theta_\alpha, T_A}$ may be presented as for the  Statement 3  as:
\begin{eqnarray}\label{equivsec1fin}
       \hspace{-0.7em} H_{\Theta_\alpha, T_A}& \hspace{-0.7em}=&\hspace{-0.7em} \big\{|\chi^0_r\rangle| \  \big(Q_r,\,\widehat{\theta}_{\bar{\alpha}},\, (\imath\hbar \partial_t -H_r)\delta^{l0},\,   gh\big) |\chi^{l}_r\rangle =(\delta|\chi^{l-1}_r\rangle ,0 ,0,-l) ,  |\chi^l_r\rangle\in H^l({Q_r})   \big\}
          \end{eqnarray}
 for $|\chi^{-1}_r\rangle =|\chi^{-1}\rangle=0$, $l=0,1,..., K_r$, $K_r\in \mathbb{N}$.

Note, the space $H_{\Theta_\alpha, T_A}$ may be equivalently presented in terms of respective $Q$-complex in the Hilbert space $H(Q)=H({Q_r})\otimes H_{gh|nm}$ with help of
 the triple of operators $\big(Q, \widehat{\theta}_{\bar{\alpha}},H_{r|\Psi}\big)$ instead of $\big(Q_r, \widehat{\theta}_{\bar{\alpha}}, H_{r}\big)$ in (\ref{equivsec1fin}) but for another natural $K\geq K_r$.

The presentation for the generating functional of Green's
functions (\ref{ZPIr}), (\ref{ZPI1r}) in terms of BRST-invariant
second-class constraints for the dynamical system in question and
Statement 6  present the main results of the appendix. In case of
topological system,i.e. for $H_0=0$, the last  result is
immediately corresponds to the case of constrained BRST--BFV
approach for the Lagrangian formulations for HS fields in the
Section~\ref{constr-Lagr} without extracted from $x^\mu$ time
coordinate and with special presentation for the invertible
supermatrix $
\|\widehat{\Delta}_{\alpha\beta}\|(\hat{\Gamma},\hat{\Gamma}_{gh|m})$
(\ref{extPB2Delta}) in terms of constrained generalized spin
operator $\sigma^i_C\equiv \widehat{\sigma}{}^i_C(g)$
(\ref{extconstr}), (\ref{solextconstr}).

\end{document}